\documentclass[%
reprint,
superscriptaddress,
nofootinbib,
amsmath,amssymb,
aps,
prd,
]{revtex4-1}

\usepackage{amsmath}
\usepackage{microtype}
\usepackage{amstext}
\usepackage{graphicx}
\usepackage{booktabs}
\usepackage{amssymb}
\usepackage{enumerate}
\usepackage{esvect}
\usepackage{verbatim}
\usepackage{bbm}
\usepackage{slashed}
\usepackage{braket}
\usepackage{color}
\usepackage{graphicx}% Include figure files
\usepackage{dcolumn}% Align table columns on decimal point
\usepackage{bm}% bold math
\newcommand{\dif}{\mathrm{d}}

\newcommand{\xdify}[2]{\frac{\mathrm{d} #1}{\mathrm{d} #2}}

\newcommand{\pdif}{\partial}

\newcommand{\xpdify}[2]{\frac{\partial #1}{\partial #2}}

\newcommand{\intxtoy}[2]{\int\limits_{#1}^{#2}}

\newcommand{\defeq}{\mathrel{\mathop:}=}
\newcommand{\eqdef}{=\mathrel{\mathop:}}

\newcommand{\ofrac}[1]{\frac{1}{#1}}

\newcommand{\mz}{\mathbb{Z}}

\newcommand{\Id}{\mathbbm{1}}

\newcommand{\cross}{\times}
\newcommand{\lra}{\Leftrightarrow}

\newcommand{\com}[2]{\left[#1,#2\right]}
\newcommand{\acom}[2]{\left\{#1,#2\right\}}

\newcommand{\sumzeroxtoy}[2]{\sum\limits_{#1=0}^{#2}}

\newcommand{\at}[1]{\Big|_{#1}}

\newcommand{\anoteq}[1]{\stackrel{#1}{=}}

\newcommand{\refe}[1]{(\ref{eq:#1})}

\newcommand{\iv}{_{\textrm{v}}}
\newcommand{\Tr}{\textrm{Tr}\,}
\renewcommand{\O}{\mathcal{O}}
\newcommand{\A}{\mathcal{A}}

\newcommand{\Tc}{\mathcal{T}_\mathcal{C}}

\newcommand{\csgn}{\textrm{sgn}_\mathcal{C}}
\newcommand{\sgn}{\textrm{sgn}}

\newcommand{\W}{\mathcal{W}}
\newcommand{\E}{\mathcal{E}}
\newcommand{\F}{\mathcal{F}}
\newcommand{\K}{\mathcal{K}}
\newcommand{\M}{\mathcal{M}}
\newcommand{\C}{\mathcal{C}}
\newcommand{\N}{\mathcal{N}}
\newcommand{\T}{\mathcal{T}}
\newcommand{\D}{\mathcal{D}}
\renewcommand{\P}{\mathcal{P}}

\newcommand{\sbraket}[1]{\braket{#1}\,^{\hspace{-0.12cm}2}}

\newcommand{\nocontentsline}[3]{}
\newcommand{\tocless}[2]{\bgroup\let\addcontentsline=\nocontentsline#1{#2}\egroup}

\begin{document}
	\interfootnotelinepenalty=10000
	
	%\preprint{APS/123-QED}
	
	\title{Collisional strong-field QED kinetic equations from first principles}% Force line breaks with \\
	
	\author{Gregor Fauth}
	% \altaffiliation[Also at ]{}%Lines break automatically or can be forced with \\
	\email{fauth@thphys.uni-heidelberg.de}
	\affiliation{%
		Institut f\"ur theoretische Physik, Universit\"at Heidelberg,
		Philosophenweg 16, D-69120 Heidelberg, Germany
	}%
	
	\author{J\"{u}rgen Berges}
	\email{berges@thphys.uni-heidelberg.de}
	\affiliation{%
		Institut f\"ur theoretische Physik, Universit\"at Heidelberg, 
		Philosophenweg 16, D-69120 Heidelberg, Germany
	}%

	\author{Antonino Di Piazza}
\email{dipiazza@mpi-hd.mpg.de}
\affiliation{
	Max Planck Institute for Nuclear Physics, Saupfercheckweg 1, D-69117, Heidelberg, Germany\\
}%
%\affiliation{
	\date{\today}% It is always \today, today,

\begin{abstract}
Starting from nonequilibrium quantum field theory on a closed time path, we derive kinetic equations for the strong-field regime of quantum electrodynamics (QED) using a systematic expansion in the gauge coupling $e$. The strong field regime is characterized by a large photon field of order $\O(1/e)$, which is relevant for the description of, e.g., intense laser fields, the initial stages of off-central heavy ion collisions, and condensed matter systems with net fermion number. The strong field enters the dynamical equations via both quantum Vlasov and collision terms, which we derive to order $\O(e^2)$. The kinetic equations feature generalized scattering amplitudes that have their own equation of motion in terms of the fermion spectral function. The description includes single photon emission, electron-positron pair photoproduction, vacuum (Schwinger) pair production, their inverse processes, medium effects and contributions from the field, which are not restricted to the so-called locally-constant crossed field approximation. This extends known kinetic equations commonly used in strong-field QED of intense laser fields. In particular, we derive an expression for the asymptotic fermion pair number that includes leading-order collisions and remains valid for strongly inhomogeneous fields. For the purpose of analytically highlighting limiting cases, we also consider plane-wave fields for which it is shown how to recover Furry-picture scattering amplitudes by further assuming negligible occupations. Known on-shell descriptions are recovered in the case of simply peaked ultrarelativistic fermion occupations. Collisional strong-field equations are necessary to describe the dynamics to thermal equilibrium starting from strong-field initial conditions.
\end{abstract}

\pacs{Valid PACS appear here}
\maketitle
\tableofcontents

\section{\label{sec:1}Introduction}

Present and upcoming laser facilities \cite{APOLLON_10P,ELI,CoReLS,XCELS} promise unprecedented insights into the strong-field regime of quantum electrodynamics (QED). Strong dynamical electromagnetic fields are also generated during the early stages in off-central collisions of heavy nuclei at the Large Hadron Collider (LHC) at CERN or the Relativistic Heavy Ion Collider (RHIC) at BNL. The presence of strong electromagnetic fields and their dynamical decay can lead to a wealth of intriguing quantum phenomena, such as related to quantum anomalies which can be probed also in condensed matter systems~\cite{Kharzeev:2013ffa}. Strong fields are also essential for the description of highly charged systems, where the net fermion charge induces strong field configurations also in equilibrium \cite{Sturm_2014}. While experiments pioneered by the Stanford Linear Accelerator Center (SLAC) \cite{PhysRevLett.76.3116, PhysRevD.60.092004, PhysRevLett.79.1626} have since been developed further \cite{PhysRevX.8.011020, PhysRevX.8.031004}, they are not yet able to enter the full strong-field QED regime by means of lasers. Meanwhile, experiments employing crystals have been found to be a competitor to laser experiments \cite{RevModPhys.77.1131, PhysRevResearch.1.033014, Wistisen:2017pgr, Baier:1998vh}.

For the weak QED coupling $\alpha = e^2/4\pi\approx 1/137$ (we use natural units with $\hbar=c=k_{\textrm{B}}=\varepsilon_0=1$), the
strong-field regime may be characterized by a photon field that is parametrically as large as
\begin{align}
\label{eq:sfcon}
\A^\mu \sim \O(1/e)\, .
\end{align}
For a laser field \cite{RevModPhys.84.1177} that is described by an electric field amplitude $\mathcal{E}$ and frequency $\omega$, the counting rule \refe{sfcon} corresponds to a large (Lorentz-invariant) non-linearity parameter \cite{Ritus1985,Baier:1998vh,RevModPhys.84.1177},
\begin{align}
\label{eq:xi}
|e|\mathcal{E}/(m\omega)\gtrsim 1\, .
\end{align}
%Strong field effects can be induced by low laser field frequencies in the lab frame according to \refe{xi}, as well as by large fermion momenta which boost the field in the fermion rest frame \cite{RevModPhys.84.1177}.  
For a macroscopic photon field that varies on the time scale of the Compton length $1/m$, the counting rule \refe{sfcon} corresponds to electric fields of the order of the critical field,
\begin{align}
\E\gtrsim m^2/|e|\eqdef \mathcal{E}_{\textrm{c}}\, ,
\label{eq:Schwinger}
\end{align}
which induces electron-positron pair creation from the vacuum \cite{PhysRev.82.664, Dunne_2014, PhysRevD.82.105026, Hebenstreit:2011cr, PhysRevSTAB.14.054401}. 

Despite the smallness of the QED coupling, the theoretical description of strong field phenomena provides important challenges. Standard simulation techniques, such as based on Monte Carlo importance sampling, cannot be applied to general nonequilibrium problems. Rigorous simulations are difficult even in equilibrium in the presence of a net fermion charge leading to non-vanishing fields. As a consequence, the development of suitable approximate treatments is indispensable.

For instance, the decay times of strong electromagnetic fields in the medium created by a heavy ion collision and the role of the fields for the subsequent nonequilibrium dynamics is still poorly understood. Even the idealized problem of how an initially supercritical homogeneous electromagnetic field approaches thermal equilibrium in QED has not been answered yet. The strong field regime at early times may be accurately described by classical statistical field theory techniques \cite{Mueller:2016aao, Martin:1973zz}, while the late time behavior at high temperature in the absence of a field is successfully described using standard kinetic theory \cite{Arnold:2002zm}. In particular, the dynamics of avalanches in which large amounts of fermions are produced can be captured by a kinetic approximation of QED \cite{Bell:2008zzb, PhysRevSTAB.14.054401, Tamburini:2017sxg, Bulanov_2010, Fedotov_2010, Elkina_2011, Nerush:2010fe, Ridgers:2013, Ridgers_2014, Lobet_2015, Gelfer_2015, Grismayer_2016, Grismayer:2017foz, Jirka_2016, Gonoskov_2017, Sampath_2018, Samsonov_2019, Slade_Lowther_2019, gaisser_engel_resconi_2016full}. However, to describe in a single approach the evolution all the way from strong fields to equilibrium, or in the presence of a net fermion density, involves the interplay of strong fields and collisions beyond state-of-the-art approximations~\cite{Berges:2020fwq}. 

As an important step in this direction, we derive in this work dynamical equations for strong fields in a kinetic description including collisional processes to order $\O(e^2)$. Our {\it ab initio} derivation starts from nonequilibrium quantum field theory on a closed time path~\cite{Schwinger:1960qe, Keldysh:1964ud}. We derive coupled equations for the spatio-temporal evolution of the field expectation value and correlation functions for commutators and anti-commutators of fields using two-particle irreducible (2PI) generating functional techniques~\cite{Luttinger:1960ua,Cornwall:1974vz}. The expectation values of field commutators (anti-commutators) for bosons (fermions) describe the spectral functions of excitations, whereas their anti-commutators (commutators) characterize their transport behavior.

Applying a gradient expansion for two-point functions, we derive a kinetic description where the strong-field scattering kernel couples the transport equations for photons and fermions to an equation for the fermion spectral function. The latter includes strong-field off-shell corrections in a self-consistent way. Our description incorporates the processes of single photon emission, electron-positron pair photoproduction, vacuum pair production, their inverse processes, medium effects and contributions from the field going beyond the so-called locally-constant crossed field approximation (LCFA) \cite{RevModPhys.84.1177}. In fact, we show that our approximation order captures already the complete explicit field-dependence of the problem. To make further contact with the literature, we also consider plane-wave fields. Plane-wave degrees of freedom are identified and it is shown how to recover Furry-picture scattering amplitudes. 

Our description extends known kinetic equations commonly used in strong-field QED of intense laser fields and can be applied, in particular, to strongly inhomogeneous field configurations. Earlier approaches include collisionless approximations, e.g.\ Refs.~\cite{PhysRevD.83.065007, PhysRevD.82.105026, Vasak:1987um, Zhuang:1995pd}, such as employed to strong-field pair production by a source term \cite{Kluger:1992gb, Kluger:1998bm}. Collisional descriptions assuming subcritical or weak fields can be found in Refs.~\cite{PhysRevSTAB.14.054401, Neitz:2014hla, Neitz:2013qba, Blaizot:1999xk, Arnold:2002zm, Boyanovsky:1998pg, PhysRevE.60.4725, Bonit:1999, Jauho:1984zz, BEZZERIDES197210, HAKIM1982230}. Fermion spectral dynamics in the presence of a macroscopic field in the non-relativistic (subcritical) regime have been used in Refs.~\cite{Jauho:1984zz, 	PhysRevE.60.4725, Bonit:1999} (see also Refs. \cite{Epelbaum:2011pc, Gelis:2007pw} for strong fields in scalar theory). Collisional approaches either based on the classical statistical approximation \cite{PhysRevD.90.025016, PhysRevD.87.105006, Shi:2018sxm}, or by the use of a 
field-independent linear (`relaxation-time') collision term \cite{Bloch:1999eu} have been given. 
There are also particle-in-cell schemes \cite{Gonoskov:2014mda}, which assume the validity of the Lorentz equation between collisions and incorporate several quantum effects by strong-field scattering amplitudes \cite{Ritus1985, Heinzl:2010vg, PhysRevA.98.012134}. 

The structure of this paper is the following. We introduce the nonequilibrium equations of motion for one- and two-point correlation functions in Sec.~\ref{sec:2}. The ingredients for a kinetic limit of these equations are discussed in Sec.~\ref{sec:3}. We establish the systematics of counting couplings and gradients in the presence of a strong field and present general strong-field transport equations in Sec.~\ref{sec:4}. In Sec.~\ref{sec:5}, we point out which additional physical assumptions are necessary to reduce the collision kernels of our transport equations to various known expressions and kinetic equations in the literature and how to describe strong-field pair production in our formalism. We conclude and give an outlook in Sec.~\ref{sec:6}.

\section{\label{sec:2}Nonequilibrium QED}

\begin{figure}[h!]
\begin{center}
	\includegraphics[scale=0.20]{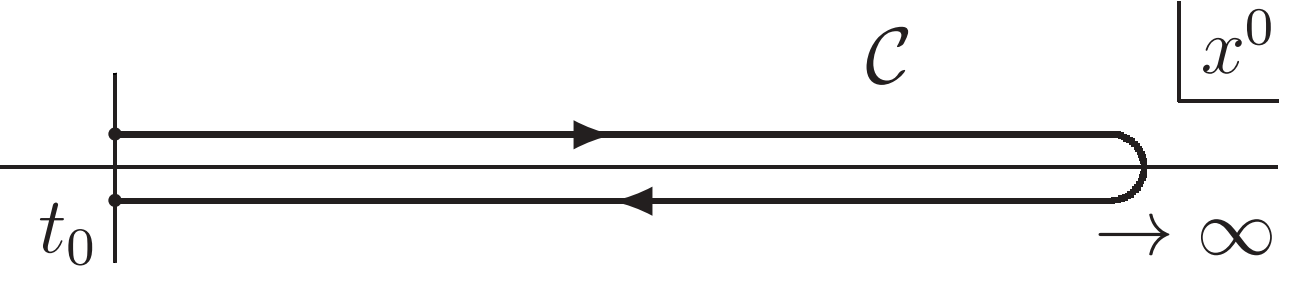}
\end{center}
	\caption{\label{fig:contour}The closed time path.}
\end{figure}

All possible information about the dynamics of quantum fields is contained in their correlation functions. The latter can be efficiently encoded in terms of a quantum effective action, which is the generating functional for time ordered field correlation functions. Here we employ the two-particle irreducible effective action $\Gamma[\mathcal{A},D,\Delta]$, which is a functional of the macroscopic field expectation value
\begin{align}
\label{eq:ensembleavg}
\A^\mu(x) = \Tr\{ \rho(t_0)A^\mu(x) \} \eqdef \braket{A^\mu(x)}\, 
\end{align}
with Heisenberg gauge field operator $A^\mu(x)$ for given density operator $\rho(t_0)$ at initial time $t_0$, as well as of the time-ordered connected two-point correlation functions
\begin{align}
\label{eq:t1}
D^{\mu\nu}(x,y)&=\braket{\Tc A^\mu(x)A^\nu(y)}-\braket{A^\mu(x)}\braket{A^\nu(y)}\, ,\\
\label{eq:t2}
\Delta(x,y)& = \braket{\Tc \Psi(x) {\bar\Psi}(y)}\, ,
\end{align}
for gauge fields and Dirac fermions with fermion field operators $\Psi$ and $\bar\Psi \defeq \Psi^\dagger \gamma^0$, where we suppress spinor indices. The expectation value of the fermion field $\Psi$ vanishes identically for the dynamics considered and plays no role in the following. The symbol $\Tc$ denotes contour time ordering on the closed time path $\C$ \cite{Keldysh:1964ud}, which starts at initial time $t_0$ and runs along the time axis and back as indicated in Fig.~\ref{fig:contour}.

\noindent Together with a non-thermal, $\rho(t_0)\neq e^{-\beta H}$, and not time-translation-invariant, $[\rho(t_0),H]\neq 0$, density matrix the contour can be used to facilitate a compact formulation of quantum field theory as an initial value problem that describes non-equilibrium physics. 

It is convenient to write the 2PI effective action as~\cite{PhysRevD.10.2428, PhysRevD.70.105010, tuprints3373, Reinosa:2009tc}
\begin{align}
\label{eq:gamman}
\Gamma[\mathcal{A},D,\Delta]&=S[\mathcal{A}]-i\Tr_\C \ln \Delta^{-1}-i\Tr_\C\Delta_0^{-1}[\A]\Delta\\
\nonumber
&+\tfrac{i}{2}\Tr_\C\ln D^{-1}+\tfrac{i}{2}\Tr_\C D_0^{-1}D+\Gamma_2[D,\Delta]\, ,
\end{align}
where $\textrm{Tr}_\C G\defeq \int_{x,\C}G(x,x)$. This identifies the pure gauge field part of the gauge-fixed classical QED action
\begin{align}
\label{eq:sdef}
S[\A]= \int_{x,\C} \Big\{ -\tfrac{1}{4}\F^{\mu\nu}(x) \F_{\mu\nu}(x)-\tfrac{1}{2\xi}\mathcal{G}^2[\A](x)\Big\}\, ,
\end{align}
with the gauge-invariant field strength tensor
\begin{align}
\label{eq:fieldtensor}
\F^{\mu\nu}(x)=\pdif^\mu\A^\nu(x)-\pdif^\nu\A^\mu(x)
\end{align}
and gauge-fixing parameter $\xi$. We use Lorenz gauge,
\begin{align}
\label{eq:Lorenz}
\mathcal{G}[\A]\defeq \pdif\cdot \A\, ,
\end{align}
and keep in mind the possibility for residual gauge-fixing. 

If computed within the 2PI loop expansion introduced below without a further kinetic limit, correlation functions such as \refe{ensembleavg} depend on the gauge-fixing parameter $\xi$ (see also Sec.~\ref{sec:4e}). This gauge-fixing dependence occurs at a higher perturbative order in the coupling than the actual approximation order \cite{PhysRevD.66.065014, Carrington:2003ut, Borsanyi:2007bf} and can be absent in the limit of on-shell photons relevant for kinetic descriptions \cite{Carrington:2007fp} (see also Eq. \refe{xilhs}). In the present paper, we discuss this in the context of Ward identities in the presence of strong fields in Sec.~ \ref{par:vertex}, where we show that the gauge-fixing parameter drops out in limiting cases.

The semi-classical or `one-loop' terms in (\ref{eq:gamman}) contain the classical photon and fermion propagators
\begin{align}
\label{eq:dzeroxi}
i D_{0,\xi}^{-1}(x,y)^{\mu\nu}&=\big[\eta^{\mu\nu} \square_x-(1-\tfrac{1}{\xi})\pdif^\mu_x\pdif^\nu_x\big]\delta_\mathcal{C}(x-y)\, ,\\
i\Delta_0^{-1}[\mathcal{A}](x,y)&=\big[i\slashed \pdif_x-e\slashed{\A}(x)-m\big]\delta_\mathcal{C}(x-y)\, 
\end{align}
in the presence of the macroscopic gauge field with $\slashed{\A} \defeq \gamma^\mu \mathcal{A}_\mu$ etc. Our metric convention is $\eta^{\mu\nu}=\textrm{diag}(+1,-1,-1,-1)$. 

\begin{figure}[h!]
	\begin{center}
		\includegraphics[scale=0.039]{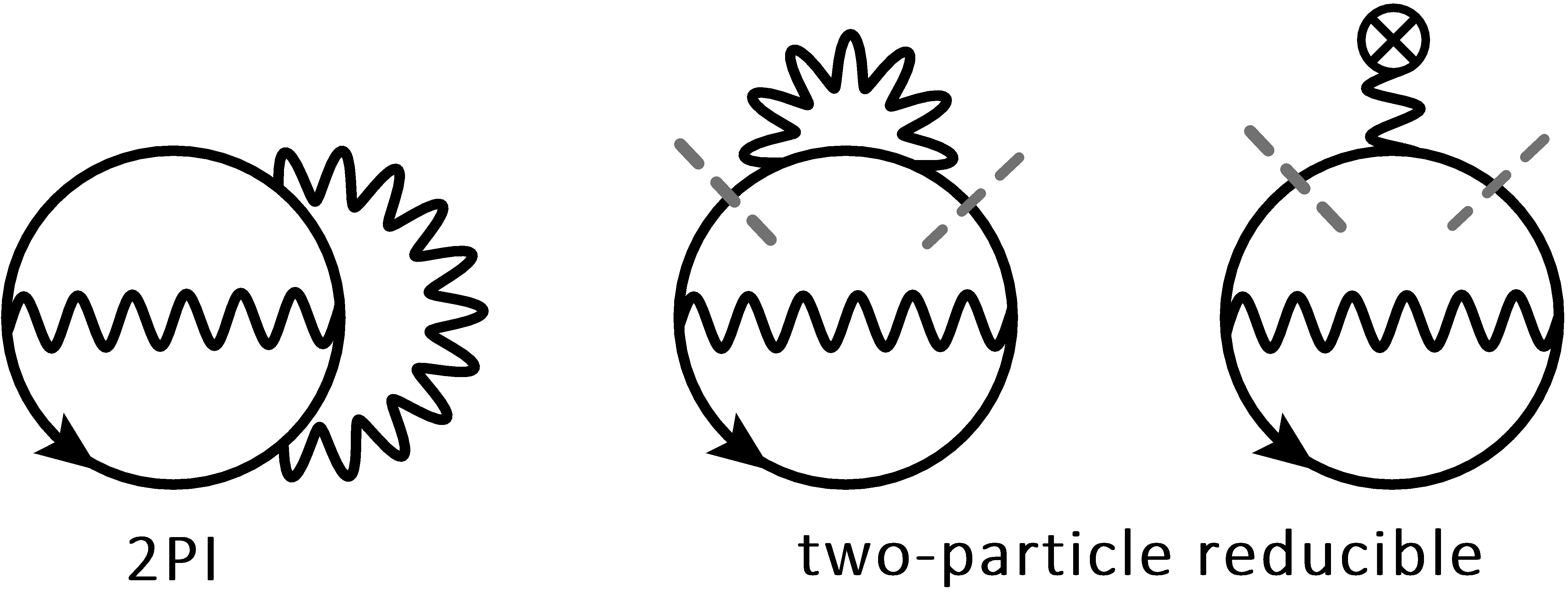}
	\end{center}
	\caption{\label{fig:2pi}Examples of 2PI and two-particle reducible diagrams.}
\end{figure}

The benefit of the decomposition identity (\ref{eq:gamman}) for the full quantum effective action $\Gamma[\mathcal{A},D,\Delta]$ is that the remaining functional $\Gamma_2[D,\Delta]$ exhibits specific properties that are very useful for the following. For QED, $\Gamma_2$ is the sum of all 2PI contributions built from the full two-point functions $D$ and $\Delta$ and there is no explicit dependence on the macroscopic field $\mathcal{A}$, which is further discussed below. A diagram is 2PI if it cannot be disconnected by cutting two propagator lines (see Fig. \ref{fig:2pi}).

\noindent The 2PI functional integral approach provides a prescription on how to close equations in terms of one- and two-point correlation functions only. Such a correlation function based description may be used to initialize the system for instance with vanishing photon and fermion particle number, described by connected two-point functions, but large electromagnetic field or vice versa. 

Furthermore, the 2PI formulation is known to facilitate a derivation of kinetic equations \cite{PhysRevD.37.2878, Hohenegger:2008zk} and may be transformed into other common formulations: Wigner transformations of 2PI two-point functions allow one to make contact with the Wigner operator formalism \cite{Vasak:1987um, Elze:1986qd, Elze:1986hq, Mrowczynski:1989bu, Mrowczynski:1992hq, Zhuang:1995pd}. In particular, equal-time Wigner functions emerge from integration over frequencies \cite{Zhuang:1995pd}. In this way one is also able to make contact with the equal-time Dirac-Heisenberg-Wigner (DHW) formalism \cite{PhysRevD.44.1825, PhysRevD.82.105026} which has been applied to the discussion of pair production from collisionless equations. Such quantum Vlasov equations \cite{Kluger:1998bm, Bloch:1999eu, Smolyansky:1997fc, Schmidt:1998vi, Blaschke:2019pnj, Hebenstreit:2008ae, PhysRevD.82.105026} emerge under the so-called `mean-field' (or `Hartree-Fock') approximation, $\Gamma_2\approx 0$. In an operator formulation, this approximation allows one to close operator equations by treating photon operators classically, at the cost of losing access to collisions. In the 2PI formulation, one can easily go beyond this mean-field order e.g.\ by means of the 2PI loop expansion as is discussed below. This way of arriving at a kinetic description starting from an effective action formulation has the additional advantage that observables derived from that effective action also become accessible under the kinetic approximation.

\subsection{\label{subsec:2c}Equations of motion}

The equations of motion for the full one- and two-point functions $\A^\mu(x)$, $D^{\mu\nu}(x,y)$, $\Delta(x,y)$ appearing in the 2PI effective action (\ref{eq:gamman}) are obtained from the stationarity conditions\footnote{These equations are valid in the absence of external source terms. Sources encoding initial conditions are stated accordingly together with the differential equations for the fields and propagators.}
\begin{align}
\label{eq:varprinz}
\frac{\delta \Gamma}{\delta \A}=0\, ,\hspace{1cm}\frac{\delta \Gamma}{\delta D}=0 \, , \hspace{1cm} \frac{\delta \Gamma}{\delta \Delta}=0\, .
\end{align}
These are coupled partial integro-differential equations for the one- and two-point functions on the closed time contour. From them emerge a Maxwell equation, and photon and electron-positron transport equations respectively.

In order to discuss the equations of motion, it is convenient to make the time ordering explicit by writing
\begin{align}
\label{eq:propdecomp1}
D^{\mu\nu}(x,y)&=F^{\mu\nu}(x,y)-\tfrac{i}{2}\rho^{\mu\nu}(x,y)\sgn_\mathcal{C}(x^0-y^0) \, ,\\
\label{eq:propdecomp2}
\Delta(x,y)&=F_\Psi(x,y)-\tfrac{i}{2}\rho_\Psi(x,y)\sgn_\mathcal{C}(x^0-y^0)\, .
\end{align}
After splitting the equations of motion into equations for  the `statistical functions' ($F$) and `spectral functions' ($\rho$), the contour $\C$ no longer appears and a clear separation into transport and spectral dynamics is achieved. These functions have distinct hermiticity properties,
\begin{align}
\label{eq:fs1}
F^{\mu\nu}(x,y)&=F^{\nu\mu}(y,x)\, ,\\
\label{eq:rs1}
\rho^{\mu\nu}(x,y)&=-\rho^{\nu\mu}(y,x)\,,\\
\label{eq:fs2}
F_\Psi(x,y)&=\gamma^0 F_\Psi^\dagger(y,x)\gamma^0 \, ,\\
\label{eq:rs2}
\rho_\Psi(x,y)&=-\gamma^0 \rho^\dagger_\Psi(y,x)\gamma^0\, .
\end{align}
These properties are related to the underlying \mbox{(anti-)commutator} representations in terms of Heisenberg field operators: 
\begin{align}
\label{eq:f1}
F^{\mu\nu}(x,y)&\defeq \tfrac{1}{2}\Braket{\acom{ A^\mu(x)}{A^\nu(y)}}-\braket{A^\mu(x)}\braket{A^\nu(y)}\,,\\
\label{eq:r1}
\rho^{\mu\nu}(x,y)&\defeq i\Braket{\com{A^\mu(x)}{A^\nu(y)}}\, ,\\
\label{eq:f2}
F_\Psi^{AB}(x,y)&\defeq \tfrac{1}{2}\Braket{\com{\Psi^A(x)}{{\bar{\Psi}^B}(y)}}\, ,\\
\label{eq:r2}
\rho_\Psi^{AB}(x,y)&\defeq i\Braket{\acom{ \Psi^A(x)}{{\bar{\Psi}^B}(y)}}
\,.
\end{align}
In particular, the equal-time (anti-)commutation rules are encoded in the spectral functions according to
\begin{align}
\label{eq:eqspec1}
\delta_\C(x^0-y^0)\rho^{\mu\nu}(x,y)&=0\, ,\\
\label{eq:eqspec2}
\delta_\C(x^0-y^0)\pdif_{x^0}\rho^{\mu\nu}(x,y)&=-\delta_\C(x-y)\eta^{\mu\nu}\, ,\\
\label{eq:eqtimespectral}
\delta_\C(x^0-y^0)i\gamma^0\rho_\Psi^{AB}(x,y)&=-\delta_\C(x-y)\delta^{AB}\, .
\end{align}
These equal-time conditions imply that spectral functions are normalized and that their initial conditions are fixed by the underlying quantum theory.

\begin{figure}[h!]
	\begin{center}
		\includegraphics[scale=0.035]{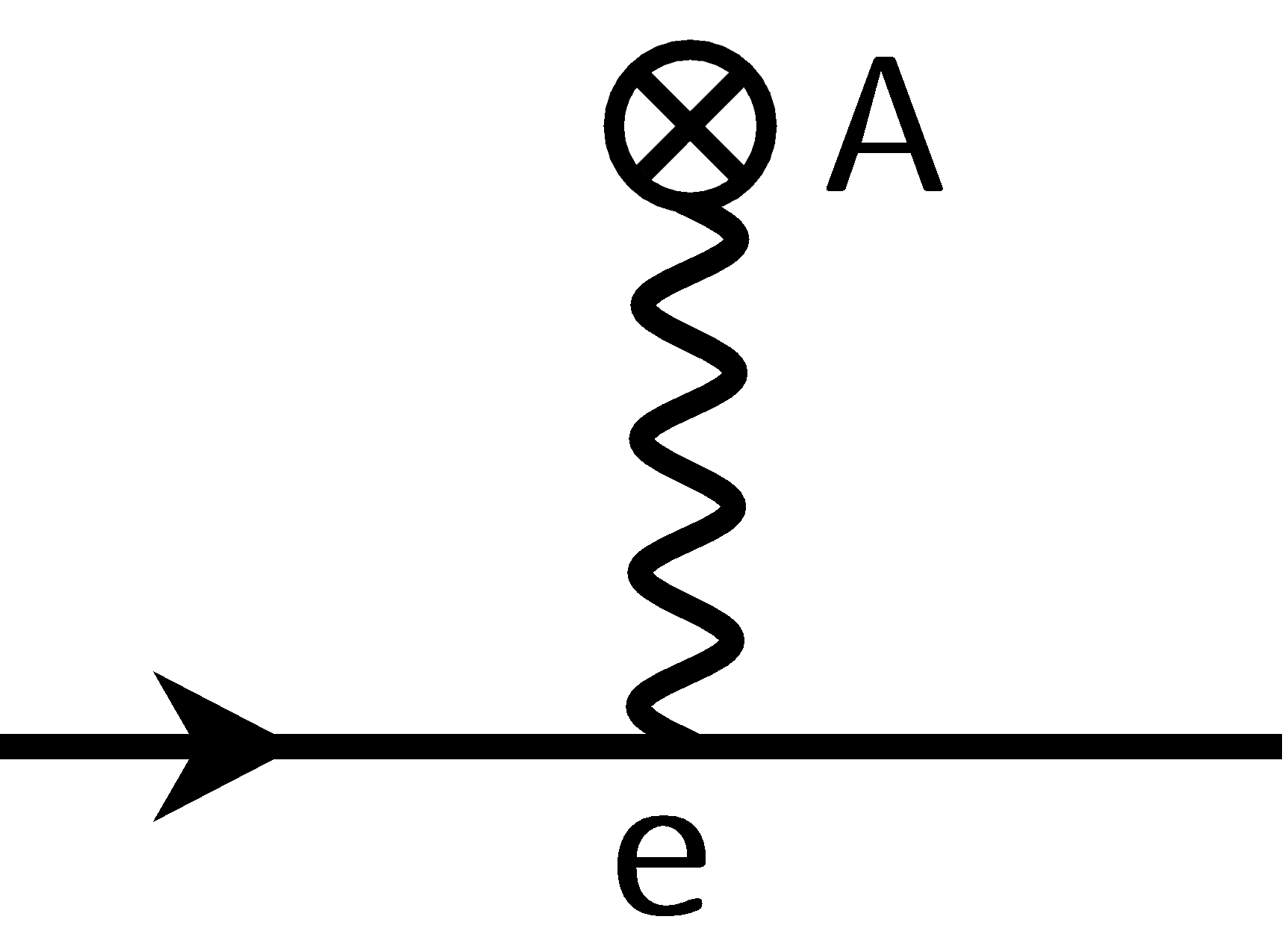}
	\end{center}
	\caption{\label{fig:vertex}The macroscopic field vertex.}
\end{figure}

An important simplification in Abelian theories such as QED occurs because of the absence of 2PI one-point function diagrams, such that $\Gamma_2[D,\Delta]$ does not explicitly depend on $\mathcal{A}$: The electromagnetic field expectation value enters the 2PI effective action for QED via the `classical vertex' term
\begin{align}
\label{eq:fieldvertex}
-ie\gamma^\mu\A_\mu(x)\delta_{\mathcal{C}}(x-y)\, ,
\end{align}
which can be depicted graphically as in Fig. \ref{fig:vertex}.

\noindent Such a contribution cannot be found in the 2PI diagrams contributing to $\Gamma_2$ since the two fermion lines emanating from the vertex could always be cut, thus making any such diagram two-particle reducible (see also Fig. \ref{fig:2pi}).

For QED, the macroscopic field therefore enters the 2PI effective action \refe{gamman} only via the classical fermion propagator $\Delta_0[\A]$ and the classical action $S[\A]$, while $\Gamma_2$ is field-independent. Since 2PI diagrams are at least two-loop, this implies that the complete explicit macroscopic field dependence enters at one-loop order of $\Gamma$,
\begin{align}
\frac{\delta\Gamma}{\delta \A}\at{F_\Psi}=0\lra \frac{\delta\Gamma^{(\textrm{1-loop})}}{\delta \A}\at{F_\Psi}=0\, .
\end{align}

Consequently, the field evolution equation always has a Maxwell-type form, i.e.
\begin{align}
\label{eq:micmax}
\Big[ {\eta^\mu}_\sigma\square_x-( 1-\tfrac{1}{\xi})\pdif^\mu_x\pdif_\sigma^x \Big]\mathcal{A}^\sigma(x)=j^\mu(x)\, ,
\end{align}
with the fermion current (see appendix \ref{app:aeom})
\begin{align}
\label{eq:j}
j^\mu(x)=-e\,\textrm{tr}\{ \gamma^\mu F_\Psi(x,x)\}\, ,
\end{align}
irrespective of the approximation order for $\Gamma_2$. This would not be the case, e.g., in QCD or self-interactring $\Phi^4$ scalar field theory, where the two-particle irreducible part of the effective action depends explicitly on the field expectation value, such that the form of the field evolution equation depends strongly on the order of approximation. Because $\Gamma_2$ is field-independent in QED, there are no further terms coming from higher order corrections. Approximations to $\Gamma_2$ affect the field evolution only implicitly via $F_\Psi$ in the fermion current \refe{j}. Furthermore, that each 2PI diagram in $\Gamma_2$ is separately gauge-invariant in QED \cite{Reinosa:2007vi} remains true in the presence of a macroscopic field due to the field-independence of $\Gamma_2$.

Notably, a vanishing field is not in general a self-consistent solution: if the system is initialized with a finite net charge density, it will develop a field from fermion fluctuations in the Maxwell equation. This field is then necessarily inhomogeneous as dictated by Gauss's law, i.e.\ the $0$-component of the Maxwell equation. Therefore, if the system equilibrates, it has to do so under this constraint for inhomogeneity.

In the equations of motion for the two-point functions, explicitly field-independent self-energies are given by
\begin{align}
\label{eq:se1}
\Sigma_\Psi[D,\Delta](x,y)&\defeq -i\frac{\delta \Gamma_2[D,\Delta ]}{\delta \Delta(y,x)}\, ,\\
\label{eq:se2}
\Sigma^{\mu\nu}[D, \Delta](x,y)&\defeq 2i\frac{\delta \Gamma_2[D,\Delta ]}{\delta D_{\mu\nu}(x,y)}\, ,
\end{align}
and can be decomposed similarly to two-point functions:
\begin{align}
\label{eq:sedecomp}
&\Sigma_{\mu\nu}(x,y)=\Sigma^{(F)}_{\mu\nu}(x,y)-\tfrac{i}{2}\Sigma^{(\rho)}_{\mu\nu}(x,y)\,\csgn(x^0-y^0)\, ,\\
\label{eq:sedecompf}
&\Sigma_\Psi(x,y)\hspace{-0.02cm}=\hspace{-0.02cm} \Sigma^{(F)}_\Psi(x,y)\hspace{-0.02cm}-\hspace{-0.02cm}\tfrac{i}{2}\Sigma_\Psi^{(\rho)}(x,y)\,\csgn(x^0-y^0)\,.
\end{align}
With these definitions, assuming Gaussian initial conditions, the stationarity conditions for the propagators in Eq.~(\ref{eq:varprinz}) can be written as\footnote{For non-Gaussian initial conditions, additional terms involving non-local interactions at initial time would appear in the equations of motion \cite{Garny:2009ni}.} \cite{Berges:2004yj}
\begin{align}
\label{eq:eom1}
&\Big[ {\eta^\mu}_\sigma\square_x-( 1-\tfrac{1}{\xi})\pdif^\mu_x\pdif_\sigma^x \Big]F^{\sigma\nu}(x,y)\\
\nonumber
&\hspace{1cm}=\int_{t_0}^{x^0}\dif^4 z\, \Sigma^{(\rho)}(x,z)^{\mu\gamma}{F(z,y)_\gamma}^\nu\\
\nonumber
&\hspace{2cm}-\int_{t_0}^{y^0}\dif^4 z\,
\Sigma^{(F)}(x,z)^{\mu\gamma}{\rho(z,y)_\gamma}^\nu\, ,\\
\label{eq:eom2}
&\Big[ {\eta^\mu}_\sigma\square_x-( 1-\tfrac{1}{\xi})\pdif^\mu_x\pdif_\sigma^x \Big]\rho^{\sigma\nu}(x,y)\\
\nonumber
&\hspace{1cm}=\int_{y^0}^{x^0}\dif^4 z\, \Sigma^{(\rho)}(x,z)^{\mu\gamma}{\rho(z,y)_\gamma}^\nu\, ,\\
\label{eq:eom3}
&\Big[ i\slashed \pdif_x-e\slashed{\mathcal{A}}(x)-m\Big]F_\Psi(x,y)\\
\nonumber
&\hspace{1cm}=\int_{t_0}^{x^0}\dif^4 z\, \Sigma_{\Psi}^{(\rho)}(x,z){F_\Psi(z,y)}\\
\nonumber
&\hspace{2cm}-\int_{t_0}^{y^0}\dif^4 z\,\Sigma^{(F)}_\Psi(x,z)\rho_\Psi(z,y)\, ,\\
\label{eq:eom4}
&\Big[i\slashed \pdif_x-e\slashed{\mathcal{A}}(x)-m\Big]\rho_\Psi(x,y)\\
\nonumber
&\hspace{1cm}=\int_{y^0}^{x^0}\dif^4 z\, \Sigma_{\Psi}^{(\rho)}(x,z){\rho_\Psi(z,y)}\, ,
\end{align}
with finite-time integrals $\int_{t_0}^{x^0}\dif ^4 z= \int_{t_0}^{x^0}\dif z^0\int_{-\infty}^{\infty}\dif^3 z$. While the structure of these equations is determined by causality, details of the underlying theory enter through the differential operators and self-energies, which couple all spectral and statistical functions to each other.

The fact that initial conditions for spectral functions are fixed by the equal-time (anti-)commutation relations \refe{eqspec1} -- \refe{eqtimespectral}, is reflected by the absence of the initial time $t_0$ in the memory integrals of their equations. In contrast, the evolution equations for the statistical functions have to be supplied by initial conditions. Non-Gaussian quantum fluctuations are built up dynamically but vanish at initial time, $x^0=y^0=t_0$, by vanishing of the memory integrals.

All equations are considered to be suitably regularized and the renormalization of the 2PI effective action for QED is discussed in detail in Ref.~\cite{Reinosa:2006cm}. Since we will finally arrive at a set of finite equations at the level of the kinetic approximation, renormalization will not be further discussed and we refer, e.g., to Refs. \cite{Kluger:1992gb, Kluger:1998bm} for details concerning dynamics.

The self-energies, encoding collisions, have leading contributions at $\Sigma,\Sigma_\Psi \sim\O(e^2)$. While self-energies have no explicit dependence on the macroscopic field by their definition in terms of the field-independent $\Gamma_2$, fermion two-point functions introduce an implicit field-dependence when evaluated from their equations of motion. As we will demonstrate, strong-field collision kernels are generated both in photon and fermion kinetic equations in this way. The macroscopic field enters via the terms $e\slashed \A(x)$, encoding in particular the Vlasov terms of fermion transport equations, which can be any order depending on the strength of $\A^\mu(x)$. By the smallness of the coupling $e$, these terms are suppressed in a naive power counting. However, in the presence of a strong field, $\A^\mu(x) \sim \O(1/e)$, these terms are effectively of order $e\A^\mu(x) \sim \O(e^0)$ such that the field-vertex \refe{fieldvertex} has to be resummed. As the macroscopic field decays \cite{Nerush:2010fe, Seipt:2016fyu, Kluger:1998bm} from its strong-field initial conditions, the system passes through different power counting scenarios that are all captured by our strong-field counting.

\subsection{\label{sec:loop}2PI loop expansion}

\begin{figure}[h!]
	\begin{center}
\includegraphics[scale=0.043]{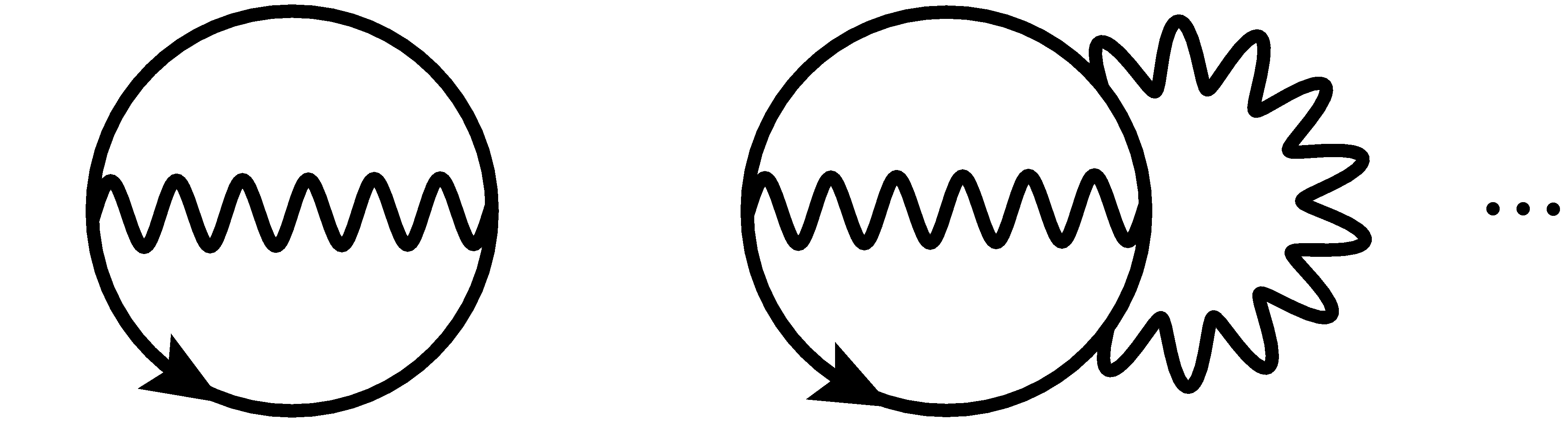}
	\end{center}
	\caption{\label{fig:ealoop}The first two 2PI loop orders, $\O(e^2)$ and $\O(e^4)$, of the effective action.}
\end{figure}

In order to close the equations \refe{eom1}–\refe{eom4} one requires explicit expressions for the self-energies \refe{se1} and \refe{se2}. This is achieved by employing a 2PI coupling or `loop' expansion, which expresses $\Gamma_2$ in terms of resummed propagators $D^{\mu\nu}$ and $\Delta$ and of free vertices. This self-consistent treatment of propagators selectively resums perturbative contributions, which helps achieving a non-secular time evolution with a valid expansion scheme at all times~\cite{Berges:2004yj, Boyanovsky:1998tg}. In such an expansion, $\Gamma_2$ can be written as
\begin{align}
\label{eq:2piloop}
&\Gamma_2[D,\Delta ]=\tfrac{i}{2}e^2D\Delta^2V_0^2+\tfrac{i}{4}e^4D^2\Delta^4V_0^4+\O(e^6)\, ,
\end{align}
where we have suppressed all indices and arguments that are contracted or integrated over. This is diagramatically depicted in Fig.~\ref{fig:ealoop}.

	\begin{figure}[h!]
	\begin{center}
		\includegraphics[scale=0.039]{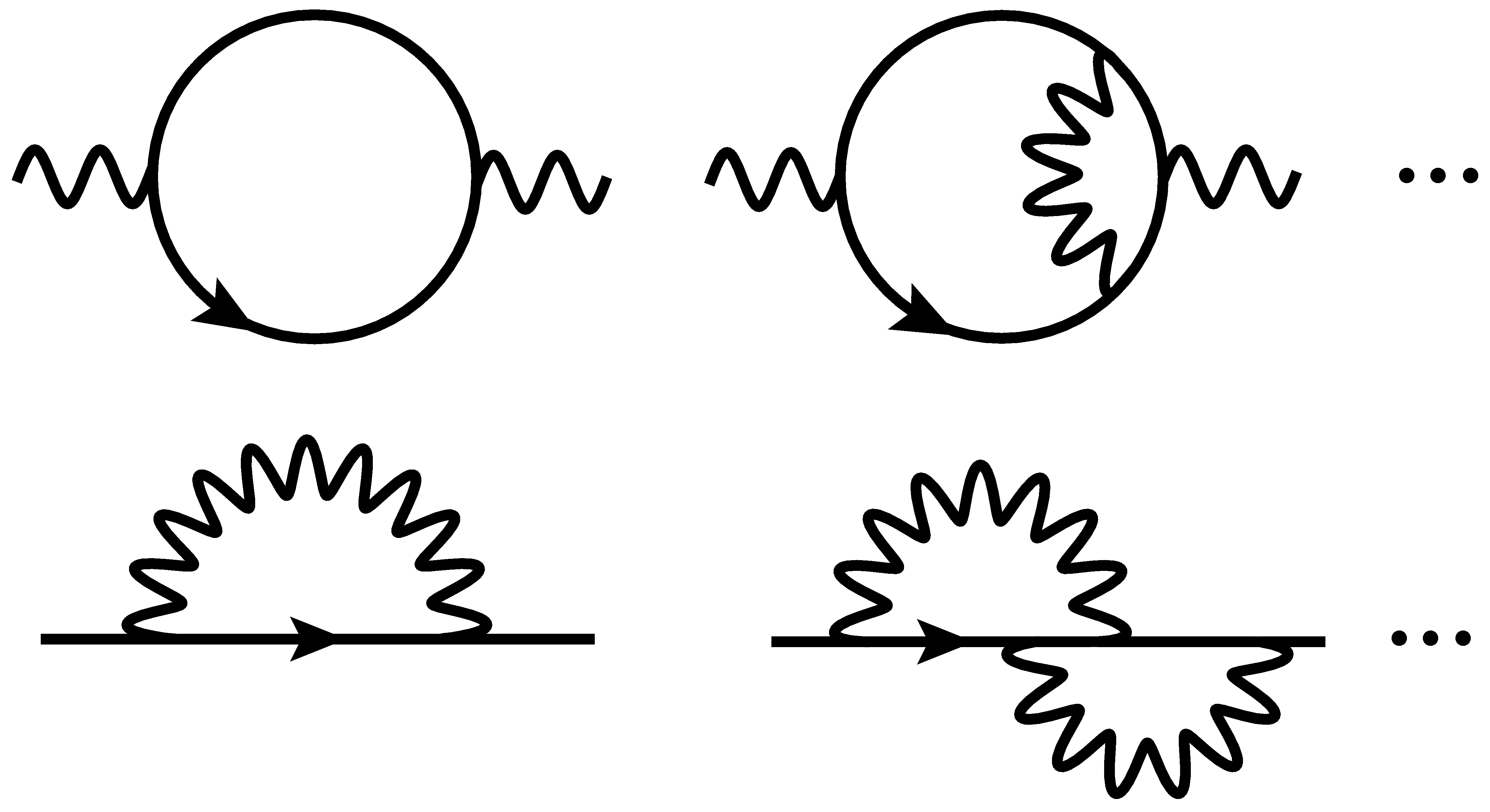}
	\end{center}
	\caption{\label{fig:seloop}The first two 2PI loop orders, $\O(e^2)$ and $\O(e^4)$, of the photon (first line) and fermion (second line) self-energy.}
\end{figure}

The explicit expressions obey Feynman rules including symmetry factors. Only the free QED vertex
\begin{align}
V_{0,AB}^\mu(x,y,z)\defeq\gamma^\mu_{AB}\delta_\C(x-z)\delta_\C(z-y)
\end{align}
appears.

Correspondingly, the 2PI loop expansion of the self-energies \refe{se1} and \refe{se2} is a series of 1PI diagrams with two amputated external legs (see Fig.~\ref{fig:seloop}). The 1PI property of the self-energies can also directly be understood from the definition of $\Gamma_2$ as the sum of all closed 2PI diagrams, from which $\Sigma,\Sigma_{\Psi}$ are obtained by opening one propagator line, i.e.\ by Eq.~\refe{se1}, \refe{se2}. 

As long as photon occupations are not too large, i.e.\ if the statistical photon two-point function obeys
\begin{align}
\label{eq:occass}
F^{\mu\nu}\ll \O(1/e^2)\, ,
\end{align}
the power counting of $e$ from vertices in a 2PI loop expansion can be expected to be a valid approximation scheme and we can truncate by virtue of the smallness of $e$. Similar conditions for the spectral functions always hold since they are normalized by the equal-time commutation relations. Since fermion occupancies are limited by Fermi-Dirac statistics there are no further corresponding constraints for the expansion scheme. The condition \refe{occass} is dynamical such that even if the system is initialized with small occupations, a kinetic description breaks down if too many photons with the same position and momentum are produced. Physically, the assumption \refe{occass} may be understood as the requirement for a sufficiently long mean free path in kinetic descriptions: The loop expansion of self-energies in the kinetic limit is an expansion in the number of particles involved in a scattering \cite{PhysRevD.74.025010, PhysRevD.71.065007, Jeon:1998zj}. The denser the medium, the smaller the mean free path, and the more likely a collision involving many particles. If the medium is too dense, collisions between arbitrarily many particles become equally likely, invalidating a truncation in an expansion of the number of particles.\footnote{In O(N) scalar theories, a far-from-equilibrium kinetic description can nevertheless be formulated on the basis of emergent degrees of freedom in this highly occupied regime \cite{Walz:2017ffj}. }

We emphasize that these considerations do not directly limit the size of the macroscopic field: Because of the field-independence of $\Gamma_2$, higher order contributions to self-energies are negligible also in the presence of strong fields and processes such as $ee\gamma\gamma$ or $eeee$ scattering do not contribute to a leading-order (LO) description (see also Ref.~\cite{Bragin:2020akq}). As long as \refe{occass} is fulfilled, the coupling remains a valid expansion parameter, no matter how large the field is at that time. Thus we may employ the leading order of self-energies to obtain a closed description that is complete at order $\O(e^2)$.

The LO of $\Gamma_2\sim \O(e^2)$, is
\begin{align}
&2i\Gamma_2[D,\Delta ]\\
\nonumber
&=-e^2\int_{xy,\C}\textrm{tr}\{\gamma^\mu\Delta(x,y)\gamma^\nu\Delta(y,x)\}D_{\mu\nu}(x,y) +\O(e^4)\, .
\end{align}
The corresponding self-energy expressions are
\begin{align}
\label{eq:se1loop}
\Sigma^{\mu\nu}(x,y)&=e^2\,\textrm{tr}\, \left\{\gamma^\mu\Delta(x,y)\gamma^\nu\Delta(y,x)\right\}+\O(e^4)\, ,\\
\label{eq:se1loopf}
\Sigma_\Psi(x,y)&=-e^2\gamma_\mu \Delta(x,y)\gamma_\nu D^{\mu\nu}(x,y)+\O(e^4)\, ,
\end{align}
where the relative sign originates from the fermion loop in $\Sigma^{\mu\nu}$. The kinetic equations derived in this paper neglect all higher orders of the 2PI loop expansion.\footnote{We expand $\Gamma_2$ to $\O(e^2)$, i.e.\ to 2PI 2-loop order, where the leading non-trivial scattering occurs in the presence of a non-vanishing field. At this order the 2PI approach coincides also with corresponding two-loop approximations for any higher $n$PI effective actions with $n>2$ \cite{PhysRevD.70.105010}.} In agreement with the coupling counting in perturbation theory, all possible crossings of $ee\gamma$ scattering terms emerge from these $\O(e^2)$ self-energies. The following sections are dedicated to understanding how effective transport and kinetic descriptions emerge from this approach.

\section{\label{sec:3}The kinetic limit of nonequilibrium QED}

To express the equations of motion in kinetic degrees of freedom, we change to center and relative space-time variables
\begin{align}
\label{eq:xsdef}
X\defeq\tfrac{1}{2}(x+y)\, ,&\hspace{1.5cm}s\defeq x-y\, .
%x=X+\tfrac{s}{2}&\hspace{1.5cm}y=X-\tfrac{s}{2}\,.
\end{align}
The four-momentum $p$ associated to $-i\pdif_s$ is the momentum that appears in kinetic equations, while $X$ is the kinetic four-position variable.

The momentum $p$ is introduced by a Wigner transform with respect to the relative coordinate $s$. For an evolution starting at time $t_0$ at which the initial conditions are given, the Wigner transform of a generic two-point function $G$ may be written as
\begin{align}
\label{eq:boundwig}
G_{t_0}(X,p)\defeq \intxtoy{-2(X^0-t_0)}{2(X^0-t_0)}\hspace{-0.2cm}\dif s^0\int\dif^3 s\, e^{ips}G(X+\tfrac{s}{2},X-\tfrac{s}{2})\, .
\end{align}
Here $t_0$ appears in the time integral as a lower boundary for all time variables. Since initially we have $X^0 = t_0$, there are no relative times to integrate in this case, which preempts a Wigner transformation starting at initial time. To nevertheless be able to talk about kinetic variables from the initial time of our kinetic description, we employ a late-time limit described in the following.

\subsection{\label{sec:latetime}Late-time limit}

For finite $t_0$ and $X^0$ the integration range for $s^0$ is always limited. Only if $t_0\rightarrow -\infty$ the relative time variable $s^0$ can be infinite, which is required for a proper introduction of Fourier frequency modes $p^0$. Of course, sending formally $t_0\rightarrow -\infty$ while still initializing the evolution at some finite time implies that a general system is initially not accurately described by these late-time equations. However, for sufficiently large $X^0$ compared to the finite initialization time, the description is expected to become accurate~\cite{Berges:2015kfa}. Therefore, instead of Eq.~(\ref{eq:boundwig}) we consider the late-time Wigner transform 
\begin{align}
\label{eq:unboundwig}
G(X,p)\defeq \int\dif^4 s\, e^{ips}G(X+\tfrac{s}{2},X-\tfrac{s}{2}) \, ,
\end{align}
which has contributions from all $s^0$ for arbitrary $X^0$.

Equal-point objects such as the fermion current \refe{j} can be expressed in terms of such late-time Wigner transforms,
\begin{align}
\label{eq:Fj}
j^\mu(X)=-e\int_p\textrm{tr}\{\gamma^\mu F_\Psi(X,p)\}\, .
\end{align}
The notation $\int_p= \int_{-\infty}^{\infty}\dif^4p/(2\pi)^4$ for momentum integrals is used throughout. The canonical equal-time anticommutator \refe{eqtimespectral} in late-time Wigner space is
\begin{align}
\label{eq:rhonorm}
-i\gamma^0\int\frac{\dif p^0}{(2\pi)}\,\rho_\Psi(X,p)&=1\, ,
\end{align}
such that the late-time vector-zero component $\tfrac{1}{4}\textrm{tr}\{\gamma^0\rho_\Psi(X,p)\}$ may be interpreted as a density of states \cite{Shen:2020jya}.

In the microscopic description, finite-time Wigner transforms \refe{boundwig} produce factors with finite-width energy-peaks on correspondingly small timescales \cite{Blaizot:2001nr} that reduce to delta peaks at late times via
\begin{align}
\label{eq:timewidth}
&\intxtoy{-2(X^0-t_0)}{2(X^0-t_0)}\hspace{-0.2cm}\dif s^0 \, e^{iP^0s^0}\xrightarrow{t_0\rightarrow -\infty} (2\pi)\delta(P^0)\, .
\end{align}
In this late-time regime, the interactions of QED may be described by those of kinetic theory in terms of degrees of freedom that carry a definite amount of energy.

Applying the late-time limit, $t_0\rightarrow -\infty$, one can write the equations of motion \refe{eom1} -- \refe{eom4} as
\begin{align}
\label{eq:raeom1}
&\Big[\eta^{\mu}{}_\sigma\square_x-( 1-\tfrac{1}{\xi})\pdif^\mu_x \pdif_\sigma^x \Big] F^{\sigma\nu}(x,y)\\
\nonumber
&=\int_z \Big[ \Sigma_{\textrm{R}}^{\mu\sigma}(x,z) {F_\sigma}^\nu(z,y)+\Sigma^{(F)}(x,z)^{\mu\sigma}D_{\textrm{A}}(z,y)_\sigma{}^\nu\Big] \, ,\\
\label{eq:raeom2}
&\Big[ {\eta^\mu}_\sigma\square_x-( 1-\tfrac{1}{\xi})\pdif^\mu_x\pdif_\sigma^x \Big]\rho^{\sigma\nu}(x,y)\\
\nonumber
&=\int_z\, \Big[ \Sigma_\textrm{R}^{\mu\sigma}(x,z){\rho_\sigma}^\nu(z,y)
+\Sigma^{(\rho)}(x,z)^{\mu\sigma}{D_\textrm{A}(z,y)_\sigma}^\nu \Big]\, ,\\
\label{eq:raeom3}
&\Big[i\slashed \pdif_x-e\slashed{\mathcal{A}}(x)-m \Big] F_\Psi(x,y)\\
\nonumber
&=\int_z\,\Big[
\Sigma_{\Psi,\textrm{R}}(x,z) F_\Psi(z,y) +\Sigma^{(F)}_\Psi(x,z) \Delta_\textrm{A}(z,y)\Big]\, ,\\
\label{eq:raeom4}
&\Big[i\slashed \pdif_x-e\slashed{\mathcal{A}}(x)-m\Big]\rho_\Psi(x,y)\\
\nonumber
&=\int_z\,\Big[ \Sigma_{\Psi,\textrm{R}}(x,z)\rho_\Psi(z,y) + \Sigma_{\Psi}^{(\rho)}(x,z){\Delta_\textrm{A}(z,y)}\Big]\, ,
\end{align}
with $\int_z=\int\dif^4 z$, where we have introduced the retarded and advanced functions for photons and fermions \refe{photonadvret} -- \refe{fermadvret} defined in appendix \ref{app:propdef}. 

Given the multitude of different nonequilibrium two-point functions, it is important to remember that there are only two independent two-point functions per field species: the statistical and spectral functions. However, this can be invalidated by approximations, in particular, by the procedure of sending $t_0 \to -\infty$ while initializing the equations at a finite time. Wigner functions that include small frequencies via \refe{unboundwig} may appear independent of each other because of spurious small frequency contributions that, in an exact description employing finite-time Wigner transforms \refe{boundwig}, do not yet exist at early times \cite{Bedaque:1994di,Greiner:1998ri}.

\subsection{\label{sec:3c}Gradient expansion}

As a next step in the derivation of kinetic equations, one considers an expansion in the Lorentz-invariant and dimensionless parameter $(s\cdot\pdif_X)$. An expansion in propagator-gradients is achieved by the late-time identity \cite{Berges:2015kfa}
\begin{align}
\label{eq:wigconv}
&\int_s \,e^{ip\cdot s}(\Sigma* G)(X+\tfrac{s}{2},X-\tfrac{s}{2})\\
\nonumber
&\hspace{1cm}=\exp\Bigg\{\frac{i}{2}\left(\xpdify{}{p_\sigma}\xpdify{}{X'^\sigma}-\xpdify{}{p'_\sigma}\xpdify{}{X^\sigma}\right)\Bigg\}\\
\nonumber
&\hspace{3cm}\cross \Sigma(X,p)G(X',p')\at{X'=X,p'=p}\, ,
\end{align}
which applies to photon and fermion convolutions
\begin{align}
(\Sigma* G)^{\mu\nu}(x,y)&\defeq \int_z \Sigma^{\mu}{}_\sigma(x,z)G^{\sigma\nu}(z,y)\, ,\\
(\Sigma_\Psi*G_\Psi)(x,y)&\defeq \int_z \Sigma_\Psi(x,z) G_\Psi(z,y)\, .
\end{align}
Expansion of the exponential in Eq. \refe{wigconv} corresponds to an expansion in $(\pdif_p\cdot \pdif_X)$, i.e.\ a gradient expansion in Wigner space. While the LO simply replaces the Wigner transform of convolutions by products of Wigner transforms, an expansion to next-to-leading order (NLO) in propagator-gradients would involve Poisson brackets,
\begin{align}
\label{eq:pbdef}
&[\Sigma,G ]_\textrm{PB}(X,p)\\
\nonumber
&\hspace{0.5cm}\defeq \xpdify{\Sigma(X,p)}{p_\sigma}\xpdify{G(X,p)}{X^\sigma}-\xpdify{\Sigma(X,p)}{X_\sigma}\xpdify{G(X,p)}{p^\sigma}\, .
\end{align}
The truncated gradient expansion leads to equations that are irreversible and local in central time $X^0$, as in the case of kinetic equations. Still, gradient expanded 2PI equations contain parts of the memory integrals of the fundamental equations and are non-local in relative time $s^0$. This allows for access to unconstrained frequency variables, which are not present in traditional kinetic descriptions as further discussed in the following sections.

The smallness of the expansion parameter $(s\cdot\pdif_X)$ can be met in several circumstances.\footnote{In the absence of a temperature far from equilibrium, no single scale may be associated to $s$. In this case, the near-equilibrium counting of dimensionful gradients $\pdif_X\sim e^2 T$ of Ref.~\cite{Blaizot:1999xk} may not be used to argue for the smallness of $(s\cdot\pdif_X)$.} Quantum field dynamics often becomes insensitive to its past, such that correlations are dominated by small $s$ \cite{Berges:2004ce, Berges:2002wr, Berges:2005ai, Berges:2001fi}. From the perspective of the spectral function, this damping of correlations in time corresponds to the emergence of a particle picture in momentum space \cite{Aarts:2001qa, PhysRevD.74.045022}. Furthermore, assuming that $(s\cdot \pdif_X)$ is small depends on what the derivative acts on: In the following, we neglect only gradients of two-point functions $G$, by dropping Poisson brackets
\begin{align}
\label{eq:poissongrad}
[\Sigma[G],G]_{\textrm{PB}}\sim\O(e^2\pdif_p\cdot\pdif_XG)\, ,
\end{align}
while formally keeping gradients of the gauge-invariant field strength tensor, \mbox{$(s\cdot\pdif_X)^j\,e\F^{\mu\nu}(X)\sim\O(e^0(s\cdot \pdif_X)^j)$}, to all orders. That is, we count field-gradients as \mbox{$(s\cdot\pdif_X)\F^{\mu\nu}\sim \F^{\mu\nu}$} and propagator-gradients as \mbox{$(s\cdot\pdif_X)G\ll G$}. This allows us to treat a large class of far-from-equilibrium initial conditions of the macroscopic field. Approximations to field-gradients are then discussed in Secs.~\ref{subsec:5a} and \ref{subsec:smallgradplane}, where we make contact with the locally-constant field approximation.

However, field-gradients may be implicit in propagator solutions (see also Sec.~\ref{subsec:smallgradplane} for the example of plane-wave fields): Given an explicit field-dependent solution for a two-point function, for example of the form 
\begin{align}
G_\Psi^{-1}[\A]\sim \slashed p-e\slashed \A-m\, ,
\end{align}
different gradients may be related via
\begin{align}
\label{eq:gradrel}
%\nu\defeq 
\frac{G_\Psi^{-1}(s\cdot \pdif_X)G_\Psi}{\slashed \A^{-1}(s\cdot\pdif_X)\slashed \A}\sim \frac{-e\slashed \A}{\slashed p-e\slashed \A-m}\, .
\end{align} 
In fact, the separation of field and propagator-gradients that is possible at the level of the equations of motion does not ensure that the ratio (\ref{eq:gradrel}) is small. Nevertheless, we can observe from Eq.~\refe{gradrel} that large fermion momenta can facilitate such a separation. When solving the kinetic equations for inhomogeneous fields derived below, the smallness of the ratio \refe{gradrel} should be be checked.

\subsection{\label{sec:3a}Distribution functions}

\subsubsection{\label{sec:redten}Reduction of tensor structures}
An identification of the linearly independent components of the fermion or photon correlation functions follows from their Lorentz transformation properties. For instance, the statistical fermion correlator can be decomposed as 
\begin{align}
F_\Psi&=F_{\Psi,\textrm{S}}+\gamma_\mu F_{\Psi,\textrm{V}}^\mu\\
\nonumber
&\hspace{0.5cm}+i\gamma^5 F_{\Psi,\textrm{P}}-\gamma^5\gamma_\mu F_{\Psi,\textrm{A}}+\tfrac{1}{2}\sigma_{\mu\nu}F^{\mu\nu}_{\Psi,\textrm{T}}\, ,
\end{align}
in terms of the scalar ($F_{\Psi,\textrm{S}}$), vector ($F_{\Psi,\textrm{V}}^\mu$), pseudo-vector ($F_{\Psi,\textrm{P}}$), axial-vector ($F_{\Psi,\textrm{A}}$) and tensor ($F^{\mu\nu}_{\Psi,\textrm{T}}$) components
\begin{align}
\label{eq:chiral1}
F_{\Psi,\textrm{S}}&\defeq \tfrac{1}{4}\textrm{tr} \left\{ \Id F_\Psi\right\}\, ,\\
\label{eq:chiral2}
F_{\Psi,\textrm{V}}^\mu&\defeq \tfrac{1}{4}\textrm{tr} \left\{ \gamma^\mu F_\Psi\right\}\, ,\\
\label{eq:chiral3}
iF_{\Psi,\textrm{P}}&\defeq \tfrac{1}{4}\textrm{tr}\left\{ \gamma^5F_\Psi\right\}\, ,\\
\label{eq:chiral4}
F_{\Psi,\textrm{A}}^\mu&\defeq \tfrac{1}{4}\textrm{tr} \left\{ \gamma^5\gamma^\mu  F_\Psi\right\}\, ,\\
\label{eq:chiral5}
F_{\Psi,\textrm{T}}^{\mu\nu}&\defeq \tfrac{1}{4}\textrm{tr}\left\{ \sigma^{\mu\nu}F_\Psi\right\}\, ,
\end{align}
with respect to the Dirac basis $\{ \Id, \gamma^\mu, \gamma^5, \gamma^5\gamma^\mu, \sigma^{\mu\nu}\}$ where $\mu<\nu$ and with $\gamma^5\defeq -\tfrac{i}{4!}\varepsilon_{\mu\nu\rho\sigma}\gamma^\mu\gamma^\nu\gamma^\rho\gamma^\sigma $ and $\sigma^{\mu\nu}\defeq \tfrac{i}{2}[\gamma^\mu,\gamma^\nu]$. Below, we often drop the label `V' for the vector component.

In the presence of chiral symmetry (facilitated by massless fermions or ultrarelativistic momenta), scalar, pseudoscalar and tensor components vanish identically \cite{Berges:2002wr}. If a description in terms of free particles is valid, the axial component of the free fermion spectral function would also vanish.

Similar comments apply to the photon distribution function and a decomposition of the Lorentz tensor structures of the photon equations of motion in the presence of a macroscopic field can be achieved with the basis discussed in Refs. \cite{Ritus:1972ky, Bajer:1975}.

With this in mind, one could write without loss of generality for each component of $F_\Psi(X,p)$:
\begin{align}
\label{eq:wkba1}
F_{\Psi,\textrm{S}}(X,p)&=-i[\tfrac{1}{2}-f_{\Psi,\textrm{S}}(X,p)]\rho_{\Psi,\textrm{S}}(X,p)\, ,\\
F_{\Psi,\textrm{V}}^\mu(X,p)&=-i[\tfrac{1}{2}-f_{\Psi,\textrm{V}}(X,p)]\rho_{\Psi,\textrm{V}}^\mu(X,p)\, ,\\
F_{\Psi,\textrm{P}}(X,p)&=-i[\tfrac{1}{2}-f_{\Psi,\textrm{P}}(X,p)]\rho_{\Psi,\textrm{P}}(X,p)\, ,\\
F_{\Psi,\textrm{A}}^\mu(X,p)&=-i[\tfrac{1}{2}-f_{\Psi,\textrm{A}}(X,p)]\rho_{\Psi,\textrm{A}}^\mu(X,p)\, ,\\
\label{eq:wkba5}
F_{\Psi,\textrm{T}}^{\mu\nu}(X,p)&=-i[\tfrac{1}{2}-f_{\Psi,\textrm{T}}(X,p)]\rho_{\Psi,\textrm{T}}^{\mu\nu}(X,p)\, .
\end{align}
The change from a description in terms of $F_{\Psi,\textrm{S}\ldots\textrm{T}}(X,p)$ to a formulation in terms of $f_{\Psi,\textrm{S}\ldots\textrm{T}}(X,p)$ is convenient because in characteristic limits $f_{\Psi,\textrm{S}\ldots\textrm{T}}(X,p)$ can be interpreted as distribution functions. 

In particular, in thermal equilibrium all distribution functions are time-independent and equal the Fermi-Dirac distribution, i.e.\ $f_{\Psi,\textrm{S}\ldots\textrm{T}}(p^0) = 1/(e^{\beta  p^0}+1)$ (and correspondingly a Bose-Einstein distribution for the photon case). For a thermal theory this is valid no matter how strong the interactions are and holds even in the absence of a dispersion relation between frequency and spatial momenta, $p^0=\omega(\vec p\,)$.

Phenomena such as the chiral magnetic effect \cite{Fukushima:2008xe, Mueller:2016ven, Mace:2016shq, Mace:2019cqo}, chiral kinetic theory \cite{Stephanov:2012ki, Gao:2017gfq, Hidaka:2016yjf, Mueller:2017arw, Huang:2018wdl} or spin transport \cite{Yang:2020hri} should become accessible from first principles by using (\ref{eq:wkba1}) -- \refe{wkba5} in the equations of motion (\ref{eq:raeom1}) -- (\ref{eq:raeom4}). However, for our current purposes of strong-field kinetic equations and to make contact with existing limiting cases in the literature, we consider a single distribution function $f_\Psi(X,p)$ for fermions and $f(X,k)$ for photons by writing~\cite{Berges:2004yj,kadanoff1962quantum}
\begin{align}
\label{eq:sKBAph}
F^{\mu\nu}(X,k)&=-i[\tfrac{1}{2}+f(X,k)]\rho^{\mu\nu}(X,k)\, ,\\
\label{eq:sKBAf}
F_\Psi(X,p)&=-i[\tfrac{1}{2}-f_\Psi(X,p)]\rho_\Psi(X,p) \, .
\end{align}

For the fermion distribution function one has Pauli's principle \cite{Berges:2002wr},
\begin{align}
\label{eq:fermsmall}
f_\Psi(X,p)\leq 1\, .
\end{align}
In order to distinguish fermion and anti-fermion distribution functions, it is convenient to define \cite{PhysRevC.97.014901}
\begin{align}
\label{eq:antidef}
f_\Psi(X,p)&\eqdef \theta(p^0)f^-
_\Psi(X, p)\\
\nonumber
&\hspace{1cm}+\theta(-p^0)[1-f^+_\Psi(X,- p)] \, .
\end{align}
In a charge conjugation invariant system, the fermion distribution function obeys \cite{tuprints3373, PhysRevD.70.105010}
\begin{align}
\label{eq:cp}
-[f_\Psi(X,-p)-1]=f_\Psi(X,p) \hspace{0.2cm}\textrm{(if CP-invariant)}\, ,
\end{align}
such that the system is charge neutral,
\begin{align}
f_\Psi^+(X,p)=f_\Psi^-(X,p) \hspace{0.4cm}\textrm{(if CP-invariant)}\, .
\end{align}
While the vacuum is CP-invariant, the general initial conditions which we want to discuss in this paper break CP-invariance by introducing a net total charge, such that $f_\Psi^+\neq f_\Psi^-$. The photon identity analogous to \refe{cp} reads~\cite{tuprints3373, PhysRevD.70.105010}
\begin{align}
\label{eq:photoncp}
-[f(X,-k)+1]=f(X,k)
\end{align}
and does not rely on CP-invariance.

\subsubsection{\label{par:particle}On-shell particle picture}

In general, the distribution functions introduced in Eqs. \refe{sKBAph} and \refe{sKBAf} depend on the off-shell frequency variable $p^0$ that is not restricted to any dispersion relation, $p^0\neq \omega(\vec p\,)$. However, they only appear in combination with the respective spectral function. As a consequence, if the physics can be approximately described by free spectral functions,
\begin{align}
\rho^{\mu\nu}_{0,\xi}(X,k)&=i(2\pi)(\eta^{\mu\nu}-(1-\xi)\tfrac{1}{k^2}k^\mu k^\nu)\sgn(k^0)\delta(k^2)\, ,\\
\label{eq:freef}
\rho_{\Psi,0}(X,p)&=i(2\pi)(\slashed p+m)\sgn(p^0)\delta(p^2-m^2)\, ,
\end{align}
then the distribution functions can be restricted to their on-shell values. Whether an on-shell description is possible is determined self-consistently by solving the equations of motion (\ref{eq:raeom2}) and (\ref{eq:raeom4}) for the spectral functions. At initial time, the photon (fermion) spectral functions are determined by the equal-time (anti)commutation rules and each subsequent time step is determined by the equations of motion. If and when on-shell spectral functions emerge depends on timescales and initial conditions for statistical propagators and the macroscopic field. As we argue in Sec.~\ref{sec:4a}, the free fermion spectral function \refe{freef} is in fact not complete at order $\O(e^2)$ in the presence of general strong fields, $\A^\mu(x)\sim 1/e$, such that a standard on-shell kinetic description breaks down. Instead, we propose in this paper a less restrictive `transport' description that includes off-shell frequencies of fermions (but not of photons) in terms of the off-shell distribution function $f_\Psi(X,p)$. The frequency dependence of this function is then determined dynamically by the equations of motion and independently of its momentum dependence $\vec p$. An electron and positron particle picture is assumed only in Sec.~\ref{par:pair} to compute particle production at asymptotic times when the field has decayed.

With this application in mind, it is instructive to compute the fermion current \refe{Fj} for the free fermion spectral function \refe{freef}, i.e.
\begin{align}
\label{eq:j0}
j^0(X)&= -2e\hspace{-0.1cm}\int\hspace{-0.1cm}\frac{\dif^3 p}{(2\pi)^3}\,\Big[f_\Psi^+(X,-\vec p\,)-f_\Psi^-(X,\vec p\,)\Big]\, ,\\
\label{eq:ji}
j^i(X)&=-2e\hspace{-0.1cm}\int\hspace{-0.1cm}\frac{\dif^3p}{(2\pi)^3}\frac{p^i}{\varepsilon(\vec p\,)}\Big[1-f_\Psi^+(X,-\vec p\,)-f_\Psi^-(X,\vec p\,)\Big]\, ,
\end{align}
with on-shell electron and positron distribution functions,
\begin{align}
\label{eq:osfdef}
f^-_\Psi(X,\vec p\,)&\defeq f^-_\Psi(X,p)\,\,\textrm{ at }\,\, p^0=\sqrt{|\vec p\, |^2+m^2}\, ,\\%=f_\Psi(X,p)\at{p^0=E_{\vec p}}
\label{eq:osafdef}
f_\Psi^+(X,-\vec p\,) &\defeq f_\Psi^+(X,-p)\,\,\textrm{ at }\,\, p^0=-\sqrt{|\vec p\, |^2+m^2}\, .
%=[1-f_\Psi(X,-p)]\at{p^0=E_{\vec p}}
\end{align}
The zero-component \refe{j0} can be interpreted in terms of the conserved electric charge 
\begin{align}
\label{eq:Qdef}
Q(X^0)\defeq \int\dif^3 X j^0(X)
\end{align}
which then reads on-shell
\begin{align}
\label{eq:charge}
Q(X^0)&=2e\int_{\vec X,\vec p}\Big[ f_\Psi^-(X,\vec p\,)-f_\Psi^+(X,-\vec p\,) \Big]\, .
\end{align} 
Similarly, on-shell, $j^i$ gives rise to the fermion pair number density
\begin{align}
\label{eq:density}
n_\Psi(X)&\defeq \int_{\vec p}\,\Big[f^-_\Psi(X,\vec p\,)+f^+_\Psi(X,-\vec p\,)\Big]\, .
\end{align}
which is related to the total pair number via
\begin{align}
\label{eq:pairnumber}
N_\Psi(X^0)\defeq \int\dif^3X\,n_\Psi(X)\, .
\end{align}
This expression will serve us to define an asymptotic particle number of strong-field systems in Sec.~\ref{par:pair}.

In contrast to the fermion case, the photon spectral function may be set to its free form also in the presence of a strong field (see Sec.~\ref{sec:photontransport}). We can then identify the on-shell photon distribution functions of kinetic theory by integrating over frequency $k^0$, i.e.
\begin{align}
\nonumber
f(X,\vec k)&\defeq \int_0^{\infty} \dif k^0\,2k^0\,f(X,k)\delta(k^2)\\
\label{eq:osf}
&=f(X,k)|_{k^0=|\vec k|}\,,
\end{align}
as we discuss below. The total number of photons is then
\begin{align}
\label{eq:photonnumber}
N(X^0)\defeq \int_{\vec X,\vec k}f(X,\vec k\,)\, .
\end{align}

\section{\label{sec:4}Strong-field QED transport equations}

We now apply the procedure of Sec.~\ref{sec:3} to the equations of motion \refe{raeom1} -- \refe{raeom4} for the statistical and spectral functions and the equation of motion \refe{micmax} for the macroscopic field. To ease the notation, we refer to the left sides of the two-point function equations as $(F\textrm{LHS})^{\mu\nu}(x,y)$, $(\rho\textrm{LHS})^{\mu\nu}(x,y)$ and  $(F\textrm{LHS})_\Psi(x,y)$, $(\rho\textrm{LHS})_\Psi(x,y)$ respectively, and similarly to the right hand sides (`RHS') or to entire equations (`EOM').

To reveal the 'gain-minus-loss' structure of collision terms, we identify the `$+-/-+$' or `Wightman functions' (defined in appendix \ref{app:propdef}) by making use of the identity
\begin{align}
\nonumber
&\Sigma^{(\rho)}(X,k)^{\mu\sigma} {F_\sigma}^\nu(X,k)-\Sigma^{(F)}(X,k)^{\mu\sigma}{\rho_\sigma}^\nu(X,k)\\
\nonumber
&\hspace{1cm}=-i\Big( \Sigma^{-+}(X,k)^{\mu\sigma}D^{+-}(X,k)_\sigma{}^\nu\\
&\hspace{2cm}-\Sigma^{+-}(X,k)^{\mu\sigma}D^{-+}(X,k)_\sigma{}^\nu \Big)\, ,
\end{align}
and an analogous identity for fermions. Then Eqs.~\refe{sKBAph} and \refe{sKBAf} can be expressed in terms of the Wightman functions as
\begin{align}
\label{eq:wightmankba1}
D^{+-}(X,k)^{\mu\nu}&=-i[1+f(X,k)]\rho^{\mu\nu}(X,k)\, ,\\
D^{-+}(X,k)^{\mu\nu}&=-if(X,k)\rho^{\mu\nu}(X,k)\, ,\\
\Delta^{+-}(X,p)&=-i[1-f_{\Psi}(X,p)]\rho_\Psi(X,p)\, ,\\
\Delta^{-+}(X,p)&=if_{\Psi}(X,p)\rho_\Psi(X,p)\, .
\end{align}
From the `$+-$' functions, one readily observes the appearance of Bose-enhancement terms $(1+f(X,k))$ for photons and Pauli-blocking terms $(1-f_\Psi(X,p))$ for fermions. In collision terms, these emerge attached to outgoing particles, while ingoing photons and fermions, associated with `$-+$' functions, are not distinguished in terms of their statistics.

\subsection{\label{sec:photontransport}Photon spectral function \& gauge-fixing independent photon drift term}

The photon transport equation is related to the evolution equation of the statistical photon propagator via 
\begin{align}
\label{eq:ftrace}
&\int\dif^4 (x-y)\, e^{ik(x-y)}\,\\
\nonumber
&\hspace{0.5cm}\cross\tfrac{1}{4}\eta_{\mu\nu} \Big[(F\textrm{EOM})^{\mu\nu}(x,y)-(F\textrm{EOM})^{\nu\mu}(y,x)\Big]\, ,
\end{align}
i.e.\  by a Wigner transformation of Lorentz-traced differences. Combined with the change of variables to $X$ and $p$, the Boltzmann derivative operator is recovered from the d'Alembertian in a Lorentz-invariant way by the identity
\begin{align}
\label{eq:photonboltz}
\square_x-\square_y&=2(\pdif_s\cdot\pdif_X)\leftrightarrow -2i(k\cdot\pdif_X)\, .
\end{align}
\indent By use of the convolution identity \refe{wigconv} at LO gradient expansion as well as of symmetry properties of the Wigner transforms given in appendix \ref{app:propdef}, one finds that Eq.~\refe{ftrace} reads 
\begin{align}
\label{eq:photonftrace}
&-i\tfrac{1}{4}\Big[2 (k\cdot\pdif_X)\eta_{\mu\sigma}-(1-\tfrac{1}{\xi})\\
\nonumber
&\hspace{1cm}\cross\Big( k_\mu \xpdify{}{X^\sigma}+k_\sigma \xpdify{}{X^\mu} \Big) \Big]F^{\mu\sigma}(X,k)\\
\nonumber
&=-i\tfrac{1}{4}\Big[ \Sigma^{-+}(X,k)^{\mu\sigma}D^{+-}(X,k)_{\sigma\mu}\\
\nonumber
&\hspace{1cm}-\Sigma^{+-}(X,k)^{\mu\sigma}D^{-+}(X,k)_{\sigma\mu} \Big]+\O(e^2\pdif_k\cdot\pdif_XG)\, .
\end{align}
The tracing over Lorentz indices reduces the ten equations for the components of $F^{\mu\nu}$ to a single scalar equation. In combination with the introduction of the distribution functions (\ref{eq:sKBAph}) and (\ref{eq:sKBAf}), which reduces the amount of independent tensor structures, \refe{photonftrace} is then sufficient to close the dynamics.

This transport equation \refe{photonftrace} is valid to all orders in the coupling of the 2PI loop expansion. To obtain a leading order collision term, we neglect terms of order $\O(e^4)$ to this equation. There are two types of such higher order terms: a) terms of order $\O(e^4)$ in $\Gamma_2$ discussed in Sec.~\ref{sec:loop}; b) terms of order $\O(e^2)$ in equations of motion for spectral functions contributing to the transport equations only at order $\O(e^4)$. Terms of the latter type appear in the analogous expression \refe{ftrace} for the photon spectral function, i.e.
\begin{align}
\label{eq:rhotrace2}
&-i\tfrac{1}{4}\Big[2 (k\cdot\pdif_X)\eta_{\mu\nu}-(1-\tfrac{1}{\xi})\\
\nonumber
&\hspace{1cm}\cross \Big( k_\mu \xpdify{}{X^\nu}+k_\nu \xpdify{}{X^\mu} \Big) \Big]\rho^{\mu\nu}(X,k)=0+\O(e^2)
\, .
\end{align}
The $\O(e^2)$ terms of this equation contribute only at order $\O(e^4)$ to the transport system, because the self-energies in Eq.~\refe{photonftrace} are already of order $\O(e^2)$ themselves, before being multiplied with the photon two-point function containing $\rho^{\mu\nu}$. It is therefore sufficient at order $\O(e^2)$ to employ the free $\O(e^0)$ solution of Eq.~\refe{rhotrace2} in transport equations. In this way, transport equations self-consistently resum statistical functions, but not spectral functions. The additional `collisional broadening' of spectral peaks, that does not enter the LO strong-field transport description explicitly, can then be estimated from its solutions, e.g.\ by evaluating spectral self-energies in terms of distribution functions or computing the decay rate. The gradient expansion further supports this special treatment of spectral functions, as we discuss in appendix \ref{app:specgrad}.

Employing the free photon spectral function \refe{freef} by this reasoning, the gauge-fixing dependence of the LHS of the photon transport equation \refe{photonftrace} drops out due to a cancellation between $(D_{0,\xi}^{\mu\nu}){}^{-1}$ and $\rho_{0,\xi}^{\mu\nu}$:
\begin{align}
\label{eq:xilhs}
&-i\tfrac{1}{4}\Big[2 (k\cdot\pdif_X)\eta_{\mu\sigma}-(1-\tfrac{1}{\xi})\\
\nonumber
&\hspace{1cm}\cross\Big( k_\mu \xpdify{}{X^\sigma}+k_\sigma \xpdify{}{X^\mu} \Big) \Big]F^{\mu\sigma}(X,k)\\
\nonumber
&=-i(2\pi)\sgn(k^0)\delta(k^2)2(k\cdot\pdif_X)f(X,k)+\O(e^2\pdif_p\cdot\pdif_XG)\, .
\end{align}

To obtain Boltzmann-type equations, one finally integrates over frequencies $k^0$, leading to the appearance of the on-shell distribution functions $f(X,\vec k)$ defined in \refe{osf},
\begin{align}
&\int_0^\infty\frac{\dif k^0}{(2\pi)}\int\dif^4 (x-y)\, e^{ik(x-y)}\,\tfrac{1}{4}\eta_{\mu\nu}\\
\nonumber
&\hspace{1cm}\cross\Big[(F\textrm{LHS})^{\mu\nu}(x,y)-(F\textrm{LHS})^{\nu\mu}(y,x)\Big]\\
\nonumber
&= -i\Bigg[\xpdify{}{X^0}+\frac{\vec k}{|\vec k|} \cdot \xpdify{}{\vec X} \Bigg]f(X,\vec k)+\O(e^2\pdif_k\cdot\pdif_XG)\, ,
\end{align}
where we have made use of \refe{xilhs}. This integration explicitly reduces the information that is redundant because an on-shell dispersion relation is valid for photons.

\subsection{\label{sec:4a}Fermion spectral function}

Similarly to the photon case discussed around Eq.~\refe{rhotrace2}, terms that are of order $\O(e^2)$ in
\begin{align}
\label{eq:volkoveq}
(i\slashed \pdif_x-e\slashed \A(x)-m)\rho_\Psi(x,y)=0+\O(e^2)\, ,
\end{align}
contribute only at order $\O(e^4)$ to the transport RHS that is already of order $\O(e^2)$ itself. Crucially, the field-dependent term in Eq.~\refe{volkoveq} is of order $\O(e^0)$ for strong fields and may thereby not be neglected in the $\O(e^2)$ transport description. In particular, this implies that a simple fermion particle picture may not exist in general strong-field systems. From a kinetic perspective, this is the essential way in which strong-field systems differ from weak-field systems that may still be described by free fermion spectral functions.

The $\O(e^0)$ solution $\bar \rho_\Psi[\A]$ of Eq.~\refe{volkoveq} has a functional dependence only on the macroscopic field $\A^\mu$. This is in contrast to the exact spectral solution which would be a functional also of $F_\Psi$, $F^{\mu\nu}$ and $\rho^{\mu\nu}$. Nevertheless, because of the field-independence of self-energies, the approximate spectral equation \refe{volkoveq} contains the complete explicit field dependence. This includes in particular infinite orders of field-gradients: For instance, the traced LHS reads
\begin{align}
&\int\dif^4(x-y)\,e^{ip(x-y)}\\
\nonumber
&\hspace{1cm}\cross \tfrac{1}{4}\textrm{tr}\Big[(\rho\textrm{LHS})_\Psi(x,y)+\gamma^0(\rho\textrm{LHS})_\Psi^\dagger(y,x)\gamma^0\Big]\\
\nonumber
&=i\xpdify{}{X^\mu}\rho_{\Psi,\textrm{V}}^\mu(X,p)\\
\nonumber
&+e\sum_{n=0}^{\infty}\frac{1}{(2n+1)!}\frac{1}{2^{2n}}\big[(i\pdif_p\cdot\pdif_X)^{2n+1}\A_\mu(X)\big]\rho_{\Psi,\textrm{V}}^\mu(X,p)\, .
\end{align}

In spacetime regions where the field vanishes one recovers from Eq.~\refe{volkoveq} the free particle description (\ref{eq:freef}), i.e.\ \mbox{$\bar \rho_\Psi[\A=0]=\rho_{\Psi,0}$}. Eq.~\refe{volkoveq} may therefore be understood as the strong-field generalization of a fermion particle picture. In particular, since the difference between $\rho_{\Psi,0}$ and $\bar \rho_\Psi[\A]$ is of order $\O(e\A)$, one would be allowed to exchange the two in a leading order description for weak fields in accordance with a near-equilibrium quasiparticle picture \cite{Blaizot:1996az, Blaizot:1996hd}.

Rephrasing the equation of motion for the fermion spectral function into an equation for the retarded propagator, $\Delta_\textrm{R}(x,y)\defeq \theta(x^0-y^0)\rho_\Psi(x,y)$, one finds that
\begin{align}
\label{eq:bw}
&[\slashed p-m-\Sigma_{\textrm{R}}(X,p)]\Delta_{\textrm{R}}(X,p)\\
\nonumber
&\simeq-1+e\int_k\slashed{\bar A}(X,k)\Delta_{\textrm{R}}(X,p+k)+\O(e^2\pdif_p\cdot\pdif_XG)\, ,
\end{align}
with $\bar A^\mu(x,y)\defeq \tfrac{1}{2}[\A^\mu(x)+\A^\mu(y)]$. In the zero-field case, this equation implies that the spectral function has a peaked shape with a `width' given by the \textit{square} of the spectral $\O(e^2)$ self-energy \cite{Blaizot:1997kw}
\begin{align}
\label{eq:polop}
-i\Sigma^{(\rho)}_\Psi(X,p)&=e^2\int_{q,k}(2\pi)^4\delta(k-p+q)\\
\nonumber
&\hspace{-1.5cm}\cross \big[1-f_\Psi(X,q)+f(X,k)\big]\gamma_\mu \rho_\Psi(X,q)\gamma_\nu\rho^{\mu\nu}(X,k)\, ,
\end{align}
which is indeed $\O(e^4)$, as anticipated by our counting of couplings in the equations of motion. In the strong-field case, the off-diagonal momentum structure of the field term in Eq.~\refe{bw} highlights the absence of a simple peak structure of general solutions of Eq.~\refe{volkoveq}. Eq.~\refe{bw} further shows that the physical reason for this more complex structure is four-momentum exchange between the retarded fermion propagator and the macroscopic field.

We give an analytical solution of Eq.~\refe{volkoveq} under the assumption of strong external plane-wave fields in Sec.~\ref{subsec:5c}, which allows us to showcase the appearance of exponentials $\exp(\O(e\A))$, that resum the field-vertex \refe{fieldvertex} as desired. By employing the solution $\bar\rho_\Psi[\A]$ of Eq.~\refe{volkoveq} in transport equations, one recovers $\O(e^2)$ strong-field scattering amplitudes in limiting cases (see Sec.~\ref{par:vertex}). A particle picture emerges only in special cases and can change with time (see Sec.~\ref{subsec:urlimit}). 

\subsection{\label{sec:4b}Strong-field photon transport equation}

\subsubsection{Collision term}

To obtain the $\O(e^2)$ strong-field photon collision term from the expression \refe{photonftrace} we need the leading order self-consistent photon self-energies, i.e.
\begin{align}
\label{eq:ph2pise1}
\Sigma^{+-}_{\mu\nu}(X,k)&=e^2\int_{p,q}(2\pi)^4\delta(k-p+q)\\
\nonumber
&\hspace{0.5cm}\cross[1-f_\Psi(X,p)]f_\Psi(X,q)\\
\nonumber
&\hspace{0.5cm}\cross \textrm{tr} \left\{\gamma_\mu\rho_\Psi(X,p)\gamma_\nu\rho_\Psi(X,q)\right\}+\O(e^4)\, ,\\
\label{eq:ph2pise2}
\Sigma^{-+}_{\mu\nu}(X,k)&=e^2\int_{p,q}(2\pi)^4\delta(k-p+q)\\
\nonumber
&\hspace{0.5cm}\cross f_\Psi(X,p)[1-f_\Psi(X,q)]\\
\nonumber
&\hspace{0.5cm}\cross \textrm{tr}\left\{ \gamma_\mu\rho_\Psi(X,p)\gamma_\nu\rho_\Psi(X,q) \right\}+\O(e^4)\, .
\end{align}
The structure of the strong-field photon transport equation is that of Eq.~\refe{ftrace} integrated over positive frequencies, $\int_0^{\infty}\dif k^0/(2\pi)$. Spectral functions are evaluated from their equations of motion with the reasoning discussed in the previous paragraphs, i.e.
\begin{align}
\rho_{\Psi}(X,p)&\rightarrow \bar \rho_{\Psi}[\A](X,p)\, ,\\
\rho^{\mu\nu}(X,k)&\rightarrow \rho^{\mu\nu}_{0,\xi}(X,k)\, ,
\end{align}
where $\bar \rho_\Psi[\A]$ denotes the solution of Eq.~\refe{volkoveq}. The $\O(e^2)$ photon transport equation then reads
\begin{align}
\label{eq:sfpkin}
\Bigg[\xpdify{}{X^0}+\frac{\vec k}{|\vec k|} \cdot \xpdify{}{\vec X} \Bigg]f(X,\vec k)&=C[\A](X,\vec k)\\
\nonumber
&\hspace{0.3cm}+\O(e^2\pdif_k\cdot\pdif_XG)+\O(e^4)\, ,
\end{align}
where the $\O(e^2)$ strong-field photon collision term is
\begin{align}
\nonumber
C[\A](X,\vec k)&=e^2\int_0^{\infty} \dif k^0\int_{p,q}(2\pi)^6 \delta(k-p+q)\\
\nonumber
&\hspace{-1.5cm}\cross\Big\{ f_\Psi(X,p)[1-f_\Psi(X,q)][1+f(X,k)]\\
\nonumber
&\hspace{-1cm}-[1-f_\Psi(X,p)]f_\Psi(X,q)f(X,k) \Big\}\\
\label{eq:photontransport}
&\hspace{-1.5cm}\cross\Big\{\P[\A](X,p,q,k)-\P_\xi[\A](X,p,q,k)\Big\}\, ,
\end{align}
with the trace
\begin{align}
\P\defeq {\P^{\mu}}_{\mu}\, ,
\end{align}
and the longitudinal projection
\begin{align}
\label{eq:pxi}
\P_\xi \defeq (1-\xi)\tfrac{1}{k^2}k_\mu k_\nu\P^{\mu\nu}
\end{align}
of the $ee\gamma$-collision kernel
\begin{align}
\label{eq:kernel}
&\P^{\mu\nu}[\A](X,p,q,k)=-(2\pi)^{-2}\delta(k^2)\sgn(k^0)\\
\nonumber
&\hspace{2cm}\cross\tfrac{1}{4}\textrm{tr}\{ \gamma^\mu i\bar \rho_{\Psi}[\A](X,p)\gamma^\nu i\bar \rho_{\Psi}[\A](X,q) \}\, .
\end{align}
\indent This general expression derived from quantum field theory plays the role of a generalized scattering amplitude squared that has its own equation of motion [Eq.~\refe{volkoveq} or equivalently Eq.~\refe{allgrad} below] and is adapted to the properties of the macroscopic field at each instance of time. This goes beyond previous approaches that have so far been restricted by additional assumptions on the macroscopic field. In particular, it provides a prescription of how to implement an inhomogeneous macroscopic field in local transport equations. We achieved this by describing collisions in terms of a dynamical strong-field fermion spectral function, which includes all leading order effects. This approach allows for many links to existing literature as we demonstrate in Sec.~\ref{sec:5}. In particular, the collision kernel \refe{kernel} may be reduced to scattering amplitudes computable from Feynman rules in strong-field QED (see Sec.~\ref{par:vertex}).

The collision term \refe{photontransport} features the factorization of interaction terms into a collision kernel and a gain-minus-loss term familiar from traditional kinetic equations. While the photon distribution functions can be reduced to on-shell distributions \refe{osf} by virtue of the delta function $\delta(k^2)$ in $\P^{\mu\nu}$, this is not in general possible for the fermion distribution function. The off-shell frequency dependence of the latter is computed dynamically by solving the transport system coupled to the fermion spectral equation \refe{volkoveq}. This allows the collision kernel, to adjust in time to a self-consistent macroscopic field as the system evolves, while still being local in the kinetic position variable $X$ without relying on locally constant fields.

\subsubsection{Strong-field photon decay rate}

By linearizing and integrating the photon transport equation over position and external momentum, one may find the field-dependent decay rate $\gamma$ of a photon with momentum $k^i$ and position $X^i$ at time $t\defeq X^0$,
\begin{align}
\label{eq:photondecayrate}
\pdif_tN(t)\simeq -\int_{\vec X,\vec k}\gamma[\A](X,\vec k)\,f(X,\vec k)\, ,
\end{align}
with the photon number \refe{photonnumber}. Such a linearization may be achieved e.g.\ under the assumption that the system is close to vacuum (i.e.\ for small distribution functions, see Sec.~\ref{subsec:smallocc}) or in linear response theory around equilibrium \cite{Weldon:1983jn, Blaizot:2001nr, Berges:2005ai}, $f(X,k)=f_{\textrm{B}}(k^0)+\delta f(X,k)$, $f_\Psi(X,p)=f_{\textrm{F}}(p^0)$ with $f_{\textrm{F}/\textrm{B}}(p^0)=1/(e^{\beta p^0}\pm 1)$. In equilibrium, gain-minus-loss terms vanish by energy conservation, $q^0=p^0-k^0$,
\begin{align}
&f_{\textrm{F}}(p^0)[1-f_{\textrm{F}}(p^0-k^0)][1+f_{\textrm{B}}(k^0)]\\
\nonumber
&\hspace{1.5cm}-[1-f_{\textrm{F}}(p^0)]f_{\textrm{F}}(p^0-k^0)f_{\textrm{B}}(k^0)\equiv 0\, ,
\end{align}
resulting in the photon equilibrium decay rate
\begin{align}
\nonumber
\gamma_{\textrm{eq}}[\A](X,\vec k)&=e^2\int_0^{\infty}\dif k^0\int_p\Big\{f_{\textrm{F}}(p^0)-f_{\textrm{F}}(p^0-k^0)\Big\}\\
\label{eq:eqrate}
&\hspace{-2.0cm}\cross (2\pi)^2\Big\{\P[\A](X,p,p-k,k)-\P_\xi[\A](X,p,p-k,k)\Big\}\, .
\end{align}

\subsection{\label{sec:4ferm}Strong-field fermion transport equation}

Here, we derive the fermion equations that close the transport system in terms of off-shell fermion and on-shell photon distribution functions.

\subsubsection{\label{sec:inv}Gauge-invariant fermion correlation functions}

The presence of a macroscopic field complicates the gauge-invariance of approximations such as the gradient expansion. This was not an issue in the case of the photon equations where the field is only implicit via $\bar \rho_\Psi[\A]$ and the photon self energies are gauge-invariant. In the following, before repeating the analogous steps for the fermion transport equation, we express all fermion equations in terms of the gauge-invariant field strength tensor $\F^{\mu\nu}=\pdif^\mu\A^\nu-\pdif^\nu\A^\mu$, or equivalently in terms of electric and magnetic fields,
\begin{align}
-\mathcal{F}^{0i}&\eqdef \E^i\, ,\\
%(\vec{\mathcal{B}})^i&\defeq
%-\tfrac{1}{2}\varepsilon^{ijk}
%\mathcal{F}_{jk}&\eqdef %\lra
-\mathcal{F}_{ij}&\eqdef \varepsilon_{ijk}\mathcal{B}^k
\, .
\end{align}
This is necessary, in particular, in order to identify a gauge-invariant fermion drift term that contains the gauge-invariant Lorentz force.

One can achieve gauge-invariance (as opposed to covariance) by introducing Wilson lines\footnote{In contrast to the operator Wilson lines e.g.\ of Refs.~\cite{Elze:1986hq, Elze:1986qd}, the Wilson line \refe{wilsongen} is built only from the one-point function, but is here employed alongside higher correlations that give rise to collisions without a mean-field (`Hartree-Fock') approximation.}
\begin{align}
\label{eq:wilsongen}
\W_\Gamma(y,x)\defeq \exp\Bigg(ie\int_\Gamma\dif z^\mu\mathcal{A}_\mu(z)\Bigg)\, ,
\end{align}
with $\Gamma$ indicating the path of integration from $y$ to $x$. The gauge transformation of a Wilson line exactly compensates the gauge transformation of fermion two-point functions, such that the quantities
\begin{align}
\hat F_{\Psi,\Gamma}(x,y)\defeq \W_\Gamma(y,x)F_\Psi(x,y)\, ,\\
\label{eq:invrho}
\hat \rho_{\Psi,\Gamma}(x,y)\defeq \W_\Gamma(y,x)\rho_\Psi(x,y)
\end{align}
are gauge-invariant (but path-dependent). It is well known that straight Wilson lines, $\W\defeq \W_{\Gamma=[x,y]}$, facilitate a derivation of gauge-invariant transport equations \cite{Vasak:1987um, Elze:1986hq, Elze:1986qd, Blaizot:1999xk}. Following this approach, we employ
\begin{align}
\W(y,x)= \exp\Bigg( ie s^\mu \int_{-\tfrac{1}{2}}^{\tfrac{1}{2}}\dif \lambda \,\A_\mu(X+\lambda s) \Bigg),
\end{align}
and express everything in terms of gauge-invariant late-time Wigner functions
\begin{align}
\hat G_\Psi(X,p)\defeq \int_se^{ips}\,\W(y,x)G_\Psi(x,y)\, .
\end{align}
Invariant and covariant Wigner functions are related by
\begin{align}
\label{eq:wigrel}
e^{iw(X,p)}\hat G_\Psi(X,p)=G_\Psi(X,p)
\end{align}
with the real differential operator
\begin{align}
w(X,p)\defeq ie\int_{-\tfrac{1}{2}}^{\tfrac{1}{2}}\dif \lambda \,[e^{-i\lambda (\pdif_p\cdot\pdif_X)}\A_\mu(X)]\xpdify{}{p_\mu}\, .
\end{align}
By virtue of 
\begin{align}
w(X,p)=ie\A_\mu(X)\pdif_p^\mu+\O(e^0\pdif_p\cdot \pdif_X)
\end{align}
this relation is simple for small field-gradients (which we discuss in Secs.~\ref{subsec:5a} and \ref{subsec:fermkinplane}) in which case it becomes the translation
\begin{align}
\label{eq:kincanwig}
\hat G_\Psi(X,p)&=G_\Psi(X,p+e\A(X))+\O(e^0\pdif_p\cdot \pdif_X)\, .
\end{align}

One now has to decide whether to identify fermion distribution functions in terms of $F_\Psi$ and $\rho_\Psi$ as in \refe{sKBAf} or in terms of $\hat F_\Psi$ and $\hat \rho_\Psi$, i.e.
\begin{align}
\label{eq:tildef}
\hat F_\Psi(X,p)&=-i[\tfrac{1}{2}-\tilde f_\Psi(X,p)]\hat \rho_{\Psi}(X,p)\, .
\end{align}
In principle, $f_\Psi$ and $\tilde f_\Psi$ are arbitrary definitions which can be translated into each other. In particular for small field-gradients one would have
\begin{align}
\label{eq:fftilde}
\tilde f_\Psi(X,p)&=f_\Psi(X,p+e\A(X))+\O(e^0\pdif_p\cdot\pdif_X)\, .
\end{align}

In photon equations, the distinction between co- and invariant fermion functions is redundant. This is because, by virtue of  
\begin{align}
\label{eq:wgrp1}
\W(x,y)\W(y,x)\equiv 1\, ,
\end{align}
one may replace co- and invariant Wigner functions in the gauge-invariant photon self-energy that features a fermion loop%\footnote{This is not simply a consequence of the gauge-invariance of the photon self-energy: There are gauge-invariant quantities, for example $\hat \Delta(x,y)\hat\Delta(y,z)\hat\Delta(z,x)$, that do not have this property.}
, i.e.
\begin{align}
\nonumber
\Sigma^{\mu\nu}(x,y)&=e^2\textrm{tr}\{\gamma^\mu\Delta(x,y)\gamma^\nu\Delta(y,x)\}\\
\label{eq:photonreplace}
&=e^2\textrm{tr}\{\gamma^\mu\hat \Delta(x,y)\gamma^\nu\hat \Delta(y,x)\}\, .
\end{align}
In Wigner space this involves two fermion momentum integrals and a delta function. In particular, the fact that
\begin{align}
\label{eq:photoninv}
&\int_{p,q}\delta(k-p+q)\,\rho_\Psi(X,p)\rho_\Psi(X,q)\\
\nonumber
&=\int\dif^4(x-y)e^{ik(x-y)}\,\hat \rho_\Psi(x,y)\hat \rho_\Psi(y,x)\, ,
\end{align}
implies that one may replace $f_\Psi$ with $\tilde f_\Psi$ if $\rho_\Psi$ is replaced with $\hat \rho_\Psi$ in the photon collision kernel \refe{kernel}. Similarly, because
\begin{align}
\label{eq:wgrp2}
\lim_{s\rightarrow 0}\W(X+\tfrac{s}{2},X-\tfrac{s}{2})\equiv 1\, ,
\end{align}
such that
\begin{align}
\nonumber
j^\mu(X)&=-e\lim_{s\rightarrow 0}\textrm{tr}\{\gamma^\mu F_\Psi(X+\tfrac{s}{2},X-\tfrac{s}{2})\}\\
\label{eq:hatj}
&=-e\lim_{s\rightarrow 0}\textrm{tr}\{\gamma^\mu \hat F_\Psi(X+\tfrac{s}{2},X-\tfrac{s}{2})\}\, ,
\end{align}
this may also be done for the current \refe{jX} in the Maxwell equation \refe{micmax}. In this way, one obtains a closed set of equations in terms of fermion distributions of the $\tilde f_\Psi$-type to any order of field-gradients. We stress that these replacements do not work in reverse (going from $\tilde f_\Psi$ to $f_\Psi$) for the fermion equations to be discussed below, such that a practicable description in terms of $f_\Psi$-type distributions would have to rely on small field-gradients by relying on Eq.~\refe{fftilde}.

\subsubsection{Gauge-invariant equations of motion:\\ 2PI vs.\ Wigner operator formalism}
Having introduced gauge-invariant correlation functions, we can express the gauge-covariant 2PI fermion equations of motion in a gauge-invariant way. We start with the equation for the fermion spectral function,
\begin{align}
\label{eq:completespectral}
&\int\dif^4(x-y)\,e^{ip(x-y)}\W(y,x)(\rho\textrm{EOM})_\Psi(x,y)\, ,
\end{align}
explicitly at our order of interest,
\begin{align}
\label{eq:allgrad}
&\Big[\tfrac{i}{2}\slashed \nabla+\slashed \Pi-m\Big]\hat \rho_\Psi(X,p)=0+\O(e^2)\, .
\end{align}
Here we have employed the commuting, real and gauge-invariant differential operators introduced in Ref.~\cite{Vasak:1987um},
\begin{align}
\label{eq:VGEop}
\nabla_\mu&\defeq \xpdify{}{X^\mu}-e\int_{-\tfrac{1}{2}}^{\tfrac{1}{2}}\dif \lambda\, [e^{-i\hbar\lambda (\pdif_p\cdot\pdif_X)}\F_{\mu\nu}(X)]\xpdify{}{p_\nu}\\
\label{eq:VGEop2}
\Pi_\mu&\defeq p_\mu-ie\int_{-\tfrac{1}{2}}^{\tfrac{1}{2}}\dif \lambda \, \lambda\,[e^{-i\hbar\lambda (\pdif_p\cdot\pdif_X)}\F_{\mu\nu}(X)]\xpdify{}{p_\nu}\, .
\end{align}
Using anti-hermiticity, Eq.~\refe{rs2}, one may verify in particular that solutions of Eq.~\refe{allgrad} satisfy
\begin{align}
\label{eq:nablarho}
i\nabla_\mu\hat\rho^{\,\mu}_\Psi(X,p)&=0\, ,\\
\label{eq:nablarho2}
\textrm{tr}\{(\slashed \Pi-m)\hat\rho_\Psi(X,p)\}&=0\, .
\end{align}
The second condition, which is satisfied by any strong-field solution, is much weaker than the on-shell condition in the absence of a field, $(\slashed p-m)\rho_{\Psi,0}(X,p)=0$.

Eq.~\refe{allgrad} is proven as in the Wigner operator formalism of Refs. \cite{Vasak:1987um, Zhuang:1995pd, PhysRevD.83.065007}. While the Wigner operator formalism has not been able to provide closed collision terms, the 2PI formalism is able to achieve this: Instead of discussing equations for the normal-ordered product, $\braket{:\hspace{-0.1cm}\Psi(x)\bar\Psi(y)\hspace{-0.1cm}:}$, resulting in real and imaginary parts with different differential operators \cite{Vasak:1987um, Zhuang:1995pd}, we distinguish real and imaginary parts of the time-ordered product $\braket{\T_\C\Psi(x)\bar\Psi(y)}$ \refe{t2}, i.e.\ statistical and spectral functions. Their 2PI equations of motion \refe{eom3} -- \refe{eom4} do not differ by their differential operators, but by the integral structure of their RHS, which automatically ensures the correct hermiticity properties of their solutions, \refe{rs1} and \refe{rs2}. Because of the absence of these RHS integrals in the approximated spectral equation \refe{volkoveq}, the anti-hermiticity \refe{rs2} of the approximate solution has to be prescribed. In fact at 1-loop, i.e.\ by neglecting collisions, the equations for $F_\Psi$ and $\rho_\Psi$ without (anti)hermiticity constraints are equivalent and the equation for the fermion statistical function alone is sufficient to discuss transport phenomena as has been done e.g.\ in Ref.~\cite{PhysRevD.82.105026}. Going to order $\O(e^2)$, the self-energy terms of the 2-loop equations for the spectral functions still do not contribute to the kinetic equations as discussed in Sec.~\ref{sec:4a}, but the self-energy terms of the statistical equations provide collision terms.

\subsubsection{\label{sec:4c}Quantum Vlasov term}
In order to obtain a gauge-invariant fermion transport equation, we consider
\begin{align}
\label{eq:fermdriftnotrace}
&\int\dif^4(x-y)\,e^{ip(x-y)}\,\W(y,x)\cross\\
\nonumber
&\hspace{1cm}\cross \Big[(F\textrm{LHS})_\Psi(x,y)-\gamma^0(F\textrm{LHS})_\Psi^\dagger(y,x)\gamma^0\Big]\\
\nonumber
&\hspace{0.5cm}=\tfrac{i}{2}\nabla_\mu\{\gamma^\mu,\hat{F}_\Psi(X,p) \}+\Pi_\mu[\gamma^\mu,\hat{F}_\Psi(X,p)]\, ,
\end{align}
where (anti-)commutators are taken in Dirac space. By building differences, the fermion mass drops out of this expression, but enters again via the spectral equation \refe{allgrad}. By taking the trace of \refe{fermdriftnotrace} we obtain the all-order in field-gradients quantum Vlasov term
\begin{align}
\label{eq:transportdrift}
\nabla_\mu\hat{F}^{\,\mu}_\Psi(X,p)&=C_\Psi(X,p)\\
\nonumber
&\hspace{-1cm}+\O(e^2\pdif_p\cdot\pdif_XG)+\O(e^4)\, ,
\end{align}
to which the commutator term with $\Pi^\mu$ does not contribute. In \refe{transportdrift} we have indicated the fermion collision term, which we compute to leading order below.

Employing Eq.~\refe{nablarho}, the fermion transport equation \refe{transportdrift} in terms of $\tilde f_\Psi$ then reads
\begin{align}
\label{eq:fermiontransport}
&\nabla_\mu[\tilde f_\Psi(X,p)i\hat \rho^{\,\mu}_\Psi(X,p)]=C_\Psi(X,p)\\
\nonumber
&\hspace{2cm}+\O(e^2\pdif_p\cdot \pdif_XG)+\O(e^4)\, .
\end{align}
The off-shell all-gradient drift term of this equation goes beyond a Lorentz force description, which it contains as its on-shell contribution (see Secs.~\ref{par:classical} and \ref{subsec:urlimit}). The emergence of this fermion drift term is distinctly different from the photon case, 
because fermion derivatives involve the macroscopic field and are first order already in the fundamental equations of motion. In particular, the momentum factor of $(p\cdot \pdif_X)$, that emerges automatically for photons via the identity \refe{photonboltz}, has to be provided by the vector component of the free fermion spectral function. Without an on-shell approximation, momentum derivatives of the spectral function in Eq. \refe{fermiontransport} are physically regulated by the macroscopic field.

\subsubsection{\label{sec:4d}Collision term \& charge conservation}

Having discussed the LHS, we now derive the gauge-invariant collision term already indicated in Eq.~\refe{fermiontransport}.

In general, gauge-invariance of the convolutions on the fermion spectral and statistical RHS is achieved by writing
\begin{align}
\label{eq:invconv}
&\W(y,x)\int_z\,  \Sigma_\Psi(x,z) G_\Psi(z,y)\\
\nonumber
&=\int_zL(x,y,z)\hat \Sigma_\Psi(x,z)\hat G_\Psi(z,y)\, ,
\end{align}
where we have identified the (triangle) Wilson loop
\begin{align}
L(x,y,z)\defeq \W(y,x)\W(x,z)\W(z,y)\, .
\end{align}
By virtue of Eqs.~\refe{wigconv} and \refe{wigrel}, the LO of the gradient expansion of this gauge invariant convolution is \cite{Blaizot:1999xk}
\begin{align}
\nonumber
&\int\dif^4(x-y)\, e^{ip(x-y)}\, \int_zL(x,y,z)\hat \Sigma_\Psi(x,z)\hat G_\Psi(z,y)\\
\label{eq:plaqgrad}
&\hspace{2cm}=\hat \Sigma_\Psi(X,p)\hat G_\Psi(X,p)\\
\nonumber
&-\tfrac{i}{2}e\F^{\mu\nu}(X)\xpdify{\hat \Sigma_\Psi(X,p)}{p^\mu}\xpdify{\hat G_\Psi(X,p)}{p^\nu}+\O\big(e^2\pdif_p\cdot\pdif_XG\big)\, .
\end{align}
For weak fields near equilibrium the additional term
	\begin{align}
	\label{eq:pterm}
	e\F^{\mu\nu}(X)\xpdify{\hat \Sigma_\Psi(X,p)}{p^\mu}\xpdify{\hat G_\Psi(X,p)}{p^\nu}
	\end{align}
	as compared to the covariant convolution, \mbox{$\int_se^{ips}(\Sigma_\Psi*G_\Psi)=\Sigma_\Psi(X,p)G_\Psi(X,p)+\O(e^2\pdif_p\cdot\pdif_XG)$}, is effectively of order $\O(e^4)$ and compatible with a kinetic description \cite{Blaizot:1999xk}. To focus on the part of the fermion RHS that contains the collision term indicated in Eq.~\refe{transportdrift},
	\begin{align}
	\label{eq:invfermcoll}
	C_\Psi(X,p)&\defeq-\tfrac{1}{4}\textrm{tr}\,\Big[ \hat \Sigma_\Psi^{-+}(X,p)\hat \Delta^{+-}(X,p)\\
	\nonumber
	&\hspace{1.5cm}-\hat \Sigma^{+-}_\Psi(X,p)\hat \Delta^{-+}(X,p)  \Big]\, ,
	\end{align}
	we drop terms of the type \refe{pterm} also in the presence of strong fields. We stress that the validity of dropping these terms in a far-from-equilibrium system requires further investigation.\footnote{As discussed in Ref.~\cite{Blaizot:1999xk} terms of the form \refe{pterm} have the effect of accounting for further off-shell corrections and replace the spatial derivative $\pdif_X\rightarrow \pdif_X-e\F^{\mu}{}_{\nu} \pdif_p^\nu$ in Poisson brackets. Alternatively, one may think of dropping these terms as setting the Wilson loop to one, $L\approx 1$. Because of the group properties \refe{wgrp1}, \refe{wgrp2} and $\W(x,z)\W(z,y)=\W(x,y)$ if $z\in [x,y]$
	this is a good approximation if the dominant contributions in $z$ are sufficiently close to the straight line $[x,y]$ because $L(x,y,z)\equiv 1$ if $z\in [x,y]$.}

At leading order, the gauge-invariant self-energies in Eq.~\refe{invfermcoll} may be written as
\begin{align}
\label{eq:f2pise1}
\hat \Sigma_\Psi^{+-}(X,p)&=e^2\int_{q,k}(2\pi)^4\delta(k-p+q)\\
\nonumber
&\hspace{0.5cm}\cross [1-\tilde f_\Psi(X,q)][1+f(X,k)]\\
\nonumber
&\hspace{0.5cm}\cross \gamma^\mu\hat \rho_\Psi(X,q)\gamma^\nu \rho_{\mu\nu}(X,k)+\O(e^4)\, ,\\
\label{eq:f2pise2}
\hat \Sigma_\Psi^{-+}(X,p)&=-e^2\int_{q,k}(2\pi)^4\delta(k-p+q)\\
\nonumber
&\hspace{0.5cm}\cross \tilde f_\Psi(X,q)f(X,k)\\
\nonumber
&\hspace{0.5cm}\cross \gamma^\mu\hat \rho_\Psi(X,q)\gamma^\nu \rho_{\mu\nu}(X,k)+\O(e^4)\, ,
\end{align}
The strong-field $\O(e^2)$ fermion collision term then reads
\begin{align}
\label{eq:fermcoll}
C_\Psi[\A](X,p)&=e^2\int_{q,k}(2\pi)^7\delta(k-p+q)\\
\nonumber
&\hspace{-1.5cm}\cross \Big\{\tilde f_\Psi(X,q)f(X,k)[1-\tilde f_\Psi(X,p)] \\
\nonumber
&\hspace{-1cm}-  [1-\tilde f_\Psi(X,q)][1+f(X,k)] \tilde f_\Psi(X,p)\Big\}\\
\nonumber
&\hspace{-1.5cm}\cross\Big\{\tilde \P[\A](X,p,q,k)-\tilde \P_\xi[\A](X,p,q,k)\Big\}+\O(e^4)\, ,
\end{align}
where $\tilde \P$ is obtained from the collision kernel \refe{kernel} by exchange of $\rho_{\Psi}\rightarrow \hat \rho_{\Psi}$ with the solution $\hat \rho_\Psi$ of Eq.~\refe{allgrad}, or at LO in field-gradients via \begin{align}
\label{eq:kernelsmallgrad}
\tilde \P(X,p,q,k)&=\P(X,p+e\A(X),q+e\A(X),k)\\
\nonumber
&\hspace{1cm}+\O(e^0\pdif_p\cdot\pdif_X)\, .
\end{align}

As anticipated in Sec.~\ref{sec:inv}, while the photon collision term is gauge-invariant also without this replacement, the fermion collision term is not. This is because gauge-invariance requires integration over \textit{both} fermion momenta according to \refe{photoninv}. Indeed, if we integrate the fermion transport equation over its external momentum, subtleties of gauge-invariance are absent and, with
\begin{align}
\label{eq:infmom}
\int_p\pdif_p^\nu\hat F_\Psi(X,p)=0\, ,
\end{align}
and using \refe{hatj}, we can recover the Maxwell current \refe{Fj} in the fermion transport equation via
\begin{align}
\label{eq:jfromF}
-4e\int_p\nabla_\mu \hat F_\Psi^\mu(X,p)= \pdif_\mu j^\mu(X)\, .
\end{align}
As a consequence of the U(1) symmetry of QED, this current is conserved by the fundamental equations, as well as by our approximate transport equations, such that the total electric charge \refe{Qdef} is constant,
\begin{align}
\label{eq:chargecoll}
\pdif_t Q(t)= -4e\int\dif^3X\int_pC_\Psi(X,p)=0\, ,
\end{align}
with $t\defeq X^0$. To see this, one may verify that the relabeling $q\leftrightarrow p$ and $k\rightarrow -k$ leaves both the delta function and the gain-minus-loss term invariant [by virtue of \refe{photoncp}], but changes the sign of the collision kernel (also without tilde), i.e.
\begin{align}
\tilde \P^{\mu\nu}(X,p,q,k)=-\tilde \P^{\nu\mu}(X,q,p,-k)\, .
\end{align}
\subsection{\label{sec:4e}Transport Maxwell equation\\\& gauge-fixing dependence}
The free photon propagator $D_{0,\xi}^{\mu\nu}$ \refe{dzeroxi} and spectral function $\rho^{\mu\nu}_{0,\xi}$ \refe{freef} introduce a gauge-fixing dependence. This $\xi$-dependence is distributed over several equations of motion by virtue of $\P_\xi$ \refe{pxi} and the solution
\begin{align}
\label{eq:axi}
\A_\xi^\mu(x)=-i\int_yD^{\mu\nu}_{0,\xi}(x,y)j_\nu(y)
\end{align}
of the Maxwell equation \refe{micmax} with the late-time current
\begin{align}
\label{eq:jX}
j^\mu(X)=2e\int_p[1-2\tilde f_\Psi(X,p)]i \hat \rho^{\,\mu}_{\Psi}(X,p)\, .
\end{align}

There are two ways in which $\xi$-dependence is controlled. Firstly, starting from the 2PI effective action, a perturbative coupling expansion shows that the total $\xi$-dependence of $\P_\xi[\A_\xi]$ is always of higher perturbative order in $e$ \cite{PhysRevD.66.065014, Carrington:2003ut, Borsanyi:2007bf}. Indeed, for a free fermion spectral function $\rho_{{\Psi},0}$, leading order collisions are trivially gauge-fixing independent,
\begin{align}
\label{eq:freegaugefix}
\delta(k-p+q)\P_\xi(X,p,q,k)\xrightarrow{\bar \rho_\Psi\rightarrow \rho_{{\Psi},0}} 0\, .
\end{align}

Secondly, the $\xi$-dependence can drop out for on-shell photons \cite{Carrington:2007fp} (see also Eq.~\refe{xilhs}). We demonstrate this also in the strong-field case by virtue of Ward identities for scattering amplitudes that emerge in the kinetic approximation and play the role of redressed 1PI vertices. We make contact with such strong-field Ward identities \cite{Boca_2010, PhysRevA.83.032106, Ilderton:2010wr} in the case of plane-wave fields in Sec.~\ref{par:vertex}, where the $\xi$-dependence then drops out in a corresponding limit. A general proof for cancellations between $(D_{0,\xi}^{\mu\nu})^{-1}$ in $\A_\xi^\mu$ and $\rho_{0,\xi}^{\mu\nu}$ in $\P_\xi$ in the self-consistent strong-field case $\P_\xi[\A_\xi]$ seems highly non-trivial.
\vspace{1cm}

\begin{figure}[h!]
	\begin{center}
		\includegraphics[scale=0.43]{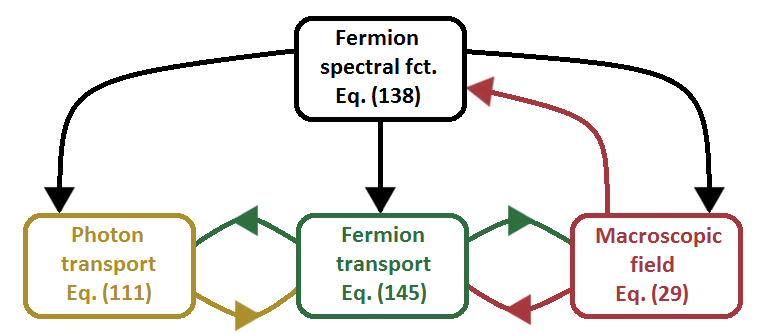}
	\end{center}
	\caption{\label{fig:1}Structure of the strong-field transport system. }
\end{figure}

A summary of the interconnections among the extended transport system which we have now arrived at is graphically presented in Fig.~\ref{fig:1}. The transport equations for photons [Eq.~\refe{sfpkin} for $f(X,\vec k)$] and fermions [Eq.~\refe{fermiontransport} for $\tilde f_\Psi(X,p)$] couple to eachother via the collision terms \refe{photontransport} and \refe{fermcoll}. They are supplemented by the Maxwell equation for the macroscopic field [Eq.~\refe{micmax} or equivalently Eq.~\refe{axi} for $\A^\mu(X)$] and the equation for the fermion spectral function [Eq.~\refe{allgrad} for $\hat \rho_\Psi(X,p)$ or equivalently Eq.~\refe{volkoveq} for $\rho_\Psi$], which couples to the Maxwell equation via the current \refe{jX}. The macroscopic field enters the fermion spectral and transport equation explicitly via the strong-field derivatives \refe{VGEop} and \refe{VGEop2}, and the photon and fermion transport equations implicitly via the strong-field fermion spectral function in the scattering kernel \refe{kernel}.

\section{\label{sec:5}Strong-field QED kinetic equations}

In this section, we investigate ways to further approximate the transport system of Sec.~\ref{sec:4} and how to reduce it to Boltzmann-type equations with scattering amplitudes by considering limiting cases of the collision kernel. To this end, we discuss various common additional approximations in strong-field QED, namely small field-gradients (locally constant fields), classical fermion propagation (Lorentz force), external plane-wave fields (Volkov states), near-vacuum physics (small occupations), as well as fermion distributions that are peaked at large momenta (ultrarelativistic limit). In particular, the ultrarelativistic limit finally allows us to make contact with fermion on-shell descriptions [e.g. Ref.~\cite{PhysRevSTAB.14.054401}], which are valid if a long-lived separation of scales exists (see Sec.~\ref{subsec:urlimit}).

\subsection{\label{subsec:5a}The case of small field-gradients}

So far, our transport equations have been infinite order in gradients of the macroscopic field. In a physical situation with small field-gradients, one can simplify the collision kernels and the fermion drift term. We demonstrate how to do this at the level of the equations for the fermion spectral and statistical functions in the following.\footnote{A collisionless discussion of field-gradients can be found in Ref.~\cite{PhysRevD.82.105026}, where it is shown that field-gradients can enhance pair production rates in particular for low momenta.} For this purpose we assume in this section that
\begin{align}
\label{eq:smallfieldgrad}
|(s\cdot \pdif_X)\F^{\mu\nu}|\ll |\F^{\mu\nu}|\,.
\end{align}
This means we only keep LO terms $\O\big(e^0(\pdif_p\cdot \pdif_X)^0\big)$ and truncate the NLO $\O(e^0\pdif_p\cdot\pdif_X)$ of gauge-invariant field-gradients (see appendix \ref{app:gaugegrad} for a comparison of approximations to invariant and covariant field-gradients).

We can simplify the fermion spectral equation of motion \refe{allgrad} and in turn the collision kernel \refe{kernel} by using \refe{smallfieldgrad}. The exponential derivatives of the differential operators \refe{VGEop} and \refe{VGEop2} allow for an expansion in terms of gradients of the field-strength tensor. Thereby one can explicitly compute the first orders of the $\lambda$ integrals, i.e.~\cite{Vasak:1987um, Zhuang:1995pd}
\begin{align}
	\nabla_\mu(X,p)&=\xpdify{}{X^\mu}-e\F_{\mu\nu}(X)\pdif_p^\nu\\
	\nonumber
	&\hspace{1cm}+\tfrac{1}{24}(\pdif_p\cdot\pdif_X)^2e\F_{\mu\nu}(X)\pdif_p^\nu\\
	\nonumber
	&\hspace{2cm}+\O\big((e^0\pdif_p\cdot\pdif_X)^4\big)\, ,\\
	\Pi_\mu(X,p)&=p_\mu-\tfrac{1}{12}(\pdif_p\cdot\pdif_X)e\F_{\mu\nu}(X)\pdif_p^\nu\\
	\nonumber
	&\hspace{2cm}+\O\big((e^0\pdif_p\cdot\pdif_X)^3\big)\, .
\end{align}
Note in particular, that the leading order of $\nabla^\mu$, 
\begin{align}
\label{eq:gradlim}
	\nabla^\mu(X,p)=\D^\mu(X,p)+\O\big((e^0\pdif_p\cdot\pdif_X)^2\big)\, ,
\end{align}
is the classical Vlasov derivative
\begin{align}
\label{eq:vlasovder}
\D^\mu(X,p)\defeq \xpdify{}{X_\mu}-e\F^{\mu\sigma}(X)\xpdify{}{p^\sigma}\, ,
\end{align}
which contains the Lorentz force as its on-shell contribution (see Sec.~\ref{par:classical}).

Neglecting gradients of the field-strength tensor, the gauge-invariant spectral equation \refe{allgrad} becomes
\begin{align}
	\label{eq:lcfaeq}
	%&i\left[\xpdify{}{X_\mu}-e\F^{\mu\sigma}(X)\xpdify{}{p_\sigma}\right]
	\Big[\tfrac{i}{2}\slashed \D+\slashed p-m\Big]\hat{\rho}_{\Psi}(X,p)=0+\O(e^2)+\O(e^0\pdif_p\cdot\pdif_X)\, .
\end{align}
Solutions of equation \refe{lcfaeq} neglect field-gradients, but are exact in the field strength. This implies in particular that, even for a constant strong field strength tensor, the fermion spectral function is not a delta peak and does not allow for a simple particle picture.\footnote{This is an essential difference to Yukawa theory \cite{Berges:2009bx} or scalar $\lambda \phi^3$ theory (which are diagramatically very similar to QED): the LO equation of motion, e.g.\ for the scalar spectral function with a strong constant scalar macroscopic field $\phi_0\sim\O(1/\lambda)$,
\begin{align}
[\square_x+M^2]\rho(x,y)=0+\O(\lambda)\textrm{ with } M^2=m^2+\lambda\phi_0\, ,
\end{align}
does have a delta-peaked particle solution
\begin{align}
\rho(X,p)=i(2\pi)\sgn(p^0)\delta(p^2-M^2)\, .
\end{align}
Moreover, the equation for the scalar statistical propagator \cite{Berges:2015kfa}
\begin{align}
\nonumber
&2(p\cdot \pdif_X)F(X,p)+\Big(\xpdify{}{X^\mu} M^2(X)\Big)\Big(\xpdify{}{p_\mu}F(X,p)\Big)\\
&\hspace{1cm}=0+\O(\lambda)+\O\big(\lambda^0(\pdif_X\cdot \pdif_p)^2\big)
\end{align}
has a force term, $\pdif_XM^2(X)$ with $M^2(X)=m^2+\lambda\phi(X)$, which is NLO of the gradient expansion and vanishes for constant fields.}

The fermion transport equation \refe{fermiontransport} for small field-gradients then reduces to
\begin{align}
\label{eq:smalldrift}
&i\hat \rho^{\,\mu}_\Psi(X,p)\D_\mu\tilde f_\Psi(X,p)%\\
%\nonumber
%&=%\tfrac{i}{4\hbar}\textrm{tr}\{ (\slashed p-m)\hat \rho_\Psi(X,p) \}\big[1-2\tilde f_\Psi(X,p)\big]
=C_\Psi(X,p)\\
\nonumber
&\hspace{1cm}+\O(e^0\pdif_p\cdot\pdif_X)+\O(e^2\pdif_p\cdot\pdif_X G)+\O(e^4)\, ,
\end{align}
where we have used the fact that in contrast to $\nabla^\mu$, which contains higher order derivatives, $\mathcal{D}^\mu$ satisfies the Leibniz product rule and that a solution of \refe{lcfaeq} satisfies
\begin{align}
\label{eq:smallschwinger}
&i\D_\mu\hat \rho^{\,\mu}_\Psi(X,p)=0\, .
%\nonumber
%&=-\tfrac{1}{2}\textrm{tr}\{ (\slashed p-m)\hat \rho_\Psi(X,p) \}+\O(e^0\pdif_p\cdot \pdif_X)\, .
\end{align}
Plugging the solution of the approximated equation for the spectral function \refe{lcfaeq} into the collision kernel \refe{kernel} one obtains photon and fermion collision terms for fields with small gradients. In Sec.~\ref{subsec:smallgradplane}, we demonstrate how the locally-constant field approximation arises from such spectral functions in the special case of plane-wave fields. There, instead of solving the approximated equation \refe{lcfaeq}, we will first solve the infinite order gradient equation \refe{volkoveq} (or equivalently \refe{allgrad}) and approximate gradients in the solution in the end.

\subsection{\label{par:pair}Asymptotic (Schwinger) pair production\\from unequal-time correlations}
In this section, as an application of the above small field-gradient approximation, we discuss how pair production is implemented in the present formalism. We start in the regime of the collisionless Schwinger pair production yield per volume $V$ and time-interval $T$ \cite{PhysRev.82.664},
\begin{align}
\label{eq:loschwinger}
N_\Psi(\infty)\simeq VT\,\frac{e^2\mathcal{E}^2}{4\pi^3}\, \exp\Big(-\frac{\pi\E_{\textrm{c}}}{\mathcal{E}}\Big)\, ,
\end{align}
i.e.\ the regime of constant fields at 1-loop, and end this section with a general collisional expression for inhomogeneous fields.

Under the asymptotic assumption
\begin{align}
\label{eq:asymptoticstate}
\lim_{X^0\rightarrow \pm\infty}\hat \rho_\Psi(X,p)=\rho_{\Psi,0}(X,p)\, ,
\end{align}
even in the presence of strong fields, one can extract for asymptotically late times from the fermion transport equation the fermion pair number \refe{pairnumber}.

At 1-loop order $\O(e^0)$ and for small field gradients, Eq. \refe{transportdrift} simply reads
\begin{align}
\label{eq:1looppair}
\D_\mu \hat F_\Psi^\mu(X,p)=0+\O(e^2)\, .
\end{align}
 In order to extract the fermion pair number \refe{pairnumber}, we integrate Eq.~\refe{1looppair} over negative and positive energies separately and subtract the resulting integrals (instead of summing them, which would instead give the trivial total charge \refe{chargecoll}), i.e.
\begin{align}
\label{eq:pair1}
\Big(\int_{-\infty}^{0}\frac{\dif p^0}{(2\pi)}-\int_{0}^{\infty}\frac{\dif p^0}{(2\pi)}\Big)\int_{X,\vec p}\D_\mu\hat F_\Psi^\mu(X,p)=0\, .
\end{align}
For the momentum derivatives $\pdif_p^i$ of $\D^\mu$ we exploit \refe{infmom}, and for its frequency derivative $\pdif_p^0$ we note that
\begin{align}
\nonumber
&\Big(\int_{-\infty}^{0}\dif p^0-\int_{0}^{\infty}\dif p^0\Big)\,\pdif_p^0\hat F_\Psi(X,p)\\
\label{eq:pzero}
&\hspace{2cm}=2\int\dif p^0\delta(p^0)\,\hat F_\Psi(X,p).
\end{align}
This term eventually acts as a source term to the asymptotic number of fermion pairs. For the position-space derivative we use
\begin{align}
\int\dif^3 X \pdif_{X}^i \hat F_\Psi^{\, i}(X,p)&=0\, .
\end{align}
Finally, the time-derivative in $\D^\mu$ allows us to identify the pair number \refe{pairnumber} in the asymptotic past and future,
\begin{align}
\nonumber
&\Big(\int_{-\infty}^{0}\frac{\dif p^0}{(2\pi)}-\int_{0}^{\infty}\frac{\dif p^0}{(2\pi)}\Big)\int_{X,\vec p}\pdif_\mu\hat F_\Psi^\mu(X,p)\\
\label{eq:numberid}
&\hspace{2cm}=\tfrac{1}{2}\big(N_\Psi(\infty)-N_\Psi(-\infty)\big)\,,
\end{align}
where we have employed the asymptotic assumption \refe{asymptoticstate} and identified the on-shell electron and positron distribution functions \refe{antidef}, \refe{osfdef} and \refe{osafdef} in the asymptotic past and future. Applying the above identities to the 1-loop transport equation \refe{pair1} gives the result \begin{align}
\label{eq:pairlo}
&N_\Psi(\infty)-N_\Psi(-\infty)\\
\nonumber
&=\int_{X,p}\big[1-2\tilde f_\Psi(X,p)\big]\,2e\E^i(X)\,(2\pi)\delta(p^0)i\hat \rho^{\, i}_\Psi(X,p)\\
\nonumber
&\hspace{1cm}+\O(e^0\pdif_p\cdot\pdif_X)+\O(e^2)\, .
\end{align}
Importantly, this expression relates pair production to self-consistent spectral and field dynamics. The asymptotic assumption \refe{asymptoticstate} only fixes a boundary condition at $X^0\rightarrow \infty$ and interacting spectral dynamics [Eq.~\refe{allgrad}] contribute to \refe{pairlo} at all finite times $X^0$. In particular, the above expression shows that pair production from the vacuum occurs off-shell at the time of the creation event: the fermion yield \refe{pairlo} vanishes for a free (on-shell) fermion spectral function, because massive fermions can not have zero energy, i.e.
\begin{align}
\label{eq:noparticle}
\delta(p^0)\delta(p^2-m^2)\equiv 0\, .
\end{align}
It is only the subsequent evolution, that brings these off-shell contributions from vacuum pairs to the on-shell regime in the asymptotic future,  where a particle number $N_\Psi(\infty)$ is well-defined. Furthermore, the expression \refe{pairlo} vanishes for $\E=0$, even if $\mathcal{B}\neq0$, in accordance with the general statement that magnetic fields can not produce fermion pairs. In our derivation, this is a consequence of the vanishing of momentum derivatives at infinity, i.e.\ Eq.~\refe{infmom}. 
The structure of the expression \refe{pairlo} is reminiscent of the time-integrated source term of the quantum Vlasov equation from which particle production at zero energy is well known \cite{Kluger:1998bm}. Such a source term is not manifest in Eqs.~\refe{1looppair}, \refe{smalldrift}, or \refe{fermiontransport},\footnote{This is similar to Ref.~\cite{Gelis:2007pw} which shows (in scalar theory) that a source term is manifest in equations for disconnected two-point functions but not for connected two-point functions such as ours.} but we have demonstrated here that vacuum pair production is nevertheless contained in these transport equations by coupling to the dynamics of the fermion spectral function. 

To recover Eq.~\refe{loschwinger} from Eq.~\refe{pairlo} one should solve the fermion spectral equation \refe{volkoveq} or equivalently \refe{allgrad} for $\E=const$, $\mathcal{B}=0$  \cite{PhysRevD.82.105026, Kluger:1998bm, Cohen:2008wz}. This can be done analytically \cite{Fradkin:1991zq}, but we will not explore it further in this paper.

Since practicable procedures at 1-loop already exist in literature, we want to stress that the significance of Eq.~\refe{pairlo} does not stem from its 1-loop practicability but from the fact that it may be systematically generalized and thereby put in the context of thermalization, while other procedures have struggled to do so: At 1-loop, where the equations for spectral and statistical functions are decoupled, one may compute the asymptotic fermion particle number by ignoring spectral dynamics and solving the complete tensorial system for the statistical function. In literature, this is often done in terms of the \textit{equal-time} `DHW' function $\hat F_\Psi(t,t,\vec x,\vec y\,)\gamma^0$, or $\int\dif p^0 \hat F_\Psi(X,p)\gamma^0$ in Wigner space. In fact, existing transport derivations of the Schwinger result typically employ such equal-time formulations \cite{PhysRevD.44.1825, Best:1992gb, Zhuang:1995pd, PhysRevD.82.105026}, in which spectral informationn such as a distinction between on and off shell contributions is not explicitly accessible due to spectral functions being constant at equal times [see Eq.~\refe{eqtimespectral}]. Equal-time equations can be closed, e.g.\ by neglecting collisions, but how to close an equal-time description for general strong fields in a controlled approximation is an open problem. From an unequal-time perspective, the equation for the fermion statistical function is not self-sufficient at $\O(e^2)$, but couples to the fermion spectral function \refe{volkoveq}, which is not on-shell for strong fields. The unequal-time approach closes by including this equation for the spectral function and is thereby systematically generalizable to higher loop orders that are essential for the approach to equilibrium.

Simply by keeping field-gradients and the collision term, i.e. starting from Eq.~\refe{transportdrift} instead of Eq.~\refe{1looppair}, one obtains
\begin{align}
\nonumber
&N_\Psi(\infty)-N_\Psi(-\infty)=\Big(\int_{-\infty}^{0}\frac{\dif p^0}{(2\pi)}-\int_{0}^{\infty}\frac{\dif p^0}{(2\pi)}\Big)\,\int_{X,\vec p} \\
\nonumber
&\cross\Bigg\{e\int_{-\tfrac{1}{2}}^{\tfrac{1}{2}}\dif\lambda\,\bigg[e^{-i\lambda(\pdif_p\cdot\pdif_X)}\E^{i}\bigg]\pdif_p^0\,\bigg[(1-2\tilde f_\Psi)i\hat\rho_\Psi^{\,i}\bigg]+C_\Psi\Bigg\}\\
\label{eq:pairgeneral}
&\hspace{1cm}+\O(e^2\pdif_p\cdot\pdif_X G)+\O(e^4)\,  .
\end{align}
Due to the presence of higher-order frequency derivatives, the identity \refe{pzero} is not sufficient to treat inhomogeneous fields, which are able to transfer momentum and produce occupations with finite energy, $p^0\neq 0$. A complete self-consistent solution of the set of equations in Fig. \ref{fig:1} is necessary to obtain a numerical result for the asymptotic pair number in this way. In particular, the collisional part of \refe{pairgeneral} contains  contributions from $0\rightarrow 3$ (2-loop vacuum pair production) and $1\rightarrow 2$ processes (`seeded cascades'), the latter of which dominate over vacuum pair production in subcritical fields \cite{Nerush:2010fe, PhysRevSTAB.14.054401, Tamburini:2017sxg}. In contrast to Eq.~\refe{pairgeneral}, the 1-loop result \refe{loschwinger} describes the effect of a constant external electric field with no feedback from the dynamics of the photon sector.

\subsection{\label{par:classical}Lorentz force \& classical propagation\\in isolated systems}

The Lorentz force,
\begin{align}
\label{eq:lorentzforce}
L^\mu(X,p)/m\defeq \tfrac{e}{m}\F^{\mu\nu}(X)p_\nu\, ,
\end{align}
emerges from the quantum Vlasov term of Eq.~\refe{fermiontransport} in the case of a free fermion spectral function and small field-gradients via
\begin{align}
\label{eq:classicaldrift}
&\nabla_\mu[\hat\rho_\Psi^{\,\mu}(X,p)\tilde f_\Psi(X,p)]\rightarrow \rho_{\Psi,0}^{\mu}\D_\mu\tilde f_\Psi\\
\nonumber
&=i(2\pi)\delta(p^2-m^2)\sgn(p^0)[(p\cdot \pdif_X)+(L\cdot \pdif_p)]\tilde f_\Psi\, ,
\end{align}
where the factor of $p^\mu$ is provided by the vector component
\begin{align}
\label{eq:freevector}
\rho_{\Psi,0}^\mu(X,p)&=ip^\mu(2\pi) \delta(p^2-m^2)\,\sgn(p^0)\, .
\end{align}
Therefore, on-shell particles may be described by the Lorentz force. However, the validity of employing a free spectral function in Eq.~\refe{classicaldrift}, i.e.\ whether on-shell particles indeed dominate the dynamics, depends on the details of the strong-field system:

Typical experiments where on-shell particles dominate the dynamics are, for example, those where an electron beam or material target is initially in a zero-field region and then collides with a strong field such as a laser beam \cite{PhysRevX.8.031004, PhysRevX.8.011020}. In such a setting fermion distribution functions are initialized with occupations only in the on-shell region and the subsequent deviations from on-shell occupations induced by the strong field often remain small even when fermion pairs are produced: This is because these systems feature a separation of time scales due to the typically very large values of the parameter $\xi=|e|\E/(m\omega)$ \cite{RevModPhys.84.1177}, implying that particles (target or produced) are transported in momentum space to relativistic energies in much less than a laser period. Thereby, the fermion distribution function is typically peaked at an ultrarelaticistic scale and far away from its equilibrium (Fermi-Dirac) shape. At such high energies, off-shell effects can be suppressed \cite{Baier:1998vh} and can \textit{remain} suppressed, if the ultrarelativistic peak in the fermion distribution function is long-lived (see Sec.~\ref{subsec:urlimit}).

In the presence of such long-lived peaked distribution functions, one may then distinguish two kinds of quantum effects \cite{Baier:1998vh}: One class is related to the recoil that a fermion experiences \textit{during} collisions (i.e., emissions of photons). This is controlled by the (spacetime and momentum dependent) parameter \cite{RevModPhys.84.1177}
\begin{align}
\label{eq:fullchidef}
\chi\defeq \hbar \sqrt{-(e\F^{\mu\nu}p_\nu)^2}/m^3\, ,
\end{align}
which may be small even for large $\xi$ or vice versa. Systems that have small $\chi$ may be described completely (both drifting and collisional interactions) in terms of the classical radiation reaction force \cite{LANDAU1975171, RevModPhys.84.1177,Burton_2014,Blackburn_2020} that includes collisional corrections to the Lorentz force \cite{PhysRevSTAB.14.054401}. The other class of quantum effects is related to how accurate a classical description is \textit{between} collisions. This is commonly discussed in terms of the de Broglie wavelength $\hbar/p^*$, which is then required to be small enough such that the quasiclassical approximation applies \cite{Landau_b_3_1977}, and smaller than the mean-free-path such that a separation between propagation and interaction is possible \cite{Arnold:2007pg}. In our context, $p^*$ is the characteristic momentum of the fermion distribution function. At higher and higher energies, the de Broglie wavelength decreases whereas the parameter $\chi$ increases, such that quantum effects remain important during collisions for ultrarelativistic fermions and no radiation reaction force description exists \cite{Baier:1998vh}. These parameters are not manifest at the level of the equations of motion, but become accessible by analysis of its solution (see e.g.\ Secs.~\ref{subsec:smallgradplane} and \ref{subsec:urlimit}). In the absence of peaked distribution functions, the medium may not be completely described by a single de Broglie wavelength and no such separation of scales may be identified.

In fact, a peaked fermion distribution describes a far-from-equilibrium situation that does not survive indefinitely in an isolated system. Thereby, systems for which an on-shell Lorentz force description is typically insufficient are those which are initialized with a supercritical field, $\E\gtrsim \E_{\textrm{c}}$, and which are then isolated and left on their own. In such systems, fermions are produced from the vacuum -- off-shell and at low energies according to Eq.~\refe{pairlo} -- and then transported in momentum space by the gain-minus-loss structure of the collision terms towards a distribution that is not sharply peaked at any single scale. To describe the evolution towards such a distribution, one requires a description that is valid over a wide range of energies. Thus, the separation of scales from the case of an external field may not be exploited to argue for a Lorentz force description of the equilibration of isolated strong-field systems.

Near equilibrium, a weak field again favors on-shell descriptions, because the field term $e\slashed \A$ in the equation of motion of the fermion spectral function \refe{volkoveq} then contributes to the transport description only at higher orders and collisions may be added to the on-shell Vlasov equation [see Eq.~\refe{vlasoveq} below] perturbatively in the field vertex \refe{fieldvertex}. However our analysis suggests that for intermediate times, at which off-shell contributions from vacuum pair production equilibrate in the presence of a depleting field, one requires a description of off-shell drifting beyond the Lorentz force. The description derived in Sec.~\ref{sec:4} can capture this evolution of off-shell contributions in $\tilde f_\Psi(X,p)$ as they move in phase space towards $p^0=\sqrt{|\vec p\,|^2+m^2}$ to become on-shell particles in the asymptotic future.

It is then instructive to follow how the Lorentz force emerges from the off-shell drift term of Eq.~\refe{smalldrift}, which contains the frequency derivative term
\begin{align}
\label{eq:offshelldrift}
\hat \rho^{\,\mu}_\Psi(X,p)e\F_{\mu 0}(X)\,\pdif_p^0\tilde  f_\Psi(X,p)\, .
\end{align}
As we have shown in Sec.~\ref{par:pair}, in the asymptotic future the effect of this off-shell frequency derivative is fermion pair production. In the on-shell regime, where pair production is forbidden via Eq.~\refe{noparticle}, this off-shell frequency derivative is controlled by the dispersion relation, \mbox{$p^0=\varepsilon(\vec p\,)$}: The term
\begin{align}
p\cdot L(X,p)=0\lra L^0=\frac{\vec L\cdot \vec p}{p^0}\, ,
\end{align}
then contains the group velocity
\begin{align}
\xpdify{\varepsilon(\vec p\,)}{\vec p}=\frac{\vec p}{\varepsilon(\vec p\,)}\, ,
\end{align}
such that, by chain rule, one may replace
\begin{align}
\label{eq:totalvlasov}
\frac{L^0}{p^0}\xpdify{}{p^0}+\frac{\vec L}{p^0}\cdot \xpdify{}{\vec p}\rightarrow 
\frac{\vec L}{\varepsilon(\vec p\,)}\cdot \xpdify{}{\vec p}
\end{align}
and recover the classical Vlasov equation
\begin{align}
	\label{eq:vlasoveq}
	&\sgn(p^0)\delta(p^2-m^2)(p\cdot\mathcal{D})\tilde f_\Psi(X,p)=0+\O(e^2)\, .
\end{align}
Making use of the fact that
\begin{align}
\label{eq:vlasovterm2}
&-ep_\mu\mathcal{F}^{\mu\sigma}(X)\xpdify{}{p^\sigma}=L^0\xpdify{}{p^0}+\vec L\cdot \xpdify{}{\vec p}\\
\nonumber
&\hspace{0.5cm}= e\vec p\cdot \vec{\mathcal{E}}(X)\xpdify{}{p^0}+e\Big[p^0\vec{\mathcal{E}}(X)+\vec p\cross \vec{\mathcal{B}}(X)\Big]\cdot \xpdify{}{\vec p} \, .
\end{align}
and applying definitions for on-shell electron and positron distribution functions $\tilde f^-_\Psi$ and $\tilde f^+_\Psi$ analogously to Eqs.~\refe{antidef}, \refe{osfdef} and \refe{osafdef}, one may then split Eq.~\refe{vlasoveq} into equations for electrons and positrons by integrating Eq.~\refe{vlasoveq} over positive or negative frequencies respectively. The positron equation obtains the opposite sign of charge $e\rightarrow -e$ from the sign $\vec p\rightarrow -\vec p$ of the momentum derivative,
	\begin{align}
		\nonumber
	&\Bigg(\xpdify{}{X^0}+\frac{\vec p}{\varepsilon(\vec p\,)}\cdot \xpdify{}{\vec X}\pm e\Big[\vec{\mathcal{E}}(X)+\frac{\vec p}{\varepsilon(\vec p\,)}\cross \vec{ \mathcal{B}}(X)\Big]\cdot\xpdify{}{\vec p}\Bigg)\\
	\label{eq:vlasov1}
	&\hspace{2cm}\cross \tilde f_\Psi^\mp(X,\vec p\,)=0+\O(e^2)\, .
	\end{align}
\indent If we interpret $X$ and $p$ as functions $X(\lambda)$ and $p(\lambda)$, then the curves along which $\tilde f_\Psi$ is constant, i.e.\ the characteristic curves
\begin{align}
	\xdify{}{\lambda}\tilde f_\Psi(X(\lambda),p(\lambda))=0\, ,
\end{align}
solve the Lorentz equation \cite{Vasak:1987um}
\begin{align}
\label{eq:lorentzeq}
	\xdify{p^\mu}{\lambda}&=L^\mu(X,p)\, ,\\
	p^\mu &=\xdify{X^\mu}{\lambda}\, ,
\end{align}
with the Lorentz force \refe{lorentzforce}. Adding collisions that are non-linear in $\tilde f_\Psi$ makes this method of characteristics inapplicable and the concept of trajectories breaks down.

We reiterate that, for general strong fields and fermion distribution functions, the limit of classical propagation \refe{classicaldrift} is not controlled by an expansion in a small parameter and a combination of the Lorentz force term with the $\O(e^2)$ collision term \refe{fermcoll} is not in general complete to leading order $\O(e^2)$. To be complete in a general situation, the Lorentz force term should be replaced by the quantum Vlasov term of Eq.~\refe{fermiontransport} (or that of Eq.~\refe{smalldrift} for small field-gradients).

\subsection{\label{subsec:5c}The case of strong external plane-wave fields}

We assume in the following that the macroscopic field is of the one-dimensional `plane-wave' form 
\begin{align}
	\label{eq:plane}
	\A^\mu(x)\simeq\A^\mu_\textrm{v}(n\cdot x)\, ,\,\,\,\,\textrm{ with }\,\,\,\,
	n^2=0\, ,
\end{align}
as originally employed by Volkov \cite{Wolkow:1935zz}.\footnote{Other integrable cases include external fields such as the Coulomb potential (leading to hydrogen levels), homogeneous magnetic fields (leading to Landau levels), constant crossed fields (leading to Airy functions) and constant non-crossed electric fields (leading to Weber parabolic cylinder functions).} We drop the label `v' where the context is clear. Assuming \refe{plane} means we suppress the parts of the dynamics of the macroscopic field that deviate from a plane-wave field form. The plane-wave approximation is widely used in studying the interaction of laser fields with matter and is valid if the laser beam is not tightly focused in space such that the wave front is approximately flat. Even under such a relatively controlled setup, but especially in isolated systems, one has to take into account that the validity of the plane-wave approximation can be limited in time. The validity time-scale then depends on the back reaction \cite{Kluger:1998bm, Bloch:1999eu} of the matter on the field via Maxwell's equation \refe{micmax}. A simple parametric estimate suggests a large range of validity up to times of $t\iv\sim\O(1/e^2)$. However it is well known \cite{Baier:1998vh} that strong macroscopic fields can further decrease this timescale. Below, we assume that the plane-wave approximation is valid for the times under consideration.

Although this assumption significantly simplifies the equations, we stress that it does not restrict the discussion of a multitude of common experimental field configurations, such as (linearly or elliptically) polarized fields, (long or short) pulses, monochromatic or polychromatic fields, and (constant or strongly varying) crossed fields.

The field strength tensor of plane-wave fields can be written as
\begin{align}
\F_{\textrm{v}}^{\mu\nu}(n\cdot x)=n^\mu \dot \A_{\textrm{v}}^\nu(n\cdot x)-n^\nu \dot \A_{\textrm{v}}^\mu(n\cdot x)\, ,
\end{align}
where a dot stands for a derivative with respect to the argument. From this it follows that plane-wave fields necessarily satisfy
\begin{align}
\label{eq:fsquared}
-\tfrac{1}{2}\eta_{\mu\rho}\eta_{\nu\sigma}\F_{\textrm{v}}^{\mu\nu}\F_{\textrm{v}}^{\rho \sigma}&=|\vec{\mathcal{E}}|^2-|\vec{\mathcal{B}}|^2=0\, ,\\
\label{eq:topological}
-\tfrac{1}{8}\varepsilon_{\mu\nu\rho\sigma}\F^{\mu\nu}_\textrm{v}\F^{\rho\sigma}_\textrm{v}&=\vec{\mathcal{E}}\cdot \vec{\mathcal{B}}=0\, .
\end{align}
Therefore, the magnetic field $\vec{\mathcal{B}}$ is always perpendicular to and of equal absolute value of the electric field $\vec \E$, such that it is sufficient to only talk about electric fields in the context of plane-waves. In particular, the topological term \refe{topological} associated with CP violation \cite{PhysRevLett.38.1440, PhysRevD.16.1791} vanishes identically. This has the implication that the pseudoscalar component of the spectral function (which we introduce in Sec.~\ref{par:volkovspectral}, see also appendix \ref{app:pseudo}) vanishes.

Plane-wave systems are most conveniently described using lightcone coordinates that use the special direction $n^\mu$ of the field,
\begin{align}
x^-&\defeq x^0-x^3=n\cdot x\, ,\\
x^+&\defeq \tfrac{1}{2}(x^0+x^3)\, ,\\
\vec x_\perp &\defeq (x^1, x^2, 0)\, .
\end{align} Lightcone coordinates have metric tensor $\eta^{+-}=\eta^{-+}=\eta_{+-}=\eta_{-+}=1$, $
\eta^{++}=\eta^{--}=\eta_{++}=\eta_{--}=0$
such that $x^+=x_-$, $x^-=x_+$ and $p\cdot s = p^+s^-+p^-s^+-\vec p_\perp \cdot \vec s_\perp$.

We work in Lorenz gauge [$\pdif\cdot\A(x)=0$] and use the residual gauge freedom to also fix temporal axial gauge [$\A^0(x)=0$]. In lightcone coordinates that use the physical direction $n^\mu$ of the field, this is equivalent (for vanishing asymptotic boundary conditions) to so-called lightfront gauge \cite{BERESTETSKII1982118}, i.e.
\begin{align}
\label{eq:planegauge1}
\A^-\iv(x^-)&=0\, ,\\
\A^+\iv(x^-)&=0\, ,
\end{align}
which is conveniently formulated in a frame in which
\begin{align}
n^\mu=(1,0,0,1)\, .
\end{align}
In this frame and gauge, the electric field is simply
\begin{align}
\label{eq:Eplane}
\vec{\mathcal{E}}(x^-)=-\dot {\vec \A}_{\perp}(x^-)\, .
\end{align}

In particular, this allows for a simple form of the (symmetric) energy momentum tensor
\begin{align}
\label{eq:planewaveT}
T^{\mu\nu}\iv \defeq \F\iv^{\mu}{}_{\sigma}\F\iv^{\sigma\nu}=n^\mu n^\nu |\vec{\mathcal{E}}|^2
\end{align}
from which the energy density of the plane-wave field
\begin{align}
\label{eq:clt}
T^{00}\iv= \tfrac{1}{2}(|\vec{\mathcal{E}}|^2+|\vec{\mathcal{B}}|^2)=|\vec{\mathcal{E}}|^2
\end{align} can be read off. A peculiarity of the plane-wave field is that the classical quantity \refe{planewaveT} coincides with the exact vacuum expectation value of the energy momentum tensor up to fermionic contributions \cite{PhysRev.82.664}.

For any function $\K(X,s^-)$ of $n\cdot s\eqdef s^-$, one has
\begin{align}
\nonumber
\int_s e^{i(p-q)s} \K(X,s^-)&=(2\pi)^3\delta(p^--q^-)\delta(\vec p_\perp-\vec q_\perp)\\
\label{eq:planedelta}
&\hspace{-1cm}\cross\int \dif s^-\,e^{i(p^+-q^+)s^-}\K(X,s^-)\, .
\end{align}
This is can be written compactly as
\begin{align}
\label{eq:lid}
&\int_s e^{i(p-q)s} \K(X,s^-)\\
\nonumber
&\hspace{1cm}=\int\frac{\dif l}{(2\pi)}(2\pi)^4\delta(p-q-ln)\,\K(X,l)\, ,
\end{align}
with the one-dimensional Wigner transform
\begin{align}
\K(X,l)\defeq \int\dif s^-e^{ils^-}\K(X,s^-)\, .
\end{align}

\subsubsection{\label{par:volkovspectral}Spectral function \& plane-wave d.o.f.}
\indent A solution of the equation for the fermion spectral function \refe{volkoveq} for plane-wave fields is\footnote{This plane-wave spectral function $\rho_{\Psi,\textrm{v}}$ is the antisymmetric part of the time ordered `Volkov propagator' \cite{PhysRev.81.115, Ritus1985, PhysRevD.97.056028} (see appendix \ref{app:volkov}). By plugging $\rho_{\Psi,\textrm{v}}$ into our transport equations we resum the \textit{symmetric} part of the fermion propagator to self-consistent 2-loop order.}
\begin{align}
\label{eq:volkovsol}
\rho_{\Psi,\textrm{v}}(x,y)&=i(2\pi)\int_q\delta(q^2-m^2)\sgn(q^0)\\
\nonumber
&\hspace{1cm}\cross R_{q}(x
)(\slashed q+m) \bar R_{q}(y)\, .
\end{align}
The field dependence enters via the Ritus matrices $R_q$, $\bar R_q$ \cite{Nikishov:1964zza, Nikishov:1964zzab, Ritus1985} which are defined as\footnote{$S_p[X(\lambda)]$ is the classical action for the trajectory $X^\mu(\lambda)$ of a test particle in a plane-wave field \cite{LANDAU1975109}. This fact gives rise to an interpretation of plane-wave scattering probabilities in terms of a stationary phase principle \cite{PhysRevA.83.032106, PhysRevD.93.085028}.}
\begin{align}
&R_p^{AB}(x)\defeq \left[ \Id+\frac{e}{2}\frac{\slashed n \slashed \A_\textrm{v}(n\cdot x)}{(n\cdot p)} \right]{}^{AB} e^{iS_p(x)}\, ,\\
\label{eq:lorentzaction}
&S_p(x)\\
\nonumber
&\defeq -p\cdot x-\frac{1}{2(n\cdot p)}\int_{-\infty}^{(n\cdot x)} \dif\lambda\big[2\A\iv(\lambda)\cdot p-e^2\A\iv^2(\lambda)\big]\, ,\\
&\bar R_p(x)\defeq%\left[ 1-\frac{e}{2}\frac{ \slashed n\slashed \A_\textrm{v}(n\cdot x)}{(n\cdot p)} \right]{}^{AB} e^{iS_p(x)}=
\gamma^0R_p^\dagger(x)\gamma^0\, .
\end{align}
The essential property of the Ritus matrices is that they translate the strong field Dirac operator in position-space into the free Dirac operator in momentum space, i.e.
\begin{align}
\label{eq:ritid}
&(i\slashed \pdif_x-e\slashed \A_\textrm{v}(n\cdot x)-m)R_p(x)=R_p(x)(\slashed p-m)\, .
\end{align}
The plane-wave spectral function contains the strong-field dressed mass \cite{Brown:1964zzb, PhysRevD.97.056028} (see Sec.~\ref{par:invspec}) and recovers the Airy-type scattering amplitudes for small field-gradients (see Sec.~\ref{subsec:smallgradplane}). For the proof that \refe{volkovsol} solves \refe{volkoveq}, and satisfies the symmetry constraint \refe{rs2}, as well as for the computation of its Dirac components, we refer to the appendices \ref{app:volkov} and \ref{app:volkovcomp}.

\begin{figure}[h!]
	\begin{center}
		\includegraphics[scale=0.035]{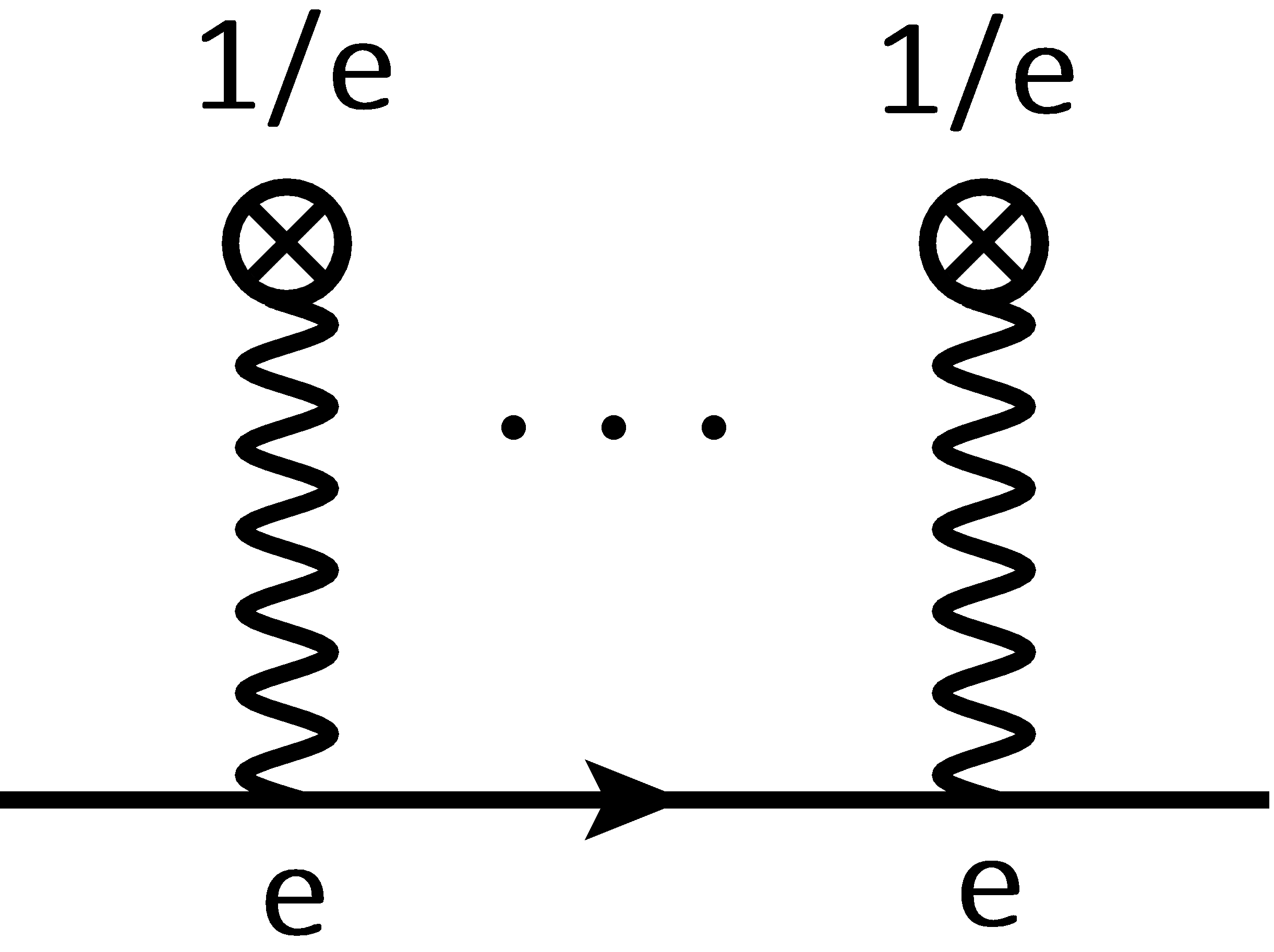}
	\end{center}
	\caption{\label{fig:resum}Resummation of the macroscopic field vertex.}
\end{figure}

The non-perturbative nature of the plane-wave spectral function can be observed from the exponential $e^{iS_p}$: The field-dependent part of the exponent is small for not too strong fields and an expansion in powers of $e$ could be truncated in that case [corresponding to perturbation theory with the vertex \refe{fieldvertex}]. However for strong fields, $\A\sim\O(1/e)$, the exponent is $\O(e^0)$ and all orders in $e$, have to be taken into account as depicted in Fig.~\ref{fig:resum}.

For the application to our transport equations we need the late-time Wigner transform
		\begin{align}
		\label{eq:volkovsolwig}
		\rho_{\Psi,\textrm{v}}(X,p)&=i(2\pi)\int_q  \delta(q^2-m^2)\sgn(q^0)\\
		\nonumber
		&\hspace{0.3cm}\cross\int_s e^{ips}R_{q}(X+\tfrac{s}{2}
		)(\slashed q+m) \bar R_{q}(X-\tfrac{s}{2})\, .
		\end{align}
		From this expression we can observe that the plane-wave spectral function captures off-shell effects: the external momentum $p$ is not restricted to on-shell values but becomes on-shell in the limit $\A\iv\rightarrow 0$, which recovers the free spectral function via
		\begin{align}
		\label{eq:nofieldlim}
		R_p^{AB}(x)&\xrightarrow{\A\iv\rightarrow 0}\delta^{AB}e^{-ipx}\, ,\\
		\bar R_p^{AB}(x)&\xrightarrow{\A\iv\rightarrow 0} \delta^{AB}e^{ipx}\, .%\\
		%\bar \rho_{\Psi,\textrm{v}}[\A_\textrm{v}]&\xrightarrow{\A\iv\rightarrow 0}\rho_{\Psi,0}\,.
		\end{align}

With the identity \refe{lid} we can discuss the emergence of plane-wave fermion degrees of freedom in strong fields by writing
 \begin{align}
   \label{eq:planeparticle}
 \rho_{\Psi,\textrm{v}}(X,p)&=i\int\dif l\,\K(X,l;p-ln)\\
  \nonumber
 &\cross \delta(p^2-m^2-2l(n\cdot p))\,\sgn(p^0-ln^0)
 \end{align}
 with the field-dependent Dirac matrix
 \begin{align}
 \label{eq:K}
 \K(X,l;q)&\defeq \int\dif s^- e^{ils^-}e^{-i\N_q(X^-,s^-)}\\
 \nonumber
 &\hspace{-2cm}\cross\left[ \Id+\frac{e}{2}\frac{\slashed n \slashed \A_\textrm{v}(X+\tfrac{s}{2})}{(n\cdot q)} \right](\slashed q+m)\left[ \Id-\frac{e}{2}\frac{\slashed n \slashed \A_\textrm{v}(X-\tfrac{s}{2})}{(n\cdot q)} \right]
 \end{align}
 and the field-dependent phase factor
 \begin{align}
 \label{eq:Ndef}
 &\N_q(n\cdot X,n\cdot s)\\
 \nonumber
 &\defeq \int_{-\tfrac{n\cdot s}{2}}^{\tfrac{n\cdot s}{2}}\dif \lambda\,\Bigg(\frac{e\A(n\cdot X+\lambda)\cdot q}{n\cdot q}-\frac{e^2\A^2(n\cdot X+\lambda)}{2(n\cdot q)}\Bigg)\, .
 \end{align}
 While the phase in terms of  $S_q$ fully depends on $x^\mu$ and $y^\mu$, the phase $\N_q$ only depends on $n\cdot X$ and $n\cdot s$ via
 \begin{align}
 S_q(x)- S_q(y)=-q\cdot s-\N_q(n\cdot X,n\cdot s)\, .
 \end{align}
 From \refe{planeparticle} we observe that, by the integration over $l$, the on-shell condition for free fermions ($l=0$) is modified to the condition (with $l$ unconstrained)
 \begin{align}
 \label{eq:epl}
p^0=l\pm\varepsilon_l(\vec p\,)
\lra p^+=\frac{|\vec p_\perp|^2+m^2}{2p^-}+l\, ,
 \end{align}
 where we have defined the plane-wave relation
 \begin{align}
 \label{eq:epl2}
 \varepsilon_l(\vec p\,)&\defeq (\varepsilon^2(\vec p\,)+l^2-2lp_z )^{1/2}\\
 \nonumber
 &=(|\vec p_\perp|^2+m^2+(p_z-l)^2)^{1/2}\, .
 \end{align}
This expression depends explicitly on the $z$-component $p_z\defeq p^3$ in which the plane-wave field varies, is positive and satisfies
 \begin{align}
 \label{eq:epl3}
 \varepsilon_l(\vec p\,)&=\varepsilon_{-l}(-\vec p\,)\, .
% \sgn(\pm \varepsilon_l(\vec p\,))&=\pm 1\, .
 \end{align}
 From the context of plane-wave collision terms one further observes in Sec.~\ref{par:planecoll} that the parameter $l$ corresponds to the energy exchanged between fermions and the macroscopic field during quantum processes. Since $l$ is integrated over, the relation $\varepsilon_l(\vec p\,)$ does not on its own restrict the external momentum of the fermion spectral function. Its interpretation as a dispersion relation is thereby not straightforward. Depending on the details of the macroscopic field, the integration over $l$ may have different effects such as broadening the peak structure or adding more peaks. The plane-wave spectral function thereby describes interactions with different $l$-modes of the macroscopic field, %and propagate with group velocity
%\begin{align}
%\xpdify{\varepsilon_l(\vec p\,)}{p^i} = \frac{p^i}{\varepsilon_l(\vec p\,)}-\delta^{3i}\frac{l}{\varepsilon_l(\vec p\,)}\, ,
%\end{align}
where the lowest mode, $l=0$, describes freely propagating particles via
\begin{align}
\label{eq:zerol}
\varepsilon_{l}(\vec p\,)\xrightarrow{l\rightarrow 0} \varepsilon(\vec p\,)\, .
\end{align}

In particular, if the macroscopic field is periodic in $s^-$ with frequency $\omega$, $\K(X,l;q)$ has support only for $l=j\omega$ with $j\in \mz$ and a countable peak structure emerges via
 \begin{align}
 \label{eq:integerl}
 \int\frac{\dif l}{(2\pi)}\rightarrow \omega \sum_{j=-\infty}^{\infty}\, ,
 \end{align}
 (see also Ref.~\cite{Heinzl:2010vg} for a similar discussion at the level of amplitudes). If $l$ is continuous, the infinitely many delta peaks may merge to form a function with finite-width peaks, such as the function computed in Ref.~\cite{PhysRevD.83.065007}.
 
As we will see in Sec.~\ref{par:planecoll}, the $l$-modes can be kept track of as individual degrees of freedom by defining appropriate distribution functions that are summed or integrated over in the collision terms. A traditional on-shell description in terms of only the $l=0$-mode is then favored as long as a separation of scales in terms of ultrarelativistic fermions exists, as we discuss in Sec.~\ref{subsec:urlimit}.
 
\subsubsection{Collision kernel}

Plugging the plane-wave spectral function \refe{volkovsolwig} into \refe{kernel} we obtain the strong-field plane-wave collision kernel
\begin{align}
&\P\iv^{\mu\nu}(X,p,q,k)\\
\nonumber
&=\delta(k^2)\sgn(k^0)\int_{s_1,s_2}\hspace{-0.2cm}e^{ips_1}e^{iqs_2}\int_{p',q'}\hspace{-0.2cm}e^{-ip's_1}e^{-iq's_2}\\
\nonumber
%\cross\\ \nonumber &\hspace{0.5cm}\cross
&\cross \delta(p'^2-m^2)\sgn(p'^0)\delta(q'^2-m^2)\sgn(q'^0)\\
\nonumber
&\hspace{1.5cm}\cross\T^{\mu\nu}_{p'q'}(X,s_1,s_2)\,e^{-i[\N_{p'}(X,s_1)+\N_{q'}(X,s_2)]}\, ,
\end{align}
where we have defined the pre-exponential
\begin{widetext}
	\begin{align}
	\label{eq:preexp}
	&-4\T^{\mu\nu}_{p'q'}(X,s_1^-,s_2^-)\\
	\nonumber
	&\defeq \textrm{tr}\Bigg\{ \gamma^\mu\Big[1+\frac{e\slashed n\slashed \A\iv(X+\tfrac{s_1}{2})}{2(n\cdot p')}\Big](\slashed p'+m)\Big[1-\frac{e\slashed n\slashed \A\iv(X-\tfrac{s_1}{2})}{2(n\cdot p')}\Big]\gamma^\nu\Big[1+\frac{e\slashed n\slashed \A\iv(X+\tfrac{s_2}{2})}{2(n\cdot q')}\Big](\slashed q'+m)\Big[1-\frac{e\slashed n\slashed \A\iv(X-\tfrac{s_2}{2})}{2(n\cdot q')}\Big] \Bigg\}\, ,
	\end{align}
	such that, together with the phase \refe{Ndef}, the trace over the Ritus matrices becomes
\begin{align}
&\tfrac{1}{4}\textrm{tr}\{ \gamma^\mu R_{p'}(X+\tfrac{s_1}{2})(\slashed p'+m)\bar R_{p'}(X-\tfrac{s_1}{2})
\gamma^\nu R_{q'}(X+\tfrac{s_2}{2})(\slashed q'+m)\bar R_{q'}(X-\tfrac{s_2}{2}) \}\\
\nonumber
&\hspace{2cm}= \T^{\mu\nu}_{p'q'}(X,s_1,s_2)\,e^{-ip's_1}e^{-iq's_2}e^{-i[\N_{p'}(X,s_1)+\N_{q'}(X,s_2)]}\, .
\end{align}
We discuss the familiar case of $s_1+s_2=0$ that emerges in the absence of a medium in Sec.~\ref{par:vertex}. For zero field, the phase $\N_p$ vanishes and the pre-exponential becomes the on-shell amplitude squared
\begin{align}
\label{eq:Tzerolimit}
&\int_{s_1,s_2}\, e^{i(p-p')s_1}\, e^{i(q-q') s_2}\,\eta_{\mu\nu}\T_{p'q'}^{\mu\nu}(X,s_1,s_2)\xrightarrow{\A\iv\rightarrow 0}-(2\pi)^4\delta(p-p')(2\pi)^4\delta(q-q')\big[-2(p\cdot q)+4m^2\big]\, , 
\end{align}
\end{widetext}
such that $p'\rightarrow p$ and $q'\rightarrow q$ as $\A\iv\rightarrow 0$. In the presence of a field, the plus-components of $p'$ and $p$, and $q'$ and $q$ do not coincide and $p^+$ and $q^+$ are not on-shell.

\subsubsection{\label{par:kin}Off-shell vs.\ on-shell kinematics}

 Classically, particle motion in a plane-wave field [described by the classical Vlasov equation \refe{vlasoveq}] is characterized by the conservation of the two transverse and the minus-component of the canonical momentum. The plus-component, that is conserved for free particles, is no longer conserved in the presence of a plane-wave field that exchanges energy with particles in this longitudinal direction.
 
We can derive this interpretation of the field as an energy reservoir from our plane-wave collision kernel also in the off-shell quantum case. By applying identity \refe{lid}, we may write
\begin{align}
\label{eq:planekernel}
&\P\iv^{\mu\nu}(X,p,q,k)=\delta(k^2)\sgn(k^0)\int\frac{\dif l_1}{(2\pi)}\int\frac{\dif l_2}{(2\pi)}\\
\nonumber
&\cross\int_{p',q'}\delta(p'^2-m^2)\sgn(p'^0)\delta(q'^2-m^2)\sgn(q'^0)(2\pi)^8\\
\nonumber
&\cross\delta(p-p'-l_1n)\delta(q-q'-l_2n)\mathcal{Q}^{\mu\nu}(X,l_1,l_2;p',q')
\end{align}
with the remaining kernel
\begin{align}
\nonumber
&\mathcal{Q}^{\mu\nu}(X,l_1,l_2;p',q')\defeq \int \dif s_1^- e^{il_1s_1^-}\int \dif s_2^-e^{il_2s_2^-}\\
\label{eq:Qkernel}
&\hspace{0.6cm}\cross \T^{\mu\nu}_{p'q'}(X,s_1^-,s_2^-)\,e^{-i[\N_{p'}(X,s_1^-)+\N_{q'}(X,s_2^-)]}\, .
\end{align}
\indent 
The collision terms therefore contain delta functions enforcing the kinematic conditions
\begin{align}
\label{eq:sfkinematic1}
k-p+q&=0\,,\\
p-p'-l_1n&=0\, ,\\
q-q'-l_2n&=0\, ,\\
p'^2-m^2&=0\, ,\\
q'^2-m^2&=0\, ,\\
\label{eq:sfkinematic2}
k^2&=0\, ,
\end{align}
where $l_1$ is the Fourier conjugate to $(n\cdot s_1)$ and $l_2$ to $(n\cdot s_2)$. An equivalent set of equations is
\begin{align}
\label{eq:fk1}
k-p'+q'&=(l_1-l_2)n\, ,\\
\label{eq:fk2}
p-p'&=l_1n\, ,\\
\label{eq:fk3}
q-q'&=l_2n\, ,\\
\label{eq:poff}
p^2-m^2&=2l_1(n\cdot p)\, ,\\
\label{eq:qoff}
q^2-m^2&=2l_2(n\cdot q)\, ,\\
\label{eq:pmq}
(p'-q')^2&=-2(l_1-l_2)(n\cdot k)\, .
\end{align}
Equations \refe{poff} and \refe{qoff} make explicit that the physical momenta $p$, $q$ (carried by the fermion distribution functions) contribute with arbitrary off-shell values, where the `off-shellness' $2l_1(n\cdot p')$ and $2l_2(n\cdot q')$ is integrated over in the collision terms. In this way, the macroscopic field provides the momenta $l_1n^\mu$ and $l_2n^\mu$, preventing the collision terms from vanishing kinematically. Furthermore, the auxiliary momenta $p'$, $q'$ are not conserved and $k-p'+q'$ is not always zero, but corresponds to the energy exchanged with the field, $(l_1-l_2)n$.

In comparison, the zero-field kinematic conditions are
\begin{align}
\label{eq:freekinematics}
k-p+q&=0\, ,\\
p^2-m^2&=0\, ,\\
q^2-m^2&=0\, ,\\
k^2&=0\, ,
\end{align}
which are `forbidden' because
\begin{align}
(k+q)^2\stackrel{!}{=}p^2&\lra (k\cdot q)\stackrel{!}{=}0\hspace{0.5cm} \textrm{ (on-shell) }\, ,\\
(p-k)^2\stackrel{!}{=}q^2&\lra (k\cdot p)\stackrel{!}{=}0\hspace{0.5cm} \textrm{ (on-shell)}
\end{align}
for massive fermions can only be fulfilled for the trivial case of $\vec k=0$, while otherwise
\begin{align}
k\cdot p=|\vec k|\varepsilon(\vec p\,)-\vec k\cdot \vec p >0\hspace{0.5cm}\textrm{ (on-shell)}\, .
\end{align}
Thereby, for vanishing macroscopic field, the delta functions have vanishing overlap and zero-field collision terms vanish at leading order $\O(e^2)$.

\subsection{\label{par:planecoll}Plane-wave photon kinetic equation}

\subsubsection{\label{par:collmax}Collision term}

Employing the plane-wave collision kernel \refe{planekernel}, the photon transport equation \refe{sfpkin} obtains the following collision term:
\begin{widetext}
	\begin{align}
	\label{eq:planephotoncoll}
	&C\iv(X,\vec k)=e^2\int_0^\infty \dif k^0 \delta(k^2)\int\frac{\dif l_1}{(2\pi)}\int\frac{\dif l_2}{(2\pi)}\,\int_{p,q}(2\pi)^6\delta(k-p+q)\,\mathcal{Q}(X,l_1,l_2,p-l_1n,q-l_2n)\\
	\nonumber
	&\hspace{0.3cm}\cross\delta(p^2-m^2-2l_1(n\cdot p))\,\,\sgn(p^0-l_1n^0)\delta(q^2-m^2-2l_2(n\cdot q))\,\,\sgn(q^0-l_2n^0)\\
	\nonumber
	&\hspace{0.3cm}\cross\Big\{ f_\Psi(X,p)[1-f_\Psi(X,q)][1+f(X,k)]-[1-f_\Psi(X,p)] f_\Psi(X,q)f(X,k) \Big\}\, .
\end{align}
We can identify the crossings of $ee\gamma$ scattering depicted in Fig. \ref{fig:coll} by taking the frequency integrals over
\begin{align}
\delta(p^2-m^2-2l(n\cdot p))= \frac{1}{2\varepsilon_l(\vec p\,)}\Big[\delta(p^0-l-\varepsilon_{\vec p\,}(l))+\delta(p^0-l+\varepsilon_{\vec p\,}(l))\Big]\, .
\end{align}
Identifying plane-wave degrees of freedom in terms of the plane-wave fermion and anti-fermion distribution functions
	\begin{align}
	\label{eq:lminus}
	f^-_\Psi(X,l,\vec p\,)&\defeq f_\Psi(X,p)\,\,\,\textrm{ at }\,\,\,p^0=l+\varepsilon_{l}(\vec p\,)\, ,\\
	\label{eq:lplus}
	f^+_\Psi(X,-l,-\vec p\,)&\defeq 1-f_\Psi(X,p)\,\,\,\textrm{ at }\,\,\,p^0=l-\varepsilon_{l}(\vec p\,)\, 
	\end{align}
	and making use of Eq.~\refe{epl3}, the plane-wave photon collision term may equivalently be written as
	\begin{align}
	\label{eq:crossphoton}
	C\iv(X,\vec k)&=e^2\frac{1}{2|\vec k|}\int\frac{\dif l_1}{(2\pi)}\int \frac{\dif^3p}{(2\pi)^3}\frac{1}{2\varepsilon_{l_1}(\vec p\,)}\int\frac{\dif l_2}{(2\pi)}\int \frac{\dif^3q}{(2\pi)^3}\frac{1}{2\varepsilon_{l_2}(\vec q\,)}(2\pi)^4\\
	\nonumber
	&\cross\Big[\delta(\vec k+\vec p-\vec q\,)\,\delta(|\vec k|+\varepsilon_{l_1}(\vec p\,)-\varepsilon_{l_2}(\vec q\,)+l_1-l_2)\,\mathcal{Q}^{e^+\rightarrow e^+\gamma}(X,l_1,l_2,\vec p,\vec q\,)\\
	\nonumber
	&\cross\Big\{ [1-f_\Psi^+(X,l_1,\vec p\,)]f_\Psi^+(X,l_2,\vec q\,)[1+f(X,\vec k)]-f_\Psi^+(X,l_1,\vec p\,)[1-f_\Psi^+(X,l_2,\vec q\,)]f(X,\vec k) \Big\}\\
	\nonumber
	&+\delta(\vec k-\vec p+\vec q\,)\,\delta(|\vec k|-\varepsilon_{l_1}(\vec p\,)+\varepsilon_{l_2}(\vec q\,)-l_1+l_2)\,\mathcal{Q}^{e^-\rightarrow e^-\gamma}(X,l_1,l_2,\vec p,\vec q\,)\\
	\nonumber
	&\cross\Big\{ f_\Psi^-(X,l_1,\vec p\,)[1-f_\Psi^-(X,l_2,\vec q\,)][1+f(X,\vec k)]-[1-f_\Psi^-(X,l_1,\vec p\,)]f_\Psi^-(X,l_2,\vec q\,)f(X,\vec k) \Big\}\\
	\nonumber
	&+\delta(\vec k-\vec p-\vec q\,)\,\delta(|\vec k|-\varepsilon_{l_1}(\vec p\,)-\varepsilon_{l_2}(\vec q\,)-l_1-l_2)\,\mathcal{Q}^{\gamma\rightarrow e^+e^-}(X,l_1,l_2,\vec p,\vec q\,)\\
	\nonumber
	&\cross\Big\{ f_\Psi^-(X,l_1,\vec p\,)f_\Psi^+(X,l_2,\vec q\,)[1+f(X,\vec k)]-[1-f_\Psi^-(X,l_1,\vec p\,)][1-f_\Psi^+(X,l_2,\vec q\,)]f(X,\vec k) \Big\}\\
	\nonumber
	&+\delta(\vec k+\vec p+\vec q\,)\,\delta(|\vec k|+\varepsilon_{l_1}(\vec p\,)+\varepsilon_{l_2}(\vec q\,)+l_1+l_2)\,\mathcal{Q}^{0\rightarrow e^+e^-\gamma}(X,l_1,l_2,\vec p,\vec q\,)\\
	\nonumber
	&\cross\Big\{ [1-f_\Psi^+(X,l_1,\vec p\,)][1-f_\Psi^-(X,l_2,\vec q\, )][1+f(X,\vec k)]-f_\Psi^+(X,l_1,\vec p\,)f_\Psi^-(X,l_2,\vec q\,)f(X,\vec k) \Big\}\Big]\, ,
	\end{align}
	
		\begin{figure}[h!]
		\begin{center}
			\includegraphics[scale=0.08]{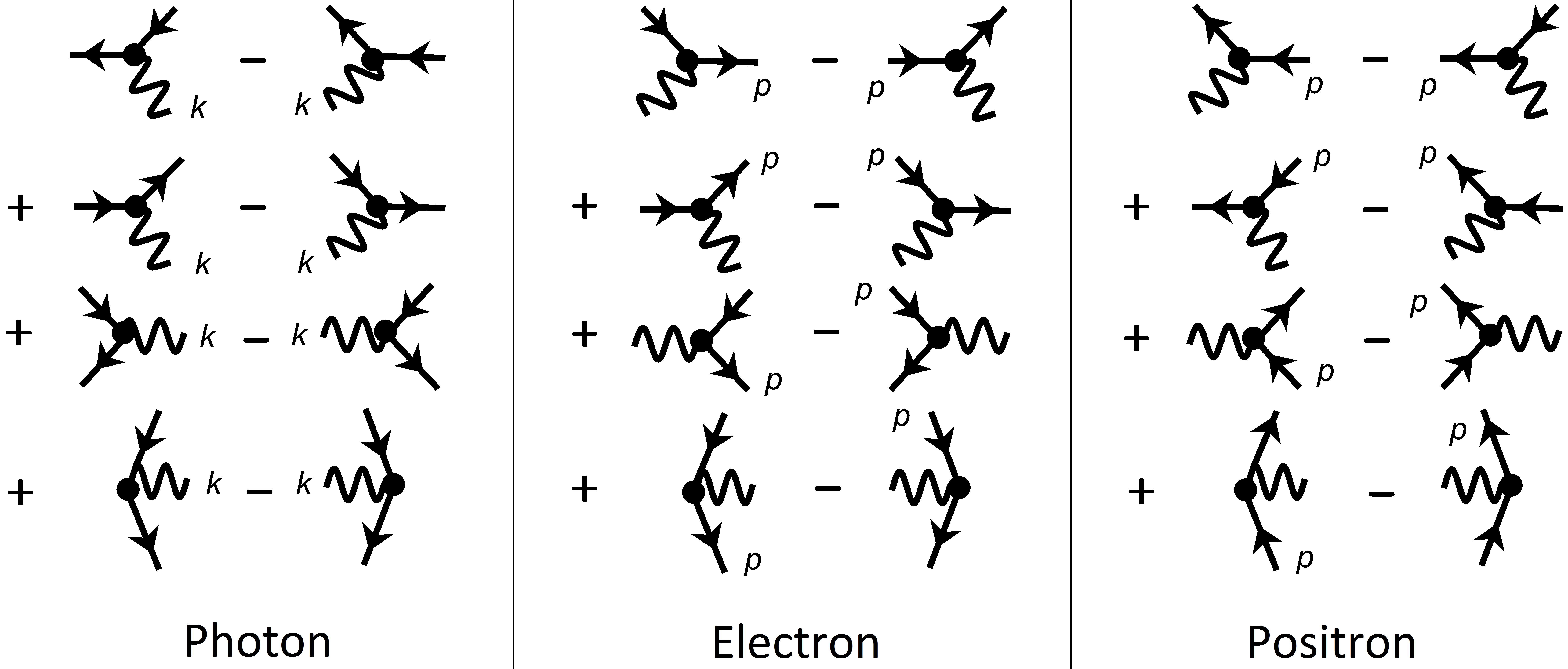}
		\end{center}
		\caption{\label{fig:coll}Diagrammatic photon, electron and positron collision terms; labeled are the respective external momenta.}
	\end{figure}
	
	\noindent with the collision kernels for the different crossings of $ee\gamma$ scattering\footnote{The sign in the kernels involving one positron recovers the positron term $(\slashed p-m)$ from the electron term $(\slashed p+m)=-(-\slashed p-m)$ after the change of sign $p\rightarrow -p$ (see also Ref.~\cite{PhysRevA.95.032121}).}
	\begin{align}
	\mathcal{Q}^{e^+\rightarrow e^+\gamma}(X,-l_1,-l_2,-\vec p,-\vec q\,)&\defeq \mathcal{Q}^{\mu}{}_\mu(X,l_1,l_2,p-l_1n,q-l_2n)\,\,\,\textrm{ at }\,\,\,p^0=l_1-\varepsilon_{l_1}(\vec p\,),\,\,q^0=l_2-\varepsilon_{l_2}(\vec q\,)\, ,\\
	\mathcal{Q}^{e^-\rightarrow e^-\gamma}(X,l_1,l_2,\vec p,\vec q\,)&\defeq\mathcal{Q}^{\mu}{}_\mu(X,l_1,l_2, p-l_1n,q-l_2n)\,\,\,\textrm{ at }\,\,\,p^0=l_1+\varepsilon_{l_1}(\vec p\,),\,\,q^0=l_2+\varepsilon_{l_2}(\vec q\,)\, ,\\
	-\mathcal{Q}^{\gamma\rightarrow e^+e^-}(X,l_1,-l_2,\vec p,-\vec q\,)&\defeq \mathcal{Q}^{\mu}{}_\mu(X,l_1,l_2, p-l_1n,q-l_2n)\,\,\,\textrm{ at }\,\,\,p^0=l_1+\varepsilon_{l_1}(\vec p\,),\,\,q^0=l_2-\varepsilon_{l_2}(\vec q\,)\, ,\\
	-\mathcal{Q}^{0\rightarrow e^+e^-\gamma}(X,-l_1,l_2,-\vec p,\vec q\,)&\defeq\mathcal{Q}^{\mu}{}_\mu(X,l_1,l_2, p-l_1n,q-l_2n)\,\,\,\textrm{ at }\,\,\,p^0=l_1-\varepsilon_{l_1}(\vec p\,),\,\,q^0=l_2+\varepsilon_{l_2}(\vec q\,) \, .
	\end{align}
\end{widetext}

As anticipated in our discussion of the plane-wave spectral function \refe{planeparticle}, we can observe from the energy conserving delta functions in the collision term \refe{crossphoton} that the Wigner variables $l_i$ correspond to the energy that is exchanged with the macroscopic field by degrees of freedom with energy $\varepsilon_{l_i}(\vec p_i)$. By means of the changes of variables $p_z\rightarrow p_z-l_1$ and $q_z\rightarrow q_z-l_2$, this energy exchange can be written in the Lorentz covariant form
\begin{align}
\label{eq:covcons}
k-p+q-(l_1-l_2)n=0\, ,
\end{align}
which clearly relates four-momentum conservation to the structure of the plane-wave field.

From the delta functions of the $0\rightarrow 3$ and $3\rightarrow 0$ processes in Eq.~\refe{crossphoton}, we observe that they are forbidden for plane-wave fields, since the combination of energy and momentum conditions,
\begin{align}
\label{eq:2loopsource}
|\vec k|+\varepsilon_{l_1}(\vec p\,)+\varepsilon_{l_2}(\vec q\,)+l_1+l_2&=0\, ,\\
\vec k+\vec p+\vec q&=0\, ,
\end{align}
can not be fulfilled. In the absence of a macroscopic field, such processes are already forbidden by energy conservation alone. For an arbitrary macroscopic field such a $0\rightarrow 3$ term would act as a source term for vacuum pair production since it does not come with any distribution function and therefore would not vanish for $f_\Psi^+=f_\Psi^-=0$. The fact that this contribution vanishes for plane-wave fields is in agreement with the general statement that plane waves are not able to produce pairs from the vacuum \cite{PhysRev.82.664, Narozhnyi:1976zs}. In general, such $0\rightarrow 3$ terms contribute to vacuum pair production at 2-loop order via Eq.~\refe{pairgeneral}.

	\subsubsection{\label{par:vertex}Emergence of a gauge-invariant vertex and gauge-fixing independent amplitude in plane-wave vacuum}
	
	Electromagnetic interactions are often described in terms of probabilities for scattering events built from S-matrix amplitudes, which are computed in terms of Feynman rules with free on-shell asymptotic states in vacuum, i.e.\ vanishing or single mode distribution functions. Such an S-matrix based formulation is not able to resolve real-time dynamics between in-medium states, for which the interaction is not adiabatically switched off. In this section, we follow the emergence of such amplitudes and thereby highlight limitations to their ability to capture collective dynamics of strong-field systems.
	
	General considerations about the trace $\T$, Eq.~\refe{preexp}, can be found e.g.\ in the reviews \cite{Mitter_1975,Ritus1985,Ehlotzky_2009,RevModPhys.84.1177} (see also Ref.~\cite{PhysRevA.98.012134}) for the special case of $s_1+s_2=0$ and with $X$ integrated over, which is needed for the computation of probabilities. In this section we identify a scattering amplitude that is local in $X$ and demonstrate that the reduction in terms of relative variables $s_1$ and $s_2$ is related to vanishing or single mode plane-wave fermion distribution functions,
	\begin{align}
	\label{eq:zeroferm}
	f^{\pm}_\Psi(X,l,\vec p\,)\rightarrow 0\, ,
	\end{align}
	which we refer to as the `plane-wave vacuum'. Importantly, such vacuum approximations to distribution functions may only be applied once the relevant degrees of freedom are separated from quantum vacuum fluctuations, because general off-shell distribution functions contain the `quantum half' which can never physically vanish [see e.g.\ the constant terms in Eqs.~\refe{cp} and \refe{photoncp}].

We start in-medium, i.e. without the assumption \refe{zeroferm}, where the collision kernel \refe{planekernel}, may be factorized in terms of Volkov spinors,
		\begin{align}
		U_{p\sigma}(x)&\defeq R_p(x)u_{p\sigma}\, ,\\
		\bar U_{p\sigma}(x)&\defeq U_{p\sigma}^\dagger \gamma^0=\bar u_{p\sigma}\bar R_p(x)\, ,
		\end{align}
		and written as a `spin sum'
		\begin{widetext}
		\begin{align}
		\label{eq:spinorkernel}
		\P\iv^{\mu\nu}(X,p,q,k)&=-\delta(k^2)\sgn(k^0)\int_{p',q'}\delta(p'^2-m^2)\sgn(p'^0)\delta(q'^2-m^2)\sgn(q'^0)\\
		\nonumber
		&\hspace{1cm}\cross\tfrac{1}{4}\sum_{\sigma\sigma'}\int_{s_1,s_2}e^{ips_1}e^{iqs_2}\Big[\bar U_{q'\sigma'}(X-\tfrac{s_2}{2})\gamma^\mu
		 U_{p'\sigma}(X+\tfrac{s_1}{2})\Big]\Big[\bar U_{p'\sigma}(X-\tfrac{s_1}{2})
		 \gamma^\nu
		 U_{q'\sigma'}(X+\tfrac{s_2}{2})\Big]\, ,
		\end{align}
		\end{widetext}
		by introducing spin labels $\sigma$ and $\sigma'$ via
	\begin{align}
\label{eq:spinsum}
\delta(p^2-m^2)\sum_{\sigma}u_{p\sigma}\bar u_{p\sigma}&=\delta(p^2-m^2)(\slashed p+m)\, .
\end{align}
By amputating the free Dirac spinors $\bar u_{p\sigma}$, $u_{p\sigma}$ of
\begin{align}
&\bar U_{q\sigma'}(X-\tfrac{s_2}{2})\gamma^\mu U_{p'\sigma}(X+\tfrac{s_1}{2})\\
\nonumber
&\hspace{2cm}\eqdef \bar u_{q\sigma'}\,V^\mu_{qp}(X,s_2,s_1)\,u_{p\sigma}\, ,
\end{align}
in Eq.~\refe{spinorkernel}, we may identify the vertex
\begin{align}
\label{eq:planevertex}
V^\mu_{pq}(X,s_1,s_2)&=\bar R_{p}(X-\tfrac{s_1}{2})\gamma^\mu R_{q}(X+\tfrac{s_2}{2})\, .
\end{align}
This expression differs from the well-known local and gauge-invariant plane-wave vertex \cite{Morozov:1981pw, Ritus1985},
\begin{align}
\label{eq:localvertex}
\Gamma^\mu_{pq}(x)\defeq\bar R_{p}(x)\gamma^\mu R_q(x)=V_{pq}^\mu(X,-s,s) \, ,
\end{align}
by its spacetime structure. This difference arises because the local vertex $\Gamma^\mu_{pq}$ is constructed from the time-ordered Volkov propagator (see appendix \ref{app:volkov}), which is a vacuum object, i.e.\ assumes the absence of a medium by vanishing distribution functions, while our vertex $V^\mu_{pq}$ is constructed in the presence of distribution functions from the antisymmetric part of the Volkov propagator alone. The additional $s$-dependence of $V^\mu_{pq}$, which is integrated over in the collision kernel thus implements the fact that the effective interaction in a strong-field medium is non-local.

While $\Gamma^\mu_{pq}$ is gauge-invariant, $V^\mu_{pq}$ is not.\footnote{The Volkov spinors transform as $U_{p\sigma}(x)\rightarrow e^{i\alpha(x)}U_{p\sigma}(x)$ and $\bar U_{p\sigma}(x)\rightarrow \bar U_{p\sigma}(x)e^{-i\alpha(x)}$ with a U(1) group element $e^{i\alpha(x)}$.} We stress that our collision term is nevertheless gauge-invariant, such that this is not a flaw of our description, but simply exhibits the physical limitations of the concept of scattering probabilities. Electromagnetic interactions in the presence of a medium, i.e.\ arbitrary distribution functions, can not in general be described by assigning probabilities to individual events. While the photon collision term \refe{crossphoton} is gauge-invariant by virtue of additional momentum integrals, the collision kernel and the vertex \refe{planevertex} are not gauge-invariant on their own. Without further assumptions, we can not identify scattering probabilities from them. As we now demonstrate, gauge-invariant amplitudes can be defined in plane-wave vacuum.

First, we investigate how $V^\mu_{pq}$ reduces to $\Gamma^\mu_{pq}$. For this we need to put the vertex back into the context of the collision term:  In general, the photon collision term is of the form
\begin{align}
\label{eq:nonfac}
\int_{p,q}\delta(k-p+q)\,g(X,p,q,k)\,\P^{\mu\nu}(X,p,q,k)\, ,
\end{align}
with the gain-minus-loss term $g$. If we now assume the absence of a medium, i.e.\ Eq.~\refe{zeroferm}, there are no other objects carrying fermion momentum dependence other than the kernel itself. We may then write
\begin{align}
\label{eq:vacuumkernel}
&\int_{p,q}\delta(k-p+q)\,\P\iv^{\mu\nu}(X,p,q,k)\\
\nonumber
&=\int_{p,q,l}\hspace{-0.1cm}\delta(p^2-m^2)\sgn(p^0)\delta(q^2-m^2)\sgn(q^0)\delta(k^2)\sgn(k^0)\\
\nonumber
&\hspace{2cm}\cross \delta(k-p+q-ln)\,\mathcal{Q}_{\textrm{vac}}^{\mu\nu}(X,l,p,q)\, ,
\end{align}
with the gauge-invariant vacuum kernel
\begin{align}
\label{eq:qvac}
&\mathcal{Q}_{\textrm{vac}}^{\mu\nu}(X,l,p,q)\\
\nonumber
&\defeq\int\dif s^-\,e^{ils^-}\,\T^{\mu\nu}_{pq}(X,s,-s)\,e^{-i[\N_p(X,s)+\N_q(X,-s)]}\, .
\end{align}
Here, the underlying structure that is simplified by the vacuum assumption is the product of Wigner space fermion spectral functions \mbox{$\rho_\Psi(X,p)\rho_\Psi(X,q)$}, that can in general not be factorized in real-space via \mbox{$\rho_\Psi(x,y)\rho_\Psi(y,x)=\hat\rho_\Psi(x,y)\hat\rho_\Psi(y,x)$} in the presence of fermion distribution functions, e.g.\ as in expression \refe{nonfac}. However, in the vacuum case, Eq.~\refe{vacuumkernel} contains such a factorization, where the $\delta(k-p+q)$ has been expressed in real space to invert the Wigner transforms of the spectral functions as in Eq.~\refe{photoninv}. Eq.~\refe{vacuumkernel} then allows us to identify the auxiliary momentum labels $p'$, $q'$ of the collision kernel with the physical fermion momenta $p$, $q$ in the vacuum case. The emerging vacuum collision kernel is gauge-invariant on its own, and has contributions only from values of $s_1$ and $s_2$ satisfying the condition $s_1+s_2=0$. The momentum labels $p$ and $q$ of the scattering kernel are now on-shell, but there are no fermion distribution functions left. Correspondingly, $k-p+q\neq 0$ because momentum is exchanged with the macroscopic field as the amplitude would otherwise vanish kinematically as in the zero-field case. In case of fermion distribution functions that vanish almost everywhere, except e.g.\ a few ultrarelativistic modes, the dominant contributions from the collision kernel still come from the region of $s_1+s_2=0$. By taking the collision kernel out of the context of the transport equation in this way, medium effects from more complex fermion distribution functions such as the non-local interaction via Eq.~\refe{planevertex}, and the difference between the on-shell labels $p'$, $q'$ and the off-shell labels $p$, $q$ are missed.

We can now make contact with the language of amplitudes by writing the vacuum collision kernel \refe{qvac} in terms of the local vertex \refe{localvertex},
\begin{align}
\label{eq:amplitudekernel}
&\int_{p,q}\hspace{-0.1cm}\delta(k-p+q)\P\iv(X,p,q,k)=-\delta(k^2)\sgn(k^0)\\
\nonumber
&\cross\int_{p,q}\hspace{-0.1cm}\delta(p^2-m^2)\sgn(p^0)\delta(q^2-m^2)\sgn(q^0)\tfrac{1}{4}\eta_{\mu\nu}\sum_{\sigma\sigma'}\\
\nonumber
&\cross\int_se^{iks}\Big[\bar u_{p\sigma}\Gamma^\mu_{pq}(X+\tfrac{s}{2})u_{q\sigma'}\Big]\Big[\bar u_{q\sigma'}\Gamma^\nu_{qp}(X-\tfrac{s}{2})u_{p\sigma}\Big]\, .
\end{align}
From this we may read off the local amplitude
\begin{align}
\label{eq:localamplitude}
\tilde \M^\mu_{\sigma\sigma'}(X,p,q,k)&=\int_s e^{iks}\,\bar u_{p\sigma}\,\Gamma^\mu_{pq}(X+\tfrac{s}{2})\,u_{q\sigma'}\, ,\\
\nonumber
[\tilde \M^\mu_{\sigma\sigma'}(X,p,q,k)]^*&=\int_s e^{iks}\,\bar u_{q\sigma'}\,\Gamma^\mu_{qp}(X-\tfrac{s}{2})\,u_{p\sigma  }\,.
\end{align}
It is tempting to go one step further and  identify the square of the well-known global amplitude \cite{Ritus1985},
\begin{align}
\label{eq:globalamplitude}
\M_{\sigma\sigma'}^\mu(p,q,k)&=\int_x\, e^{ikx}\,\bar u_{p\sigma}\,\Gamma^\mu_{pq}(x)\,u_{q\sigma'}\, ,
\end{align}
by integrating over all $X$ and returning to microscopic position variables via
\begin{align}
\int\dif^4X\int_se^{iks}=\int\dif^4xe^{ikx}\int\dif^4ye^{-iky}\, .
\end{align}
However, it is important to note that such an integration over all times $X^0$ generally includes late times outside of the range of validity of both the plane-wave field and the plane-wave vacuum approximation. Even if one makes assumptions such as \refe{plane} and \refe{zeroferm} at initial time, the macroscopic field does not remain a plane wave and distribution functions do not in general remain negligible, but backreact on the field, such that a self-consistent description of both becomes essential to describe equilibration. What type of interactions take place in a strong-field system depends also on the details of distribution functions and is a time (and space) dependent questions. To determine this time dependence, one has to solve the transport system including the dynamics of distribution functions away from the plane-wave vacuum. Instead, a common approach in literature is to rely on the S-matrix in the Furry picture \cite{PhysRev.81.115}, which takes amplitudes out of the context of in-medium evolution equations. To then extract local probabilities from this S-matrix, the LCFA is a necessary approximation, as otherwise what is supposed to be the local probability density may turn out to be negative (see e.g.\ Ref.~\cite{PhysRevA.98.012134}). This problem occurs because the probability that a scattering will take place in an external field and in the absence of a medium at any time is not a self-consistent concept in general. The gauge-invariant amplitudes such as \refe{globalamplitude} are not observable and probabilities for individual scattering processes need not exist to compute statistical observables such as electrical conductivity \cite{Carrington:2007fp} or pressure \cite{Borsanyi:2007bf}. No further approximations such as the plane-wave vacuum or locally constant fields are necessary to compute also e.g.\ the photon decay rate \refe{photondecayrate} from the equations discussed in Sec.~\ref{sec:4}.

While the physical interpretation of the strong-field amplitude \refe{globalamplitude} is problematic, that object is very useful to understand the $\xi$-dependence of our equations. The amplitude \refe{globalamplitude} is known to obey the modified Ward identity \cite{Boca_2010, PhysRevA.83.032106, Ilderton:2010wr}
\begin{align}
\nonumber
k\cdot \M_{\sigma\sigma'}&=\bar u_{p\sigma}\slashed n u_{q\sigma'}\int_l(2\pi)^4\,\delta(k-p+q-ln)\\
\label{eq:3pointward}
&\hspace{1cm}\cross\int\dif(n\cdot x) \xpdify{\big(e^{i\Phi(n\cdot x)}\big)}{(n\cdot x)}\, ,
\end{align}
with the phase
\begin{align}
\nonumber 
\Phi(n\cdot x)&\defeq l\,n\cdot x+\int_{-\infty}^{n\cdot x}\dif\lambda\Big(\frac{e\A(\lambda)\cdot q}{n\cdot q}-\frac{e^2\A^2(\lambda)}{2(n\cdot q)}\Big)\\
&\hspace{0.5cm}-\int_{-\infty}^{n\cdot x}\dif\lambda\Big(\frac{e\A(\lambda)\cdot p}{n\cdot p}-\frac{e^2\A^2(\lambda)}{2(n\cdot p)}\Big)\, ,
\end{align}
relating gauge-fixing to boundary terms at $n\cdot x=\pm\infty$. Vanishing boundary terms then lead to gauge-fixing independence, $\P_\xi\equiv 0$.
\vspace{1cm}

\subsection{\label{subsec:fermkinplane}Plane-wave fermion kinetic equation}

Because the fermion collision term \refe{fermcoll} relies on the gauge-invariant fermion spectral function $\hat \rho_\Psi$ (as opposed to the covariant function $\rho_\Psi$), we start this section by investigating this function for plane-wave fields. The well-known plane-wave momentum and dressed mass emerge automatically in this function. These gauge-invariant expressions then serve us to approximate field-gradients in a gauge-invariant manner, equivalently to Sec.~\ref{subsec:5a}, but at the level of the solution rather than the equation of motion.
 
\subsubsection{\label{par:invspec}Gauge-invariant spectral function:\\plane-wave momentum \& dressed mass}

The covariant plane-wave spectral function \refe{planeparticle} transforms as any other fermion two-point function. The ambiguity \cite{Elze:1986qd, Vasak:1987um, Blaizot:1999xk} for the choice of the path of integration in the Wilson line is not present in the plane-wave case because there is only one path in one dimension from $n\cdot x$ to $n\cdot y$. Thereby, the Wilson line automatically emerges with a straight path of integration,
\begin{align}
\label{eq:planewilson}
&\W\iv(y,x)=\exp\Bigg(ie\frac{s^\mu}{s^-}\int_{-\tfrac{s^-}{2}}^{\tfrac{s^-}{2}}\dif \lambda \, \A_\mu(X^-+ \lambda)\Bigg)\, ,
\end{align}
despite the $3+1$ dimensional nature of the underlying theory. Defining the phase-average
\begin{align}
\label{eq:phaseavgdef}
\braket{a}(X^-,s^-)\defeq\frac{1}{(n\cdot s)}\int_{-\tfrac{n\cdot s}{2}}^{\tfrac{n\cdot s}{2}}\dif \lambda\, a(n\cdot X+\lambda)
\end{align}
(not to be confused with the ensemble average \refe{ensembleavg}) for any plane-wave function $a(n\cdot x)$, we can make the gauge-invariance of $\hat \rho_{\Psi,\textrm{v}}$ manifest. By employing Eq.~\refe{phaseavgdef}, the plane-wave Wilson line \refe{planewilson} can be written as
\begin{align}
\W\iv(y,x)=e^{ies^\mu\braket{\A_\mu}(X,s)}\,.
\end{align}

The Lorentz equation for plane-wave fields is solved by the gauge-invariant momentum of an electron in a plane-wave \cite{LANDAU1975109}
\begin{align}
\label{eq:kinmom}
&\pi^\mu_q(n\cdot X)\\
\nonumber
&\defeq q^\mu -e\A^\mu(X)+n^\mu\Bigg(\frac{e\A(X)\cdot q }{(n\cdot q)} -\frac{e^2\A^2(X)}{2(n\cdot q)}\Bigg)\, ,
\end{align}
which is related to the Lorentz action \refe{lorentzaction} via
\begin{align}
(i\pdif_x^\mu-e\A\iv^\mu(n\cdot x))\,e^{iS_q(x)}=\pi_q^\mu(n\cdot x)\,e^{iS_q(x)}\, .
\end{align}
The plane-wave momentum obeys
\begin{align}
\pi^2_q=q^2 \,\,\,\,\textrm{ and }\,\,\,\,n\cdot \pi_q=n\cdot q
\end{align}
and is related to the free mass $m$ and the gauge-invariant dressed mass \cite{Brown:1964zzb, PhysRevD.97.056028} 
\begin{align}
\nonumber
\tilde m^2(X,s^-)&\defeq m^2-\frac{e^2}{(n\cdot s)}\int_{-\tfrac{n\cdot s}{2}}^{\tfrac{n\cdot s}{2}}\dif \lambda\, \A^2(n\cdot X+\lambda)\\
\label{eq:dressedmass}
&\hspace{-0.5cm}+\frac{e^2}{(n\cdot s)^2}\Big[ \int_{-\tfrac{n\cdot s}{2}}^{\tfrac{n\cdot s}{2}}\dif \lambda \,\A(n\cdot X+ \lambda) \Big]^2
\end{align}
 via \cite{PhysRevD.97.056028}
\begin{align}
\label{eq:dressedkin}
\braket{\pi_q^2}=m^2\,\,\,\,\textrm{ and }\,\,\,\,
\sbraket{\pi_q}=\tilde m^2
\end{align}
for any $q$ with $q^2=m^2$ (which in our context is ensured by the delta function under the integral e.g.\ in Eq.~\refe{invariantvolkov}).

We can identify this plane-wave momentum in the exponent of the gauge-invariant spectral function via 
\begin{align}
q\cdot s-es^\mu\braket{\A_\mu}+\N_q=\braket{\pi^\mu_q}s_\mu\, ,
\end{align}
such that an exact solution of Eq.~\refe{allgrad} in the plane-wave case may be written as
\begin{align}
\label{eq:invariantvolkov}
\hat \rho_{\Psi,\textrm{v}}(X,p)&=i(2\pi)\int_q\delta(q^2-m^2)\sgn(q^0)\\
\nonumber
&\hspace{0.5cm}\cross\int_s e^{i(p-\braket{\pi_q}(X,s))s}\, \, \tilde \K_q(X,s)\, ,
\end{align}
with the gauge-invariant Dirac matrix
\begin{align}
&\tilde \K_q(X,s)\\
\nonumber
&\defeq \left[ 1+\frac{e}{2}\frac{\slashed n \slashed \A_\textrm{v}(X+\tfrac{s}{2})}{(n\cdot q)} \right](\slashed q+m)\left[ 1-\frac{e}{2}\frac{\slashed n \slashed \A_\textrm{v}(X-\tfrac{s}{2})}{(n\cdot q)} \right]\, .
\end{align}
While the covariant spectral function \refe{planeparticle} makes manifest the energy exchange with the field and facilitates a formulation in the plane-wave degrees of freedom \refe{lminus} and \refe{lplus}, the invariant function \refe{invariantvolkov} makes manifest the solution of the Lorentz equation \refe{kinmom}.

The scalar and pseudoscalar components of $\tilde K_q$ are
\begin{align}
\tfrac{1}{4}\textrm{tr}\{\tilde \K_q(X,s)\}&=m\, ,\\
\tfrac{1}{4}\textrm{tr}\{\gamma^5\tilde \K_q(X,s)\}&=0\, .
\end{align}
The vanishing of the pseudoscalar component is a direct consequence of the crossed nature of plane-wave fields, i.e.\ Eq.~\refe{topological}. The vector component, which plays a crucial role in the quantum Vlasov term, contains the plane-wave momentum also in the pre-exponential and is given in Sec.~\ref{par:planedrift}. The axial and tensor components can be found in the appendix \ref{app:volkovcomp}. The tensor and scalar components vanish for massless fermions in agreement with chiral symmetry.

Similarly to the identity \refe{lid} one has
\begin{align}
&\int_se^{i(p-\braket{\pi_q}(X,s^-))s}\, \tilde \K(X,s^-)=\int\dif s^- (2\pi)^3\\
\nonumber
&\hspace{0.5cm}\cross\delta\big(p^--q^-\big)\delta\big(\vec p_\perp-\braket{\vec \pi_{\perp,q}}(X,s^-)\big)\\
\nonumber
&\hspace{1.5cm}\cross e^{i\big[p^+-\braket{\pi_q^+}(X,s^-)\big]s^-}\, \tilde \K(X,s^-)\, .
\end{align}

Computation of the scalar component (see appendix \ref{app:scalar}) results in
\begin{align}
\hat \rho_{\Psi,\textrm{v},\textrm{S}}(X,p)
\label{eq:svolk}
&= i\,m\,\frac{1}{2p^-}\\
\nonumber
&\hspace{-0.5cm}\cross\int\dif s^-\exp\Bigg\{i\Big(p^+-\frac{|\vec p_\perp|^2+\tilde m^2(X,s^-)}{2p^-}\Big)s^-\Bigg\}\, .
\end{align}
The corresponding symmetric component has been computed in Ref.~\cite{PhysRevD.83.065007} (see also Ref.~\cite{Hebenstreit:2011cr}) for various choices of plane-wave fields, in the context of scalar QED.

\subsubsection{\label{subsec:smallgradplane}Plane-wave fields with small gradients}

In this section, we investigate the gauge-invariant approximation of field-gradients using the example of the scalar spectral component \refe{svolk}.

For plane-wave fields, the gradient expansion becomes an expansion in longitudinal gradients via
\begin{align}
(s\cdot\pdif_X)^j\A_\mu(n\cdot X)= (n\cdot s)^j\,\A_\mu^{(j)}(n\cdot X)\, ,
\end{align}
where $\A_\mu^{(j)}$ is the $j$th derivative with respect to $n\cdot X$.

In the scalar component \refe{svolk}, field-gradients are carried only by the gauge-invariant mass
\begin{align}
\nonumber
\tilde m^2(X,s^-)&=m^2-e^2\tfrac{1}{12}(n\cdot s)^2\dot{\A}^2(X)-e^2\tfrac{1}{720}(n\cdot s)^4\\
\label{eq:dressedmasslo}
&\hspace{-1.7cm}\cross \Big[ 3\A^{(3)}(X)\cdot \dot \A(X)+\ddot \A^2(X)\Big]+ \O\big((e^0s\cdot\pdif_X)^{5}\big)\, ,
\end{align}
whose expansion is gauge-invariant order by order \cite{PhysRevD.97.056028}.

Similar to the fact that the equation of motion \refe{lcfaeq} has contributions from constant gauge-invariant fields, the second term of the dressed mass is also non-trivial for constant electric fields and generally not small compared to unity. In fact, introducing the dimensionless and Lorentz-invariant quantities \begin{align}
\label{eq:xidef}
\xi_0&\defeq \frac{m}{\omega}\frac{\F_0}{{\mathcal{E}_\textrm{c}}}\, ,\\
\label{eq:chidef}
\chi_0(p)&\defeq \frac{n\cdot p}{m}\frac{\F_0}{\mathcal{E}_\textrm{c}}\, ,\\
\varphi&\defeq \omega(n\cdot s)\, ,
\end{align}
with the Schwinger critical field $\mathcal{E}_{\textrm{c}}=m^2/|e|$ and a characteristic field amplitude $\F_0$ and frequency $\omega$, we can write the constant-field contribution from this exponent as
\begin{align}
\label{eq:xichi}
-e^2\frac{1}{12}\frac{\dot{\A}^2(X)}{2(n\cdot p)}(n\cdot s)^3=\frac{1}{24}\frac{\xi_0^3}{\chi_0}\frac{\E^2(X)}{\F_0^2}\varphi^3\, ,
\end{align}
with the notation $\E(X)\defeq |\vec\E(X)|$. Equation \refe{xichi} reveals the significance of $\xi_0^3/\chi_0$ for locally constant fields, which is well known in laser physics \cite{Fedotov:2016afw, PhysRevD.99.076004, Ilderton:2019kqp}. All higher order gradient contributions to the exponent of the gauge-invariant spectral function from the dressed mass, e.g.\ the next order terms
\begin{align}
 e^2\frac{1}{720}\frac{1}{2(n\cdot p)}\Big[ 3\ddot{\vec\E}(X)\cdot \vec\E(X)+\dot \E^2(X)\Big](n\cdot s)^5\, ,
\end{align}
are suppressed by gauge-invariant gradients. Similarly, one may explicitly verify that under this locally-constant field approximation, the Wilson line can be approximated in such a way that the relation between covariant and invariant fermion two-point functions, Eq.~\refe{kincanwig}, and the relation between $f_\Psi$- and $\tilde f_\Psi$-type fermion distribution functions, Eq.~\refe{fftilde}, indeed holds.

Keeping the LO of the dressed mass, we find the gauge-invariant LO scalar component
\begin{align}
\label{eq:scalarlcfa}
\hat \rho_{\Psi,\textrm{v},\textrm{S}}(X,p)
&= i\,m\,\frac{1}{2p^-}\\
\nonumber
&\hspace{-2.0cm}\cross\int\frac{\dif \varphi}{\omega}\exp\Bigg\{i\Big(p^+-\frac{|\vec p_\perp|^2+ m^2}{2p^-}\Big)\frac{\varphi}{\omega}-i\frac{1}{24}\frac{\xi_0^3}{\chi_0}\frac{\E^2(X)}{\F_0^2}\varphi^3\Bigg\}\\
\nonumber
&\hspace{1cm}+\O(e^0\pdif_p\cdot\pdif_X)\, .
\end{align}
The $\varphi$ integral leads to the Airy function\footnote{The Airy function $\textrm{Ai}(x)\defeq \frac{1}{2\pi}\int\dif u\exp(i(xu +u^3/3))$  solves the differential equation $\textrm{Ai}''(x)-x\textrm{Ai}(x)=0$.
}
\begin{align}
\label{eq:scalarairy}
&\hat\rho_{\Psi,\textrm{v},\textrm{S}}(X,p)\\
\nonumber
&=\frac{im}{m^2\chi^{2/3}(X,p)}\,\textrm{Ai}\Bigg(\hspace{-0.08cm}-\frac{p^2-m^2}{m^2\chi^{2/3}(X,p)}\Bigg)+\O(e^0\pdif_p\cdot\pdif_X)\, ,
%&\textrm{ with }\kappa(X,p)\defeq (n\cdot p)e\E(X)\equiv m^3\chi(X,p)
\end{align}
where the local parameter $\chi$ defined in \refe{fullchidef} amounts to $\chi_0$ with $\F_0$ replaced by $\E(X)$ for plane-wave fields (for the computation of Eq.~\refe{scalarairy} see appendix \ref{app:scalar}).

The LCFA strong-field scattering probabilities \cite{Nikishov:1964zza, Nikishov:1964zzab} that are used as input in the kinetic equations e.g.\ of Ref.~\cite{PhysRevSTAB.14.054401} also feature such Airy functions. As anticipated in Sec.~\ref{par:classical}, these functions may be further reduced to on-shell delta-peaks by virtue of the identity $\lim_{\chi\rightarrow 0}\tfrac{1}{\chi}\textrm{Ai}(x/\chi)=\delta(x)$, consistent with a classical radiation reaction regime.

\subsubsection{\label{par:planedrift}Quantum Vlasov term}

To discuss the quantum Vlasov term for small field-gradients of Eq.~\refe{smalldrift} for plane-wave fields it is useful to switch to lightcone coordinates,
\begin{align}
\label{eq:planeqv}
\hat \rho_{\Psi}^{\,\mu}\D_\mu\tilde f_\Psi= (\hat \rho_{\Psi}^+\D^-+\hat \rho_{\Psi}^-\D^+-\hat \rho_{\Psi,\perp}^{\,i}\D_{\perp}^{\,i})\,\tilde f_\Psi\, .
\end{align} For plane-wave fields, the lightcone components of the Vlasov derivative simplify to 
\begin{align}
\D^-\iv&=\pdif^-=\xpdify{}{X^+}\, ,\\
\D^+\iv&=\xpdify{}{X^-}+e \mathcal{E}_\perp^i(X)\xpdify{}{ p_\perp^{\,i}}\, ,\\
\D_{\perp,\textrm{v}}^i&=\xpdify{}{ X_{\perp,i}}-e\mathcal{E}_\perp^i(X) \xpdify{}{p^+}\, , 
 \end{align}
with $\pdif/\pdif X^-=(\pdif/\pdif X^0-\pdif/\pdif X^3)/2$ and \mbox{$\pdif/\pdif X^+=\pdif/\pdif X^0+\pdif/\pdif X^3$}, and analogous definitions for momentum derivatives. A $p^-$ derivative is absent as it comes with $\F_{\mu\nu}n^\nu$, which vanishes for plane-wave fields.

The all-order in field-gradients plane-wave spectral vector component is
	\begin{align}
	\label{eq:invvec}
	&\hat \rho^{\,\mu}_{\Psi,\textrm{v}}(X,p)\\
	\nonumber
	&=i(2\pi)\int_q\delta(q^2-m^2)\sgn(q^0)\int_s\, e^{i(p-\braket{\pi_q}(X,s))s}\\
	\nonumber
	&\hspace{0.7 cm}\cross\Bigg[\bar \pi^\mu_q(X,s)-n^\mu \frac{1}{8}\frac{(n\cdot s)^2e^2\sbraket{\E}(X,s)}{(n\cdot q)}\Bigg]\, ,
	\end{align}
where $\bar\pi_q$ is the plane-wave momentum in the field \mbox{$\tfrac{1}{2}[\A^\mu(x)+\A^\mu(y)]$} explicitly stated by Eq.~\refe{pibar} in the appendix. The computation of the pre-exponential makes use of the fact that
\begin{align}
\label{eq:diffavg}
\A^\mu(X+\tfrac{s}{2})-\A^\mu(X-\tfrac{s}{2})=(n\cdot s)\braket{\dot \A^\mu}(X,s)
\end{align}
and can also be found in the appendix \ref{app:vector}, alongside the leading order in field-gradients. These are all the ingredients one needs for the quantum Vlasov term for locally constant plane-wave fields. In principle, with Eq.~\refe{invvec} available, the drift term of the all-order field-gradient equation \refe{fermiontransport} is also accessible.

The lightcone components of the vector spectral function \refe{invvec} obey
\begin{align}
\label{eq:scalarvec}
\hat\rho^-_{\Psi}(X,p)&=\frac{n\cdot p}{m}\,\,\hat \rho_{\Psi,\textrm{S}}(X,p)\, ,\\
\hat \rho^{\,i}_{\Psi,\perp}(X,p)&=\frac{p_\perp^{\,i}}{m}\,\,\hat \rho_{\Psi,\textrm{S}}(X,p)+\O(e^0\pdif_p\cdot\pdif_X)\, .
\end{align}
These identities are particularly useful in the classical radiation reaction regime $(\chi\rightarrow 0)$, to recover the on-shell Lorentz force drift term of the classical Vlasov equation \refe{vlasoveq} by expressing the vector component in terms of the scalar component and then using $\lim_{\chi\rightarrow 0}\tfrac{1}{\chi}\textrm{Ai}(x/\chi)=\delta(x)$. In systems with a long-lived separation of scales in terms of ultrarelativistic fermions, it is possible to reduce Eq.~\refe{planeqv} to a Lorentz force term also without sending $\chi\rightarrow 0$ (see Secs.~\ref{par:classical} and \ref{subsec:urlimit}).

\subsubsection{\label{par:planecollfermion}Electron and positron collision terms}

	Inserting the plane-wave collision kernel \refe{planekernel} into the fermion collision term \refe{fermcoll} and making use of identities \refe{fftilde} and \refe{kernelsmallgrad} we may write the fermion collision term for small field-gradients as
	\begin{widetext}
	\begin{align}
	\label{eq:planefermioncoll}
	&C_{\Psi,\textrm{v}}(X,p)=e^2\int\frac{\dif l_1}{(2\pi)}\int\frac{\dif l_2}{(2\pi)}\int_{q,k}(2\pi)^7\delta(k-p+q)\,{\mathcal{Q}}(X,l_1,l_2,p+e\A-l_1n,q+e\A-l_2n)\\
	\nonumber
	&\hspace{0.3cm}\cross\delta(k^2)\sgn(k^0)\,\,\delta((p+e\A)^2-m^2-2l_1(n\cdot p))\sgn(p^0-l_1n^0)\,\,\delta((q+e\A)^2-m^2-2l_2(n\cdot q))\sgn(q^0-l_2n^0)\\
	\nonumber
	&\hspace{0.3cm}\cross \Big\{ \tilde f_\Psi(X,q)f(X,k)[1- \tilde f_\Psi(X,p)]-  [1-\tilde f_\Psi(X,q)][1+f(X,k)]\tilde f_\Psi(X,p)\Big\}\, .
	\end{align}
	\end{widetext}
	
	Here, we have relied on small field-gradients to write the invariant spectral function of the fermion collision term \refe{fermcoll} in terms of the plane-wave delta function of the covariant specrtal function \refe{planeparticle},
	\begin{align}
\label{eq:rhoapprox}
\hat \rho_{\Psi,\textrm{v}}(X,p)=\rho_{\Psi,\textrm{v}}(X,p+e\A)+\O(e^0\pdif_p\cdot\pdif_X)\, .
\end{align}
Defining electron and positron collision terms,
	\begin{align}
	\label{eq:sff}
	\tfrac{1}{2}C_{\Psi,\textrm{v}}^{-}(X,\vec p\,)&\defeq\int_0^{\infty} \frac{\dif p^0}{(2\pi)}\,C_{\Psi,\textrm{v}}(X,p)\, ,\\
	\label{eq:sfaf}
	-\tfrac{1}{2}C_{\Psi,\textrm{v}}^{+}(X,-\vec p\,)&\defeq\int_{-\infty}^0 \frac{\dif p^0}{(2\pi)}\,C_{\Psi,\textrm{v}}(X,p)\, ,
	\end{align}
	the frequency delta functions in Eq.~\refe{planefermioncoll} then allow for explicit computation of the frequency integrals and to recover the structure in terms of the strong-field scattering processes depicted in Fig.~\ref{fig:coll}.
	
	The sign in the definition \refe{sfaf} accounts for a sign that arises when substituting $\vec p\rightarrow -\vec p$. The factors of $\tfrac{1}{2}$ account for the absence of a factor $2$ in the identity for the first-order derivatives of fermions $i(\slashed\pdif_x+\slashed \pdif_y)=i\slashed \pdif_X$ as compared to the identity for the second-order d'Alembertians for photons \refe{photonboltz}.
	
	The appearance of $p+e\A$ in $\mathcal{Q}$ is resolved in the vacuum limit, where scattering kernels become gauge-invariant on their own as discussed in Sec.~\ref{par:vertex}. Since the fermion self-energy is not gauge-invariant, the emergence of gauge-invariant scattering amplitudes with no Wilson lines is far from obvious. However, in the vacuum case, a gauge-invariant fermion loop emerges from the product of the fermion self-energy and the fermion propagator under an additional momentum integral. The ultrarelativistic limit discussed below in Sec.~\ref{subsec:urlimit} then resolves any remaining obstructions to a description in terms of on-shell distribution functions.
	
\subsection{\label{subsec:smallocc}The case of small occupations}

The complexity of collisional kinetic equations is largely due to the nonlinearity in distribution functions of collision terms. However, many physical situations allow for an assumption of small distribution functions, i.e.
\begin{align}
\label{eq:smallocc}
f_\Psi^\pm(X,l,\vec p\,)\ll 1 \,\,\,\textrm{ and }\,\,\, f(X,\vec k) \ll 1\, ,
\end{align} implicit for example in the kinetic equations of Refs. \cite{PhysRevSTAB.14.054401, gaisser_engel_resconi_2016full}. For such settings close to vacuum, one may drop $2\rightarrow 1$ and $3\rightarrow 0$ processes entirely, since they contain no linear terms and are therefore suppressed\footnote{See also `phase space suppression' arguments in terms of integral measures and kinematic restrictions from the field, e.g.\ in Ref.~\cite{Gonoskov:2014mda}.} by
\begin{align}
f_\Psi^\pm f_\Psi^\pm f,\,\,\,f_\Psi^\pm f_\Psi^\pm,\,\,\, f_\Psi^\pm f\, \ll f_\Psi^\pm, \,\,\,f\, .
\end{align}
Thereby, strong-field systems with small occupations single out a direction in time -- the direction of energy transport from the macroscopic field to the particle sector by $1\rightarrow 2$ and $0\rightarrow 3$ processes -- even if the corresponding scattering matrix elements and the fundamental equations of motion are symmetric under time reversal. Similarly, one may simplify all Bose-enhancement or Pauli-blocking terms in $1\rightarrow 2$ and $0\rightarrow 3$ processes via $1+f(X,\vec k)\approx 1$ and $1-f_\Psi^{\pm}(X,l,\vec p\, )\approx 1$. In this way, small distribution functions lead to a linearization of collision terms. In contrast to a linearization around equilibrium \cite{Blaizot:2001nr} which keeps thermal distributions as in Eq.~\refe{eqrate}, collision terms linearized by small occupations violate detailed balance and are thereby no longer able to describe the approach to thermal distribution functions. Charge conservation [Eq.~\refe{chargecoll}] is still exact.

The linearized near-vacuum plane-wave photon collision  term \refe{crossphoton} then reads [after a substition to recover covariant energy conservation as described around Eq.~\refe{covcons}],
\begin{widetext}
\begin{align}
\label{eq:smalloccphoton}
C\iv(X,\vec k)&=e^2\frac{1}{2|\vec k|}\int\frac{\dif l_1}{(2\pi)}\int \frac{\dif^3p}{(2\pi)^3}\frac{1}{2\varepsilon(\vec p\,)}\int\frac{\dif l_2}{(2\pi)}\int \frac{\dif^3q}{(2\pi)^3}\frac{1}{2\varepsilon(\vec q\,)}(2\pi)^4\\
\nonumber
&\hspace{-1cm}\cross\Big[\delta(k-p+q-(l_1-l_2)n)\,\mathcal{Q}^{e^+\rightarrow e^+\gamma}(X,l_2,l_1,\vec q+l_2\vec n,\vec p+l_1\vec n\,)\,f_\Psi^+(X,l_1,\vec p+l_1\vec n\,)\\
\nonumber
&\hspace{-1cm}+\delta(k-p+q-(l_1-l_2)n)\,\mathcal{Q}^{e^-\rightarrow e^-\gamma}(X,l_1,l_2,\vec p+l_1\vec n,\vec q+l_2\vec n\,)\,f_\Psi^-(X,l_1,\vec p+l_1\vec n\,)\\
\nonumber
&\hspace{-1cm}-\delta(k-p-q-(l_1+l_2)n)\,\mathcal{Q}^{e^+e^-\rightarrow\gamma}(X,l_1,l_2,\vec p+l_1\vec n,\vec q+l_2\vec n\,)\,f(X,\vec k)\Big]\, .
\end{align}
\end{widetext}

The electron collision term \refe{sff} under the same approximation reduces to the three $1\rightarrow 2$ scattering processes, $e^-\rightarrow e^-\gamma$ with ingoing momentum $\vec p\,$, $e^-\rightarrow e^-\gamma$ with outgoing momentum $\vec p\,$, and $\gamma\rightarrow e^-e^+$. Analogously, the linearized positron collision term \refe{sfaf} contains the processes $e^+\rightarrow e^+\gamma$ and $\gamma\rightarrow e^+e^-$. In all near-vacuum collision terms, each process is weighted linearly by the distribution function of the ingoing particle as in the equations of Ref.~\cite{PhysRevSTAB.14.054401}.

We emphasize that for general macroscopic fields, these near-vacuum collision terms would all additionally contain $0\rightarrow e^+e^-\gamma$ source terms with no distribution function, contributing to vacuum pair production at 2-loop $\O(e^2)$ precision.

\subsection{\label{subsec:urlimit}The case of ultrarelativistic fermions\\ \& on-shell strong-field descriptions}

Many of the approximations discussed in previous sections are tied together in an ultrarelativistic setting: strong macroscopic fields accelerate fermions to ultrarelativistic energies within small regions of space. Once accelerated, any macroscopic field appears like a plane-wave field in the Lorentz rest frame of an ultrarelativistic fermion \cite{DiPiazza:2013vra}. Therefore, plane-wave fields represent generic qualities of strong fields in an ultrarelativistic setting. Furthermore, ultrarelativistic fermions facilitate chiral symmetry, which in turn leads to a reduction of tensor structures, which is assumed by our definition of the fermion distribution function as discussed in Sec.~\ref{sec:3a}. Additionally, large fermion momenta can facilitate that field-gradients are numerically separated from propagator-gradients as we have seen in Eq.~\refe{gradrel}. Moreover, ultrarelativistic fermions have a small de Broglie wavelength facilitating classical propagation in-between quantum processes like the emission of photons. From an analysis of the classical propagation of fermions one then finds that ultrarelativistic fermions emit radiation along their instantaneous velocity, within a cone of angular aperture  $\sim m/\varepsilon(\vec p\,)$ \cite{LANDAU1975171, Baier:1998vh}. If the particle is ultrarelativistic and its energy is the largest scale in the system, its motion has a pronounced directionality. In strong-field vacuum, i.e.\ for vanishing occupations, and if the transverse momenta are much larger than $m$, one can then show that only small patches of their trajectory contribute to scattering amplitudes \cite{Baier:1998vh, PhysRevD.93.085028} (which is the assumption of the LCFA).

There are several notions of ultrarelativistic limits for fermions in literature. They range from assumptions on kinematic restrictions \cite{PhysRevSTAB.14.054401} to expansions in terms of $\vec p_\perp/(n\cdot p)$ \cite{DiPiazza:2013vra} or $1/\gamma= m/\varepsilon(\vec p\,)$ \cite{Baier:1998vh}. In the language of the present paper, an ultrarelativistic system is defined by a fermion distribution function that is peaked at an ultrarelativistic scale $p^*$. Such a distribution function then gives meaning to single particle concepts such as the de-Broglie wavelength $\hbar/p^*$ also in many-body systems. In particular, the structure of the fermion spectral function of such a system only matters for characteristic momenta as it always appears in a product with the fermion distribution function which approximately vanishes away from the characteristic scale.

We now assume that the characteristic scale $p^*$ of fermion distribution functions is well separated from the characteristic scale $l^*$ of the strong-field spectral kernel \refe{K}, i.e.
\begin{align}
|p_z^*|\gg |l^*|\, .
\end{align}
In such a situation, the strong-field spectral function only contributes with on-shell values, because
\begin{align}
\label{eq:urapprox}
\varepsilon_{l^*}(\vec p\,^*)\simeq \sqrt{|\vec p\,^*|^2+m^2}
\end{align}
becomes the on-shell dispersion, independent of $l^*$. As anticipated by our discussion in Sec.~\ref{par:classical}, this implies that ultrarelativistic fermions may indeed be described by on-shell particles whose energy $\varepsilon(\vec p\,^*)$ then satisfies
\begin{align}
\label{eq:lenergy}
|l^*|/\varepsilon(\vec p\,^*)\ll 1\, .
\end{align}
The ultrarelativistic limit \refe{urapprox} leaves the strong-field properties of the spectral function intact, simplifies kinematic restrictions, and favors a description in terms of free distribution functions.

However, in general, there is no mechanism that dynamically controls this approximation, i.e. it may become invalid during the evolution of the system even if it is valid at initial time. An important effect that explicitly breaks the validity of an ultrarelativistic approximation is vacuum pair production, which generates off-shell contributions to $\tilde f_\Psi(X,p)$ at zero frequency according to Eq.~\refe{pairlo}.

Indeed, a kinetic description in terms of only on-shell distribution functions is suggested in Ref.~\cite{PhysRevSTAB.14.054401} for ultrarelativistic fermions in strong (but subcritical $\E\ll \E_{\textrm{c}}$) fields with small gradients. Our off-shell transport description of Sec.~\ref{sec:4} reduces to that description under the following approximations: a) an approximation of field-gradients (see Sec.~\ref{subsec:5a}); b) an approximation of collision terms for small occupations to neglect medium effects (see Secs.~\ref{par:vertex} and \ref{subsec:smallocc}); c) an assumption of ultrarelativistic simply peaked fermion distribution functions and subcritical fields to replace the quantum Vlasov term with the Lorentz force term of the classical Vlasov equation \refe{vlasov1} (see also Sec.~\ref{par:classical}) and to justify the on-shell limit of collision terms,
\begin{align}
\label{eq:energydist}
&\int\dif l_1\int \dif l_2 \,g(l_1,l_2)\mathcal{Q}(l_1,l_2)\\
\nonumber
&\hspace{1cm}\simeq g(0,0)\int\dif l_1\int \dif l_2\,\mathcal{Q}(l_1,l_2)\, ,
\end{align}
where $g$ indicates the gain-minus-loss terms and $\mathcal{Q}$ includes the delta-functions such that energy conservation is treated exactly. Together with the Lorentz force term, this closes the strong-field description in terms of the traditional on-shell particle distribution functions \refe{osfdef} and \refe{osafdef} which emerge from plane-wave distribution functions via
\begin{align}
\label{eq:fur}
f_\Psi^\pm(X,l,\vec p\,)&\xrightarrow{l\rightarrow 0}f_\Psi^\pm(X,\vec p\,)\, .
\end{align}
With all these approximations combined, also subtleties regarding gauge-invariance both of the scattering kernels and the distribution functions are resolved: The scattering kernels become gauge-invariant objects in the vacuum limit \refe{localvertex} and a distinction between the $\tilde f_\Psi$- and $f_\Psi$-type fermion distribution functions is not important after the ultrarelativistic limit \refe{fur} for distribution functions that are always only occupied in terms of a few on-shell particle modes for which gauge-invariance is assured.

Dropping all these assumptions is possible by employing the gauge-invariant off-shell equations discussed in Sec.~\ref{sec:4}. Starting from this off-shell description, it would be interesting to investigate whether collisional or inhomogeneous contributions to the particle yield \refe{pairgeneral} can invalidate an on-shell description also for subcritical fields on some significant time scale on the way to equilibrium.

\section{\label{sec:6}Conclusions \& outlook}

Our work demonstrates how to systematically derive transport and kinetic equations including collisions for general supercritical fields. The equations to order $\O(e^2)$ include local scattering kernels for strong fields that can also be inhomogeneous. This is achieved by off-shell transport equations that include non-local relative times and all field-gradients in the fermion spectral function, while retaining the gain-minus-loss structure of traditional kinetic equations. To investigate our equations analytically and to make contact to limiting cases in the literature, we have also considered plane-wave fields.

We have shown that the inclusion of fermion spectral dynamics is essential to describe collisions and fermion drifting in the presence of general strong fields. Existing derivations of strong-field Wigner descriptions in the literature have neglected spectral dynamics by limiting themselves to the collisionless regime, in which equations for spectral functions decouple from transport equations. In general, however, the macroscopic field enters the collision kernel \refe{kernel} via the fermion spectral function \refe{volkoveq}. This resums infinitely high perturbative orders of the coupling that all become relevant for sufficiently large macroscopic fields. The macroscopic field itself is governed by a Maxwell equation in the presence of a fermion current involving the quantum corrections. The general form of this Maxwell equation turns out to be valid to arbitrary loop and gradient order in our framework. Our approach paves the way for investigations of the thermalization process starting from strong field initial conditions, which requires to go beyond collisionless approximations.

We have pointed out a connection between asymptotic pair production and spectral dynamics. While 1-loop results such as the Schwinger pair production rate \refe{loschwinger} assume the macroscopic field to be external and constant in time, our 1-loop result \refe{pairlo} is fully dynamical and generalizable to the expression \refe{pairgeneral}, which in principle includes collisions to 2-loop order and all orders in field-gradients. Our description in terms of distribution functions does not rely on asymptotic expressions, such as total particle numbers or total probabilities in order to compute time-dependent observables such as the strong-field photon decay rate \refe{eqrate}.

We solved the LO equation for the fermion spectral function for the special case of an external plane-wave macroscopic field, $\A^\mu(x)\simeq \A\iv^\mu(n\cdot x)$ with a null vector $n^2=0$. This reduces the transport description to only two equations for the off-shell fermion and the on-shell photon distribution function. The plane-wave spectral function is the antisymmetric part of the well-known time-ordered Volkov fermion propagator. By employing only its antisymmetric part in the $\O(e^2)$ transport equations, we self-consistently resum quantum fluctuations to 2-loop order. Thereby, a solution of our equation goes beyond the statistical component of the \mbox{1-loop} Volkov propagator that implicitly assumes vanishing distribution functions.

Employing the all-order field-gradient plane-wave spectral function in the collision kernel reproduces expressions which are similar to the Furry picture, but have the advantage of being automatically local in the kinetic position variable $X$ while containing contributions from inhomogeneous fields not limited to the vicinity of $X$. In particular, we have demonstrated that plane-wave scattering kernels emerge with a space-time structure that is more general than the one of local scattering amplitudes that are known from laser applications. The more general scattering kernels reduce to known expressions only if relative times are restricted to certain values. We have recognized this condition as the implicit assumption that the system is in plane-wave vacuum, i.e.\ that fermion plane-wave distribution functions are negligible or have only single occupied modes. This means that medium effects are missed if a strong-field scattering kernel is obtained from Feynman rules for the Furry picture S-matrix. For negligible distribution functions, known gauge-invariant global scattering amplitudes emerge by integrating over all $X$. To employ external-field vacuum amplitudes in an isolated dynamical setting is typically inconsistent because it includes times outside the range of validity of external field and vacuum approximations as non-negligible distribution functions develop dynamically and backreact on the field. Nevertheless, these emergent amplitudes allowed us to highlight connections to Ward identities, which remove the gauge-fixing dependence of the 2PI formulation of QED in the corresponding limit. 

Furthermore, the plane-wave fermion spectral function allowed us to use the energy exchange with the macroscopic field as a parameter $l$ to label strong-field degrees of freedom with energy $\varepsilon_l(\vec p\,)$, which enable a continuous connection to the free particle degrees of freedom of an on-shell description: When this spectral function is multiplied by a fermion distribution function that is peaked on an ultrarelativistic scale $p^*$ that is well separated from the characteristic value of $l$, its dispersion relation becomes independent of $l$ and reduces to that of free fermions, $\varepsilon_l(\vec p\,^*)\simeq \sqrt{|\vec p\,^*|^2+m^2}$. This facilitates an on-shell description despite the presence of strong fields. Thereby, strong-field systems in which such a clearly separated scale $p^*$ exists for long times may be accurately captured by on-shell descriptions that combine collisions with classical Lorentz force drifting. Since any field appears as a plane wave in the rest frame of a single ultrarelativistic fermion, we expect that most statements that we arrived at under the assumption of an external plane-wave field also hold for more general fields, as long as fermion distribution functions are dominated by a few ultrarelativistic particle-modes.

In contrast, in isolated systems with supercritical fields, initial characteristic scales are dynamically affected by pair production (which occurs off-shell and is largest at zero frequency) and by the transport of fermion occupations towards an equilibrium distribution (which has its maximum at low energies and is not sharply peaked). For such isolated systems, we argued that an initial ultrarelativistic separation of scales is not long-lived, such that an on-shell Lorentz force description introduces an error larger than our desired accuracy of $\O(e^2)$. In the absence of a long-lived separation of scales, one needs instead a description that remains valid over a wide range of energies to describe the evolution of off-shell contributions induced by vacuum pair production towards the on-shell regime of the asymptotic future. The gauge-invariant fermion transport equation Eq.~\refe{fermiontransport} with its all-gradient off-shell drift and collision term constitutes such a description by coupling to the fermion spectral equation, the photon transport equation and the Maxwell equation summarized in Fig.~\ref{fig:1}.

\textit{Outlook.} The leading order equations may give insight into the largely unexplored late-time behaviour of isolated QED systems with finite net charge. If such a system equilibrates, its late-time state can not be the traditional homogeneous thermal equilibrium, because the Gauss constraint for finite net charge prevents the initial field both from decaying completely and from becoming fully homogeneous. The possible approach to such a charged time-translation invariant state may be completely described by our equations, if the equilibrium field induced by the net charge turns out to be sufficiently large.

Such a numerical computation, in particular of the self-consistent strong-field fermion spectral function, will also allow for a more detailed study of the collision kernel and the spectral peak structure. This would, e.g., enable an analysis of spectral widths and to establish under what circumstances they are small. To obtain insight into specific controlled experimental settings, one may employ other external fields in such a computation, as we have done for the plane-wave spectral function with laser fields in mind. Possible other choices of external fields include non-crossed constant electric fields, homogeneous magnetic fields and Coulomb fields.

In the future, dropping our assumption of reduced tensor structures \refe{sKBAf} with the help of Ref.~\cite{Vasak:1987um} could clarify the significance of chiral dynamics \cite{Fukushima:2008xe, Mueller:2016ven, Mace:2016shq, Mace:2019cqo} and spin transport \cite{Yang:2020hri}, and extend chiral kinetic theory \cite{Stephanov:2012ki, Gao:2017gfq, Hidaka:2016yjf, Mueller:2017arw, Huang:2018wdl} to the collisional regime. To access the transport dynamics of the axial current $j_5^\mu(X)=-e\textrm{tr}\{\gamma^5 \gamma^\mu F_\Psi(X,X) \}$, an interacting spectral function that has a non-vanishing axial component such as the strong-field spectral function employed in this paper is required (see the expression for plane-wave fields in appendix \ref{app:axial}). A leading-order collisional description including all tensor structures in the presence of a macroscopic field is now in reach and would open up diverse applications on chiral dynamics reaching from astrophysics \cite{Brandenburg:2017rcb, Masada:2018swb} to semiconductors \cite{Li:2016vlc, Haug2008}.

\textit{Acknowledgements.} This work is supported by the Deutsche Forschungsgemeinschaft (DFG, German Research Foundation) – Project-ID 273811115 – SFB 1225.

\appendix
\tocless\section{\label{app:propdef}Identities for QED two-point functions}
The following hermiticity properties of photon and fermion two-point functions are used in the main text:\\
The photon two-point functions have the properties
\begin{align}
\label{eq:photonsym}
[\rho^{\mu\nu}(x,y)]^*&=\rho^{\mu\nu}(x,y)\, ,\\
\rho^{\mu\nu}(x,y)&=-\rho^{\nu\mu}(y,x)\, ,\\
[F^{\mu\nu}(x,y)]^*&=F^{\mu\nu}(x,y)\, ,\\
F^{\mu\nu}(x,y)&=F^{\nu\mu}(y,x)\, ,
\end{align}
i.e.\ $\rho^{\mu\nu}(x,y)$ is real and antisymmetric and $F^{\mu\nu}(x,y)$ is real and symmetric. The definitions for the advanced and retarded propagators used in Sec.~\ref{sec:3} are
\begin{align}
\label{eq:photonadvret}
D_\textrm{R}^{\mu\nu}(x,y)&\defeq \theta(x^0-y^0)\rho^{\mu\nu}(x,y)\, ,\\ D_\textrm{A}^{\mu\nu}(x,y)&\defeq -\theta(y^0-x^0)\rho^{\mu\nu}(x,y)\, ,\\
\label{eq:fermadvret}
\Delta_\textrm{R}(x,y)&\defeq \theta(x^0-y^0)\rho_\Psi(x,y)\, ,\\
\Delta_\textrm{A}(x,y)&\defeq -\theta(y^0-x^0)\rho_\Psi(x,y)\, ,
\end{align}
and the same for the self-energies. These functions obey
\begin{align}
D_{\textrm{A}}^{\mu\nu}(x,y)&=D_{\textrm{R}}^{\nu\mu}(y,x)\, ,\\
\Delta_{\textrm{A}}(x,y)&=\gamma^0\Delta_{\textrm{R}}^\dagger(y,x)\gamma^0\, .
\end{align}
They are related to the spectral functions via
\begin{align}
\rho^{\mu\nu}(x,y)&=D_{\textrm{R}}^{\mu\nu}(x,y)-D_{\textrm{A}}^{\mu\nu}(x,y)\, ,\\
\rho_\Psi(x,y)&=\Delta_{\textrm{R}}(x,y)-\Delta_{\textrm{A}}(x,y)\, .
\end{align}
The definitions for the Wightman functions employed in Sec.~\ref{sec:photontransport} are
\begin{align}
D^{+-}(x,y)^{\mu\nu}&\defeq F^{\mu\nu}(x,y)-\tfrac{i}{2}\rho^{\mu\nu}(x,y)\, ,\\
D^{-+}(x,y)^{\mu\nu}&\defeq F^{\mu\nu}(x,y)+\tfrac{i}{2}\rho^{\mu\nu}(x,y)\, ,\\
\Delta^{+-}(x,y)&\defeq F_\Psi(x,y)-\tfrac{i}{2}\rho_\Psi(x,y)\, ,\\
\Delta^{-+}(x,y)&\defeq F_\Psi(x,y)+\tfrac{i}{2}\rho_\Psi(x,y)\, ,
\end{align}
and the same for the self-energies. These Wightman functions are sometimes denoted as $G^{-+}= G^<$ and $G^{+-}= G^>$ in literature. The superscripts indicate on which part of the Keldysh contour their arguments lie and can be obtained from the general functions \refe{propdecomp1}, \refe{propdecomp2} and \refe{sedecomp}, \refe{sedecompf} by explicit use of the sign functions $\sgn_\C$. Similarly to the retarded and advanced functions they obey
\begin{align}
\rho_{\mu\nu}(x,y)&=i\Big(D^{+-}_{\mu\nu}(x,y)-D^{-+}_{\mu\nu}(x,y)\Big)\, ,\\
\rho_\Psi(x,y)&=i\Big(\Delta^{+-}(x,y)-\Delta^{-+}(x,y)\Big)\, .
\end{align}
In Wigner space, one may alternatively exploit the Wigner transform of the Heaviside function,
\begin{align}
\theta(x^0-y^0)&=\lim_{\varepsilon\rightarrow 0}\int\frac{\dif \omega}{(2\pi)}e^{-i\omega s^0}\frac{i}{\omega+i\varepsilon}\, ,
\end{align}
to obtain a Källén-Lehmann representation
\begin{align}
D^{\mu\nu}_{\textrm{R}}(X,k)&=\lim_{\varepsilon \rightarrow 0}\int\frac{\dif \omega}{(2\pi)}\frac{i\rho^{\mu\nu}(X,\omega,\vec k\,)}{k^0-\omega+i\varepsilon}\, ,\\
\Delta_{\textrm{R}}(X,p)&=\lim_{\varepsilon \rightarrow 0}\int\frac{\dif \omega}{(2\pi)}\frac{i\rho_\Psi(X,\omega,\vec p\,)}{p^0-\omega+i\varepsilon}
\end{align}
and similarly for the advanced functions with
\begin{align}
\theta(y^0-x^0)&=\lim_{\varepsilon\rightarrow 0}\int\frac{\dif \omega}{(2\pi)}e^{-i\omega s^0}\frac{-i}{\omega-i\varepsilon}\, .
\end{align}
The self-energies obey  completely analogous identities. We stress that any singularity associated to the $\varepsilon$-prescription does not arise in an exact (early-time) description that employs Wigner transforms \refe{boundwig} instead of the late-time Wigner transforms \refe{unboundwig} (see also Refs. \cite{Bedaque:1994di,Greiner:1998ri}).\\
\indent The photon Wigner functions have the properties
\begin{align}
F^{\mu\nu}(X,k)&=F^{\nu\mu}(X,-k)\, ,\\
\rho^{\mu\nu}(X,k)&=-\rho^{\nu\mu}(X,-k)\, ,\\
D_{\textrm{A}}^{\mu\nu}(X,k)&=D_{\textrm{R}}^{\nu\mu}(X,-k)\, .
\end{align}
Similarly, fermion Wigner functions obey
\begin{align}
F_{\Psi}(X,p)&=\gamma^0F_\Psi^\dagger(X,p)\gamma^0\,,\\
\rho_{\Psi}(X,p)&=-\gamma^0\rho_\Psi^\dagger(X,p)\gamma^0\, ,\\
\Delta_{\textrm{A}}(X,p)&=\gamma^0\Delta_{\textrm{R}}^\dagger(X,p)\gamma^0\, .
\end{align}
Given all this, it should be remembered that there are only two independent two-point functions per field species (see also our comment at the end of Sec.~\ref{sec:latetime}).\\
\indent 
	The LO $\O(e^2)$ 2PI loop expansion of the Wightman self-energies is
	\begin{align}
	\Sigma_{\mu\nu}^{+-}(x,y)&=e^2\,\textrm{tr}\, \left\{\gamma_\mu\Delta^{+-}(x,y)\gamma_\nu\Delta^{-+}(y,x)\right\}\, ,\\
	\Sigma_{\mu\nu}^{-+}(x,y)&=e^2\,\textrm{tr}\, \left\{\gamma_\mu\Delta^{-+}(x,y)\gamma_\nu\Delta^{+-}(y,x)\right\}\, ,\\
	\Sigma_\Psi^{+-}(x,y)&=-e^2\gamma^\mu \Delta^{+-}(x,y)\gamma^\nu D_{\mu\nu}^{+-}(x,y)\, ,\\
	\Sigma_\Psi^{-+}(x,y)&=-e^2\gamma^\mu \Delta^{-+}(x,y)\gamma^\nu D_{\mu\nu}^{-+}(x,y)\, .
	\end{align}
	
\tocless\section{\label{app:aeom}2PI field equation of motion}

	As discussed around Eq.~\refe{fieldvertex}, the only objects in $\Gamma[\mathcal{A},D,\Delta]$ that depend on $\mathcal{A}$ are the classical action $S[\mathcal{A}]$ and its second derivative $i\Delta_0^{-1}[\mathcal{A}]$ such that
\begin{align}
0\anoteq{!}\frac{\delta \Gamma[\mathcal{A},D,\Delta]}{\delta \mathcal{A}^\mu}=\frac{\delta}{\delta \mathcal{A}^\mu}\Big(S[\mathcal{A}]-i\Tr\Delta_0^{-1}\Delta\Big)
\end{align}
These two terms are the variation of the classical action (where boundary terms are dropped as initial conditions)
\begin{align}
\frac{\delta}{\delta \mathcal{A}^\mu(x)}S[\mathcal{A}]
=\left[ \eta_{\mu\sigma} \square_x-(1-\tfrac{1}{\xi})\pdif_\mu\pdif_\sigma \right]\A^{\sigma}(x)\, ,
\end{align}
and the 1-loop term
\begin{widetext}
\begin{align}
-i\Tr\Delta_0^{-1}[\mathcal{A}]\Delta
&=-i\int_{\mathcal{C},yz}\textrm{tr}\,\left\{(i\slashed \pdif_y-e\slashed{\mathcal{A}}(y)-m)\delta_{\mathcal{C}}(y,z)\Delta(z,y)\right\}\\
\nonumber
&=-\int_{\mathcal{C},yz}\textrm{tr}\,\left\{(i\slashed \pdif_y-e\slashed{\mathcal{A}}(y)-m)\delta_{\mathcal{C}}(y,z)\left(\theta_\C(z,y)\Delta^{+-}(z,y)+\theta_\C(y,z)\Delta^{-+}(z,y)\right)\right\}\\
\nonumber
&=-\int_{t_0}^\infty \dif^4 y\,\textrm{tr}\,\left\{(i\slashed \pdif_y-e\slashed{\mathcal{A}}(y)-m)F_\Psi(y,y)\right\}
\end{align}
where we used that $\theta(x-y)+\theta(y-x)=1$ and that on the backward branch of the contour, $x,y\in\C^-$, $\delta_\C(x,y)=-\tfrac{1}{2}\delta(x-y)$, which results in in the fact that only the statistical function $F_\Psi(x,y)=\tfrac{1}{2}(\Delta^{+-}(x,y)+\Delta^{-+}(x,y))$ contributes to this term (see also Ref.~\cite{PhysRevD.90.025016}). The variation of this term then gives the Maxwell current 
\begin{align}
\frac{\delta}{\delta \mathcal{A}^\mu(x)}\Big( -i\Tr\Delta_0^{-1}[\mathcal{A}]\Delta \Big)
=e\,\textrm{tr}\,\left\{\gamma_\mu F_\Psi(x,x)\right\}\, .
\end{align}

\tocless\section{\label{app:specgrad}On the gradient expansion of spectral equations of  motion}
To NLO in propagator-gradients, the RHS of the tensorial equation for the fermion spectral function consists of commutators in Dirac space and Poisson-brackets,
\begin{align}
\label{eq:fermrhocom}
&\int\dif^4(x-y)\,e^{ip(x-y)}\Big[(\rho\textrm{RHS})_\Psi(x,y)+\gamma^0(\rho\textrm{RHS})_\Psi^\dagger(y,x)\gamma^0\Big]\\
\nonumber
&=\tfrac{1}{2}[\Sigma_{\Psi}^{(\Omega)},\rho_\Psi](X,p)+\tfrac{1}{2}[\Sigma^{(\rho)}_\Psi,\Omega_\Psi](X,p)+\tfrac{i}{2}[\Sigma_\Psi^{(\Omega)},\rho_\Psi]_{\textrm{PB}}(X,p)+\tfrac{i}{2}[\Sigma_\Psi^{(\rho)},\Omega_\Psi]_{\textrm{PB}}(X,p)+\O\big(e^2(\pdif_p\cdot\pdif_X)^2G\big)\, .
\end{align}
Here we introduced the hermitian, i.e.\ in the sense of $\gamma^0\Omega_\Psi^\dagger(X,p)\gamma^0=\Omega_\Psi(X,p)$, parts of retarded components
\begin{align}
\label{eq:omegadef}
\Sigma_\Psi^{(\Omega)}(X,p)&\defeq \Sigma_{\Psi,\textrm{R}}(X,p)+\Sigma_{\Psi,\textrm{A}}(X,p)\\
\label{eq:omegadef2}
\Omega_\Psi(X,p)&\defeq \Delta_{\textrm{R}}(X,p)+\Delta_{\textrm{A}}(X,p)\, .
\end{align}
An analogous expression is true for the RHS of the photon spectral equation of motion, where commutators in Dirac space are replaced by $\com{A}{B}^{\mu\nu}\defeq A^{\mu\sigma}{B_{\sigma}}^\nu-B^{\mu\sigma}{A_\sigma}^\nu$ and likewise for Poisson-brackets.

Kinetic equations describe physics for which occupations evolve decoupled from the spectrum of the theory. Eq.~\refe{fermrhocom} shows that this happens to all orders of the coupling for small propagator-gradients and sufficiently simple tensor structures, i.e. vanishing commutators: The Poisson brackets are NLO in propagator-gradients, $\O(e^2\pdif_k\cdot\pdif_XG)$, and the Dirac commutators can vanish for simple tensor structures such as those that reduce dynamics to a single distribution function as discussed in Sec.~\ref{sec:redten}. The interaction terms in the \textit{trace} of Eq.~\refe{omegadef} are always suppressed by propagator-gradients, such that the traced equation coincides with the free equation of motion. In fact, the RHS of the traced Eq.~\refe{rhotrace2} is strictly $\O(e^2(\pdif_p\cdot\pdif_X)G)$ [and not just $\O(e^2)$]. The same thing does not happen in the traced equations of motion for statistical functions, e.g.\ Eq.~\refe{photonftrace}, whose leading order in propagator-gradients does not vanish, but provides the collision terms.

\tocless\section{\label{app:gaugegrad}Covariant vs.\ invariant expansion in field-gradients}

Simply expanding the gauge-non-invariant one-point function in its gradients via
\begin{align}
\A^\mu(X+\tfrac{s}{2})-\A^\mu(X-\tfrac{s}{2})&=\sumzeroxtoy{n}{\infty}\frac{1}{(2n+1)!}\frac{1}{2^{2n}}(s\cdot \pdif_X)^{2n+1}\A^\mu(X)\, ,\\
\A^\mu(X+\tfrac{s}{2})+\A^\mu(X-\tfrac{s}{2})&=\sumzeroxtoy{n}{\infty}\frac{1}{(2n)!}\frac{1}{2^{2n-1}}(s\cdot \pdif_X)^{2n}\A^\mu(X)
\end{align}
to NLO gives the following left hand side of the covariant fermion transport equation,
\begin{align}
\label{eq:fermflhs}
&\int\dif^4(x-y)\,e^{ip(x-y)}\,\tfrac{1}{4}\textrm{tr}\Big[(F\textrm{LHS})_\Psi(x,y)-\gamma^0(F\textrm{LHS})_\Psi^\dagger(y,x)\gamma^0\Big]\\
\nonumber
&\hspace{0.5cm}= i \left [\xpdify{}{X^\mu}+e\,\Big(\xpdify{}{X_\sigma}\A_\mu(X)\Big)\xpdify{}{p^\sigma}\right]F^\mu_{\Psi}(X,p)+\O\big((e^0\pdif_p\cdot\pdif_X)^{3}\big)\, .
\end{align}
We now change to the gauge-invariant statistical propagator by introducing Wilson lines. For small fields with small gradients, we may expand the straight Wilson exponent via
\begin{align}
\int_y^x\dif z^\mu \A_\mu(z)=s^\mu \sumzeroxtoy{n}{\infty}\frac{1}{(2n+1)!}\ofrac{2^{2n}}(s\cdot\pdif_X)^{2n}\A_\mu(X)\, .
\end{align}
The leading order of straight Wilson lines, $\W(x,y)=e^{-ies\cdot\A(X)}+\O\big((e^0s\cdot\pdif_X)^{2}\big)$ (first order vanishes), is $\O(e^0)$ for strong fields and produces the missing term 
\begin{align}
\label{eq:xid}
&\W(X-\tfrac{s}{2},X+\tfrac{s}{2})\xpdify{}{X_\mu}\W(X+\tfrac{s}{2},X-\tfrac{s}{2})=-ie s_\sigma \xpdify{}{X_\mu}A^\sigma(X)+\O\big((e^0s\cdot\pdif_X)^{2}\big)\, ,
\end{align}
that is necessary to identify the gauge-invariant field strength tensor. Changing the prescription \refe{fermflhs} how to derive fermion kinetic equations to include a Wilson line, we recover the Vlasov term via
\begin{align}
\label{eq:vlasovterm}
&\int\dif^4(x-y)\,e^{ip(x-y)}\W(y,x)\,\tfrac{1}{4}\textrm{tr}\Big[(F\textrm{LHS})_\Psi(x,y)-\gamma^0(F\textrm{LHS})_\Psi^\dagger(y,x)\gamma^0\Big]\\
\nonumber
&\hspace{1cm}= i\left[\xpdify{}{X^\mu}-e\F_{\mu\sigma}(X)\xpdify{}{p_\sigma}\right]\hat F^\mu_{\Psi}(X,p)+\O\big((e^0\pdif_p\cdot\pdif_X)^{3}\big)\, .
\end{align}

\tocless\section{\label{app:volkov}Symmetric and antisymmetric parts of the Volkov propagator}

By virtue of Eq.~\refe{ritid}, $(i\slashed \pdif_x-e\slashed \A_\textrm{v}(x)-m)\E_q(x)(\slashed q+m)\bar \E_q(y)= \E_q(x)(q^2-m^2)\bar \E_q(y)$, such that indeed the plane-wave spectral function solves
\begin{align}
(i\slashed \pdif_x-e\slashed \A_\textrm{v}(x)-m)\rho_{\Psi,\textrm{v}}(x,y)
&=i(2\pi)\int_q\delta(q^2-m^2)\sgn(q^0)(i\slashed \pdif_x-e\slashed \A_\textrm{v}(x)-m)\E_q(x)(\slashed q+m)\bar \E_q(y)=0\, .
\end{align}
The Volkov spectral function is antisymmetric because
\begin{align}
\gamma^0\rho_{\Psi,\textrm{v}}(x,y)^\dagger\gamma^0
=-i(2\pi)\int_q \,\delta(q^2-m^2)\sgn(q^0) \gamma^0\bar \E_q^\dagger(y)(\slashed q+m)^\dagger\E_q^\dagger(x)\gamma^0
=-\rho_{\Psi,\textrm{v}}(y,x)\, .
\end{align}
We may put this spectral function into the context of canonical quantization in the Furry picture \cite{PhysRev.81.115}, which is achieved for plane-wave fields by means of the Volkov states \cite{PhysRevD.97.056028}
\begin{align}
\Psi_\textrm{v}(x)&=\sum_s\int\frac{\dif^3 p}{(2\pi)^3}\frac{1}{\sqrt{2\varepsilon(\vec p\,)}}\left[ c_s(\vec p\,)U_{p,s}(x)+ d_s^\dagger(\vec p\,)V_{p,s}(x)\right]\, ,\\
\bar \Psi_\textrm{v}(x)&=\sum_s\int\frac{\dif^3 p}{(2\pi)^3}\frac{1}{\sqrt{2\varepsilon(\vec p\,)}}\Big[ d_s(\vec p\,) \bar V_{p,s}(x) + c_s^\dagger(\vec p\,)\bar U_{p,s}(x) \Big]\, ,
\end{align}
via the canonical commutation relations
\begin{align}
\acom{c_{r}(\vec p\,)}{c_s^\dagger(\vec q\,)}=\acom{d_{r}(\vec p\,)}{d_s^\dagger(\vec q\,)}&=\delta_{rs}(2\pi)^3\delta(\vec p-\vec q\,)
\end{align}
for the ladder operators with $c_s(\vec p\,)\ket{0_\textrm{v}}=0$: The plane-wave spectral function can be written as the expectation value of the anticommutator of Volkov states with respect to the strong field vacuum $\ket{0_\textrm{v}}$, such that
\begin{align}
\rho_{\Psi,\textrm{v}}(x,y)=i\Braket{\acom{ \Psi_\textrm{v}(x)}{{\bar{\Psi}}_\textrm{v}(y)}}
=i(2\pi)\int\frac{\dif^4 q}{(2\pi)^4}\delta(q^2-m^2)\Big[\theta(q^0)  -\theta(-q^0)\Big]R_q(x)(\slashed q+m) \bar R_q(y)\, .
\end{align}
The symmetric part of the Volkov propagator (that has been discussed e.g.\ in Ref.~\cite{PhysRevD.83.065007}) is the commutator
\begin{align}
F_{\Psi,\textrm{v}}(x,y)=\tfrac{1}{2}\Braket{\com{\Psi_\textrm{v}(x)}{\bar{\Psi}_\textrm{v}(y)}}
&=\pi\int\frac{\dif^4 q}{(2\pi)^4}\delta(q^2-m^2)\Big[\theta(q^0)  +\theta(-q^0)\Big]R_q(x)(\slashed q+m) \bar R_q(y)\, .
\end{align}
The Volkov propagator \cite{PhysRevD.97.056028} is then built via the standard asymptotic state identity
\begin{align}
\Delta_\textrm{v}(x,y)=\rho_{\Psi,\textrm{v}}(x,y)-\tfrac{i}{2}F_{\Psi,\textrm{v}}(x,y)\,\sgn(x^0-y^0) =\Braket{0_\textrm{v}|\T \Psi_\textrm{v}(x) \bar{\Psi}_\textrm{v}(y)|0_\textrm{v} }\, ,
\end{align}
where $\T$ denotes ordinary time-ordering.

\tocless\section{\label{app:volkovcomp}Computation of plane-wave spectral components in lightcone gauge}

\tocless\subsection{\label{app:scalar}Computation of the scalar component and the dressed mass phase factor}

We compute the scalar component by first computing the the traces 
\begin{align}
\tfrac{1}{4}\textrm{tr}\{ (\slashed q +m)\}&=m\, ,\\
\tfrac{1}{4}\textrm{tr}\{ \slashed n\slashed \A(X+\tfrac{s}{2})(\slashed q+m)\}&=
\tfrac{1}{4}\textrm{tr}\{ (\slashed q+m)\slashed \A(X-\tfrac{s}{2})\slashed n \}=
\tfrac{1}{4}\textrm{tr}\{ \slashed n\slashed \A(X+\tfrac{s}{2})(\slashed q+m)\slashed \A(X-\tfrac{s}{2})\slashed n \}=0\, ,
\end{align}
where we have used that $n\cdot \A=0$. The scalar component of the plane-wave spectral function is therefore
\begin{align}
&\rho_{\textrm{S},\textrm{v}}(X,p)=m\, i(2\pi)\int_q\delta(q^2-m^2)\sgn(q^0)\int_s\, e^{i(p-q)s}e^{-i\N_q(X,s)}
\end{align}
with $\N_q$ defined by Eq.~\refe{Ndef}. Next we compute the integrals. The exponent in lightcone gauge, $\A^+=\A^-=0$, is
\begin{align*}
&S_q(X+\tfrac{s}{2})-S_q(X-\tfrac{s}{2})= -q\cdot s-\N_q(X,s)\\
&=-q^+s^--q^-s^++\vec q_\perp\cdot \vec s_\perp- \frac{1}{2q^-}\int_{-\tfrac{n\cdot s}{2}}^{\tfrac{n\cdot s}{2}}\dif \lambda \Big[-2e\vec \A_\perp(n\cdot X+\lambda)\cdot \vec q_\perp+e^2|\vec \A_\perp(n\cdot X+\lambda)|^2 \Big]\, .
\end{align*}
Since this expression is under the integral with
 \begin{align}
\label{eq:lightdelta}
\delta(q^2-m^2)=\frac{1}{|2q^-|}\delta\Big(q^+-\frac{|\vec q_\perp|^2+m^2}{2q^-}\Big)\, ,
\end{align}
 we can set $q^+=(|\vec q_\perp|^2+m^2)/2q^-$, such that under the integral
\begin{align*}
&\int\dif q^+[q\cdot s+\N_q(X,s)]\delta\Big(q^+-\frac{|\vec q_\perp|^2+m^2}{2q^-}\Big)\\
&=\frac{m^2s^-}{2q^-}+q^-s^+-\vec q_\perp\cdot \vec s_\perp+\frac{s^-}{2q^-}\Bigg\{|\vec q_\perp|^2-2e\frac{\vec q_\perp}{s^-}\cdot \int_{-\tfrac{n\cdot s}{2}}^{\tfrac{n\cdot s}{2}}\dif \lambda \Big[\vec \A_\perp(n\cdot X+\lambda)\Big]+\frac{e^2}{s^-}\int_{-\tfrac{n\cdot s}{2}}^{\tfrac{n\cdot s}{2}}\dif \lambda\Big[|\vec \A_\perp(n\cdot X+\lambda)|^2 \Big]\Bigg\}\, .
\end{align*}
Next we complete the square via 
\begin{align*}
|\vec q_\perp|^2-2e\frac{\vec q_\perp}{s^-}\cdot \int_{-\tfrac{s_-}{2}}^{\tfrac{s_-}{2}}\dif \lambda \,\vec \A_\perp(n\cdot X+\lambda)=\Big[\vec q_\perp-\frac{e}{s^-}\int_{-\tfrac{s_-}{2}}^{\tfrac{s_-}{2}}\dif \lambda \,\vec \A_\perp(n\cdot X+\lambda)\Big]^2-\frac{e^2}{(s^-)^2}\Big[ \int_{-\tfrac{s_-}{2}}^{\tfrac{s_-}{2}}\dif \lambda \,\vec \A_\perp(n\cdot X+\lambda) \Big]^2\, .
\end{align*}
Since $\vec q_\perp$ is also under the integral, we can simply substitute 
\begin{align}
\label{eq:perpsub}
\vec q_\perp\rightarrow \vec q_\perp+\frac{e}{(n\cdot s)}\int_{-\tfrac{n\cdot s}{2}}^{\tfrac{n\cdot s}{2}}\dif \lambda \,\vec \A_\perp(n\cdot X+\lambda)
\end{align}
without changing the $\dif^4 q$ measure or boundaries, such that
\begin{align*}
&\int\dif q^+[q\cdot s+\N_q(X,s)]\delta\Big(q^+-\frac{|\vec q_\perp|^2+m^2}{2q^-}\Big)= q^-s^+-\Big(\vec q_\perp+\frac{e}{s^-}\int_{-\tfrac{s_-}{2}}^{\tfrac{s_-}{2}}\dif \lambda \,\vec \A_\perp(n\cdot X+\lambda)\Big)\cdot \vec s_\perp+\frac{|\vec q_\perp|^2s^-}{2q^-}\,+\\
&\hspace{2cm}+\frac{s^-}{2q^-}\Bigg(m^2+\frac{e^2}{s^-}\int_{-\tfrac{n\cdot s}{2}}^{\tfrac{n\cdot s}{2}}\dif \lambda\Big[|\vec \A_\perp(n\cdot X+\lambda)| \Big]^2-\frac{e^2}{(s^-)^2}\Big[ \int_{-\tfrac{s_-}{2}}^{\tfrac{s_-}{2}}\dif \lambda \,\vec \A_\perp(n\cdot X+\lambda) \Big]^2\Bigg)\\
&=q^-s^+-\vec q_\perp\cdot \vec s_\perp+e\int_{X-\tfrac{s}{2}}^{X+\tfrac{s}{2}}\dif z^\mu \A_\mu(n\cdot z)+\frac{|\vec q_\perp|^2+\tilde m^2(X,s)}{2q^-}s^-\, ,
\end{align*}
where we have identified the dressed mass \refe{dressedmass} and the exponent of the Wilson line \refe{planewilson}. The substitution \refe{perpsub}, together with taking the $q^+$ integral over the delta function, changes the argument of the sign function to
\begin{align*}
\sgn(q^0)\rightarrow \sgn\Big(\tilde q^0(X,s)\Big)= \sgn(q^-)\, ,\hspace{0.3cm}\textrm{ with }\tilde q^0(X,s)\defeq \Big(\Big|\vec q_\perp+\frac{e}{(n\cdot s)}\int_{-\tfrac{n\cdot s}{2}}^{\tfrac{n\cdot s}{2}}\dif \lambda \,\vec \A_\perp(n\cdot X+\lambda)\Big|^2+m^2\Big)/2q^-+\tfrac{1}{2}q^-\, .
\end{align*}
Since here the field appears only under the absolute value, this is simply the sign function $\sgn(q^-)$ familiar from lightcone quantization and compensates the absolute value in the identity \refe{lightdelta}. The exact scalar plane-wave spectral function in position-space thereby is
\begin{align}
&\rho_{\Psi,\textrm{v},\textrm{S}}(X+\tfrac{s}{2},X-\tfrac{s}{2})=i\,m\,(2\pi)\int_q\delta(q^2-m^2)\sgn(p^0)\,e^{-i[q\cdot s+\N_q(X,s)]}\\
\nonumber
&=i\,m\,\W(X+\tfrac{s}{2},X-\tfrac{s}{2})\int\frac{\dif q^-}{(2\pi)}\frac{1}{2q^-}\int\frac{\dif^2 q_\perp}{(2\pi)^2}\,\exp\Bigg\{-i\Bigg(\frac{|\vec q_\perp|^2+\tilde m^2(X,s)}{2q^-}s^-+q^-s^+-\vec q_\perp\cdot \vec s_\perp\Bigg)\Bigg\}\, .
\end{align}
With this we can immediately identify the gauge-invariant part via $\rho_\Psi(x,y)=\W(x,y)\hat \rho_\Psi(x,y)$. Without gradient expansion, the Wilson line is exactly canceled and no additional substitution of $p\rightarrow p+e\A$ is necessary. The Wigner transform can easily be computed up to the $s^-$ integral via
\begin{align*}
\hat \rho_{\Psi,\textrm{v},\textrm{S}}(X,p)
=im\int\hspace{-0.1cm}\frac{\dif q^-}{(2\pi)}\frac{1}{2q^-}\int\hspace{-0.1cm}\frac{\dif^2 q_\perp}{(2\pi)^2}\int\dif s^- e^{ip^+ s^-}\hspace{-0.1cm}\exp\Bigg\{\hspace{-0.1cm}-i\Bigg(\frac{|\vec q_\perp|^2+\tilde m^2(X,s)}{2q^-}s^-\Bigg)\Bigg\}
(2\pi)\delta(p^--q^-) (2\pi)^2\delta(\vec p_\perp-\vec q_\perp)\, ,
\end{align*}
where we have used that $\tilde m$ only depends on $s^-$ not on $s^+, \vec s_\perp$. The result \refe{svolk} mentioned in the main text, follows after taking the trivial integrals over the delta functions. Next we prove that the leading order in gauge-invariant field-gradients of the function \refe{scalarlcfa} is equivalent to the Airy expressions \refe{scalarairy}. For this purpose we make use of
\begin{align}
\label{eq:airyid}
\int\dif \varphi\, e^{ia\varphi-b\varphi^3}=(3b)^{-1/3}\textrm{Ai}\big(-a(3b)^{-1/3}\big)\, .
\end{align}
We may apply this identity to Eq.~\refe{scalarlcfa} with
\begin{align}
\label{eq:abdef}
a=\frac{1}{\omega}\Big(p^+-\frac{|\vec p_\perp|^2+m^2}{2p^-}\Big)=\frac{1}{\omega}\frac{p^2-m^2}{2(n\cdot p)}\, ,\hspace{2cm} b=\frac{1}{24}\frac{\xi_0^3}{\chi_0(p)}\frac{\E^2(X)}{\F_0}
\end{align}
such that the prefactor and argument of the result \refe{scalarairy} are obtained via
\begin{align}
\frac{im}{\omega}\frac{(3b)^{-1/3}}{2(n\cdot p)}=im\Big((n\cdot p)e\E(X)\Big)^{-2/3}\, ,\hspace{1cm}
-a(3b)^{-1/3}=-(p^2-m^2)\Big((n\cdot p)e\E(X)\Big)^{-2/3}\, .
\end{align}
The free scalar component is obtained for $\E=0$ via
\begin{align}
\label{eq:urexp}
\frac{1}{2p^-}\int\frac{\dif\varphi}{\omega}\exp\Bigg\{i\Big(p^+-\frac{|\vec p_\perp|^2+  m^2}{2p^-}\Big)\frac{\varphi}{\omega}\Bigg\}
=(2\pi)\delta(p^2-m^2)\sgn(p^0)\,.
\end{align}

\tocless\subsection{\label{app:vector}Computation of the vector component}

We compute the vector component by first computing the traces
\begin{align}
\tfrac{1}{4}\textrm{tr}\{ \gamma^\mu(\slashed q +m) \}&=q^\mu\, ,\\
\tfrac{1}{4}\textrm{tr}\{ \gamma^\mu\slashed n\slashed \A(X+\tfrac{s}{2})(\slashed q+m) \}
&=n^\mu \A(X+\tfrac{s}{2})\cdot q-\A^\mu(X+\tfrac{s}{2})(n\cdot q)\, ,\\
\tfrac{1}{4}\textrm{tr}\{ \gamma^\mu(\slashed q+m)\slashed \A(X-\tfrac{s}{2})\slashed n \}
&=n^\mu \A(X-\tfrac{s}{2})\cdot q-\A^\mu(X-\tfrac{s}{2})(n\cdot q)\, ,\\
\tfrac{1}{4}\textrm{tr}\{ \gamma^\lambda\slashed n\slashed \A(X+\tfrac{s}{2})(\slashed q+m)\slashed \A(X-\tfrac{s}{2})\slashed n \}&=-2n^\lambda\A(X+\tfrac{s}{2})\cdot \A(X-\tfrac{s}{2})(n\cdot q)\, .
\end{align}
The vector component of the plane-wave spectral function therefore is
\begin{align}
&\rho^\mu_{\Psi,\textrm{v},\textrm{V}}(X,p)=i(2\pi)\int_q\delta(q^2-m^2)\sgn(q^0)\int_s\, e^{i(p-q)s}e^{-i\N_q(X,s)}\\
\nonumber
&\cross\Big\{ q^\mu -\frac{e}{2}[\A^\mu(X+\tfrac{s}{2})+\A^\mu(X-\tfrac{s}{2})]+n^\mu\frac{1}{2(n\cdot q)} e[\A(X+\tfrac{s}{2})+\A(X-\tfrac{s}{2})]\cdot q-n^\mu\frac{1}{2(n\cdot q)}e^2\A(X+\tfrac{s}{2})\cdot \A(X-\tfrac{s}{2})\Big\}
\end{align}
The last term involves a product of two fields which we can write as
\begin{align}
\A(X+\tfrac{s}{2})\cdot \A(X-\tfrac{s}{2})=\tfrac{1}{4}\big[\A(X+\tfrac{s}{2})+\A(X-\tfrac{s}{2})\big]^2-\tfrac{1}{4}\big[\A(X+\tfrac{s}{2})-\A(X-\tfrac{s}{2})\big]^2\, ,
\end{align}
such that we can identify the plane-wave momentum \refe{kinmom}
 at the field $\tfrac{1}{2}[\A^\mu(X+\tfrac{s}{2})+\A^\mu(X-\tfrac{s}{2})]$,
\begin{align}
\label{eq:pibar}
\bar\pi_q^\mu(X,s)\defeq q^\mu -e\tfrac{1}{2}[\A^\mu(X+\tfrac{s}{2})+\A^\mu(X-\tfrac{s}{2})]+n^\mu\Bigg(\frac{e\tfrac{1}{2}[\A(X+\tfrac{s}{2})+\A(X-\tfrac{s}{2})]\cdot q }{(n\cdot q)} -\frac{e^2\tfrac{1}{4}[\A(X+\tfrac{s}{2})+\A(X-\tfrac{s}{2})]^2}{2(n\cdot q)}\Bigg)
\end{align}
in the pre-exponential and make use of Eq.~\refe{diffavg} to write
\begin{align}
\tfrac{1}{4}\textrm{tr}\{ \gamma^\mu \K(X,l,p-ln)\}=\int\dif s^- e^{ils^-}e^{-i\N_{p}(X,s^-)}\Bigg[ \bar \pi^\mu_p(X,s)-ln^\mu-n^\mu\frac{1}{8}\frac{(n\cdot s)^2e^2\sbraket{\E}(X,s)}{(n\cdot p)}\Bigg]\, ,
\end{align}
from which the gauge-invariant vector spectral component \refe{invvec} used in the main text follows. To leading order in field gradients, we may drop the gauge-invariant higher orders of
\begin{align}
\tfrac{1}{2}[\A^\mu(X+\tfrac{s}{2})+\A^\mu(X-\tfrac{s}{2})]&=\A^\mu(X)+\tfrac{1}{8}(n\cdot s)^2\ddot\A^\mu(X)+\O\big(e^0(s\cdot \pdif_X)^3\big)\, ,\\
\A^\mu(X+\tfrac{s}{2})-\A^\mu(X-\tfrac{s}{2})&=(n\cdot s)\dot \A^\mu(X)+\tfrac{1}{24}(n\cdot s)^3{\A}^{\mu,(3)}(X)+\O\big(e^0(s\cdot \pdif_X)^4\big)\, ,
\end{align}
such that $\bar \pi_q(X,s)=\pi_q(X)+\O(e^0s\cdot\pdif_X)$ and
\begin{align}
\label{eq:kinvec}
\tfrac{1}{4}\textrm{tr}\{\gamma^\mu\tilde \K_q(X,s)\}&=\pi_q^\mu(X)-n^\mu\frac{1}{8}\frac{(n\cdot s)^2e^2\E^2(X)}{(n\cdot q)}+\O(e^0s\cdot\pdif_X)\, .
\end{align}

\tocless\subsection{\label{app:pseudo}Computation of the pseudoscalar component}

We compute the pseudoscalar component by first computing the traces
\begin{align}
\tfrac{1}{4}\textrm{tr}\{ \gamma^5(\slashed q +m)\}=
\tfrac{1}{4}\textrm{tr}\{ \gamma^5\slashed n\slashed \A(X+\tfrac{s}{2})(\slashed q+m)\}=
\tfrac{1}{4}\textrm{tr}\{ \gamma^5(\slashed q+m)\slashed \A(X-\tfrac{s}{2})\slashed n \}&=0\, ,\\
\tfrac{1}{4}\textrm{tr}\{ \gamma^5\slashed n\slashed \A(X+\tfrac{s}{2})(\slashed q+m)\slashed \A(X-\tfrac{s}{2})\slashed n \}=-i\varepsilon_{\mu\nu\sigma\rho}n^\mu\A^\nu(X+\tfrac{s}{2})\A^\sigma(X-\tfrac{s}{2})n^\rho&=0\, .
\end{align}
The pseudoscalar component therefore vanishes identically, $\rho_{\Psi,\textrm{v},\textrm{P}}(X,p)\equiv 0$ because $\varepsilon_{\mu\nu\sigma\rho} n^\mu n^\rho=0$.

\tocless\subsection{\label{app:axial}Computation of the axial component}

We compute the axial component by first computing the traces
\begin{align}
\tfrac{1}{4}\textrm{tr}\{ \gamma^5\gamma^\mu(\slashed q +m)\}&=0\, ,\\
\tfrac{1}{4}\textrm{tr}\{ \gamma^5\gamma^\mu\slashed n\slashed \A(X+\tfrac{s}{2})(\slashed q+m)\}
&=-in^\rho \A^\sigma(X+\tfrac{s}{2})q^\nu\varepsilon_{\mu\rho\sigma\nu}\,,\\
\tfrac{1}{4}\textrm{tr}\{ \gamma^5\gamma^\mu(\slashed q+m)\slashed \A(X-\tfrac{s}{2})\slashed n \}
&=-in^\rho \A^\sigma(X-\tfrac{s}{2})q^\nu\varepsilon_{\mu\nu\sigma\rho}\, ,\\
\tfrac{1}{4}\textrm{tr}\{ \gamma^5\gamma^\mu\slashed n\slashed \A(X+\tfrac{s}{2})(\slashed q+m)\slashed \A(X-\tfrac{s}{2})\slashed n \}
&=2in^\mu n^\lambda \A^\nu(X+\tfrac{s}{2})q^\rho\A^\sigma(X-\tfrac{s}{2})\varepsilon_{\lambda \nu \rho\sigma}\, .
\end{align}
The axial component of the plane-wave spectral function is therefore\\
\begin{align}
\rho_{\Psi,\textrm{v},\textrm{A}}^\mu(X,p)&=(2\pi)\int_q\delta(q^2-m^2)\sgn(q^0)\int_s\, e^{i(p-q)s}\, e^{-i\N_q(X,s)}\\
\nonumber
&\hspace{-1cm}\cross \Bigg\{\frac{e}{2}\frac{1}{n\cdot q}  \varepsilon^{\mu}{}_{\nu\rho\sigma}n^\rho q^\nu\Big[\A^\sigma(X+\tfrac{s}{2})-\A^\sigma(X-\tfrac{s}{2})\Big]-\frac{e^2}{4}\frac{1}{(n\cdot q)^2}2n^\mu n^\lambda \A^\nu(X+\tfrac{s}{2})q^\rho\A^\sigma(X-\tfrac{s}{2})\varepsilon_{\lambda \nu \rho\sigma}\Bigg\}\, .
 \end{align}
 
\tocless\subsection{Computation of the tensor component}

We compute the tensor component by first computing the traces
\begin{align}
\tfrac{1}{4}\textrm{tr}\{ \sigma^{\mu\nu}\slashed n\slashed \A(X+\tfrac{s}{2})(\slashed q+m)\slashed \A(X-\tfrac{s}{2})\slashed n \}&=\tfrac{1}{4}\textrm{tr}\{ \sigma^{\mu\nu}(\slashed q +m)\}=0\, ,\\
\tfrac{1}{4}\textrm{tr}\{ \sigma^{\mu\nu}\slashed n\slashed \A(X+\tfrac{s}{2})(\slashed q+m)\}&=im\Big[ n^\nu \A^\mu(X+\tfrac{2}{2})-n^\mu \A^\nu(X+\tfrac{s}{2}) \Big]\, ,\\
\tfrac{1}{4}\textrm{tr}\{ \sigma^{\mu\nu}(\slashed q+m)\slashed \A(X-\tfrac{s}{2})\slashed n \}&=-im\Big[ n^\nu \A^\mu(X-\tfrac{2}{2})-n^\mu \A^\nu(X-\tfrac{s}{2}) \Big]\, .
\end{align}
The tensor component of the plane-wave spectral function is therefore
\begin{align}
\rho_{\Psi,\textrm{v},\textrm{T}}^{\mu\nu}(X,p)&=m\,(2\pi)\int_q\delta(q^2-m^2)\sgn(q^0)\int_se^{i(p-q)s}e^{-i\N_q(X,s)}\\
\nonumber
&\cross\frac{1}{2(n\cdot q)}e\Big\{n^\mu\Big[\A^\nu(X+\tfrac{s}{2})-\A^\nu(X-\tfrac{s}{2})\Big]-n^\nu\Big[\A^\mu(X+\tfrac{s}{2})-\A^\mu(X-\tfrac{s}{2})\Big]\Big\}
\end{align}
To leading order in field gradients, the gauge-invariant tensor component is
\begin{align}
\hat \rho_{\Psi,\textrm{v},\textrm{T}}^{\mu\nu}(X,p)
&=m\,e\F^{\mu\nu}(X)\frac{1}{(2p^-)^2}\int\frac{\dif \varphi}{\omega}\,\frac{\varphi}{\omega}\,\exp\Bigg\{i\Big(p^+-\frac{|\vec p_\perp|^2+ m^2}{2p^-}\Big)\frac{\varphi}{\omega}-i\frac{1}{24}\frac{\xi_0^3}{\chi_0}\frac{|\vec {\mathcal{E}}(X)|^2}{\F_0^2}\varphi^3\Bigg\}+\O(e^0\pdif_p\cdot\pdif_X)\, .
\end{align}

\tocless\section{\label{app:planemaxwell}Maxwell current for plane-wave fields}
For the case of an external plane-wave field, the Maxwell current \refe{jX} can be written solely in terms of the plane-wave degrees of freedom \refe{lminus} and \refe{lplus}:
\begin{align}
\label{eq:jplane}
j_\mu(X)&=-4e\int\frac{\dif l}{(2\pi)}\int\frac{\dif^3p}{(2\pi)^3}\frac{1}{2\varepsilon_{\vec p\,}(l)}\\
\nonumber
&\cross\Big\{\tfrac{1}{2}\Big[\mathcal{K}^-_\mu(X,l,\vec p\,)+\mathcal{K}^+_\mu(X,l,\vec p\,)\Big]+\mathcal{K}^-_\mu(X,l,\vec p\,)f_\Psi^-(X,l,\vec p\,)+\mathcal{K}^+_\mu(X,l,\vec p\,)f_\Psi^+(X,l,\vec p\,)\Big\}
\end{align} 
with the fermion and anti-fermion drift kernels%of \refe{Ke} and \refe{Kp}.
\begin{align}
\label{eq:Ke}
\K^-_\mu(X,l,\vec p\,)&\defeq \tfrac{1}{4}\textrm{tr}\{\gamma_\mu\K(X,l,p-ln)\}\,\,\,\textrm{ at }\,\,\, p^0=l+\varepsilon_{l}(\vec p\,)\, ,\\
\label{eq:Kp}
\K^+_\mu(X,-l,-\vec p\,)&\defeq \tfrac{1}{4}\textrm{tr}\{\gamma_\mu\K(X,l,p-ln)\}\,\,\,\textrm{ at }\,\,\, p^0=l-\varepsilon_{l}(\vec p\,)\, .
\end{align}
The zero-field current \refe{j0} of Sec.~\ref{sec:3a} may be obtained as the special case of
\begin{align}
\label{eq:Kveczerolim}
\K_\mu^\mp(X,l,\vec p\,)&\xrightarrow{\A\iv\rightarrow 0}\pm(2\pi)\delta(l)\,p_\mu \,\textrm{ at }\,p^0=\varepsilon(\vec p\,)\, .
\end{align}
\end{widetext}
\bibliographystyle{ieeetr}
\bibliography{bibref}
\end{document}